# The Reaction between Atomic Carbon and Molecular Nitrogen as a Source of Cyanamide and Carbodiimide on Interstellar Ices

Kevin M. Hickson,*,[1] Jean-Christophe Loison[1] and Audrey Coutens[2]

[1] Univ. Bordeaux, CNRS, Bordeaux INP, ISM, UMR 5255, F-33400 Talence, France

[2] Institut de Recherche en Astrophysique et Planétologie, Université de Toulouse, CNRS, CNES, 9 av. du Colonel Roche, 31028 Toulouse Cedex 4, France

Email: kevin.hickson@u-bordeaux.fr

**Abstract**

Reactions occurring on the ice-covered surfaces of interstellar dust grains are considered to be among the most important sources of complex species in the interstellar medium. Despite this, molecules such as cyanamide, $NH_2CN$, are largely underpredicted by current astrochemical models suggesting that the network of reactions currently used to describe this species and its tautomer carbodiimide, HNCNH, are incomplete. Here, we performed a theoretical investigation of the reaction of ground state atomic carbon $C(^3P)$ with molecular nitrogen $N_2$ in both the gas-phase and on the surface of amorphous solid water (ASW) clusters to examine its potential importance in the formation of $NH_2CN$ and HNCNH. We show that the reaction of gas-phase C-atoms with $N_2$ molecules already present on the ASW surface results in the barrierless formation of CNN. Following exothermic hydrogenation reactions, the N-N bond of the C-N-N bearing intermediates is broken allowing the formation of molecules with N-C-N backbones through cyclic intermediates over low barriers. To test the importance of these processes to $NH_2CN$ and HNCNH formation, these reactions were included in a three-phase astrochemical model of low-mass protostellar evolution employing a reaction network that was updated to better describe the formation and destruction pathways of related small nitrogen bearing molecules. These simulations demonstrate that the ice surface reaction between C and $N_2$ represents by far the dominant source of $NH_2CN$ and HNCNH in protostellar environments and in dense clouds.

# 1 Introduction

Molecular nitrogen, $N_2$, is predicted to be an important reservoir for elemental nitrogen in the dense interstellar medium (ISM), with several gas-phase reactions involving atomic nitrogen in its ground $^4S$ electronic state (hereafter referred to as N) and small nitrogen bearing radicals thought to control the partitioning between N and $N_2$.[1-3] In its molecular form, nitrogen is mostly unreactive in the gas-phase ISM, participating only in ion-molecule type reactions with cationic species such as $H_3^+$ and $He^+$ to yield $N_2H^+$ and $N^+$ respectively.[4] As a result, elementary nitrogen in the form of $N_2$ is predicted to contribute very little to the overall chemistry of these regions. In current models, which adopt an elementary nitrogen abundance of around $1.2 \times 10^{-4}$ with respect to $H_2$, gas-phase $N_2$ abundances account for approximately 10 % of the total nitrogen at times corresponding to typical dense could ages ($2 \times 10^5$ years). Nevertheless, as $N_2$ depletes from the gas-phase through collisions with dust grains, we also need to consider the fraction of $N_2$ that is present on interstellar ices; a figure that is supplemented by the recombination of mobile nitrogen atoms on the surface. Indeed, modelled ice phase $N_2$ abundance levels reach values as high as $2.6 \times 10^{-5}$, accounting for approximately 40 % (two nitrogen atoms for each $N_2$) of elemental nitrogen in this form. While $N_2$ is particularly unreactive in the gas-phase, previous work on other stable interstellar reservoir species, such as CO, has shown that the ice surface can have a dramatic effect on the overall reactivity. Indeed, the activated reaction of CO with atomic hydrogen on interstellar ices has been shown to be an important source of HCO radicals, with the water ice surface (I) playing a minor role in lowering the activation barrier (II) stabilizing the energized HCO radical and (III) promoting reaction by tunneling.[5] In this way, sequential hydrogenation reactions of CO have been shown to lead to both $H_2CO$ and $CH_3OH$ formation,[6] with this pathway representing the major source of methanol in current models.[5] Similarly, experiments[7] and theory[8] have shown that the reaction between CO and atomic carbon in its ground electronic state, $C(^3P)$, (hereafter C) is barrierless and leads to the formation of CCO which can then participate in hydrogenation reactions to form larger organic species such as ketene, $H_2CCO$ and acetaldehyde, $CH_3CHO$ on the ice surface. It should be noted here that CCO formation by the C + CO reaction is also expected to occur in the gas-phase, but in the absence of exothermically accessible bimolecular exit channels, the CCO molecule formed with more than 200 kJ/mol of internal energy[9] will simply redissociate to C + CO. The crucial difference when reaction occurs on the ice surface is the capacity of the ice to allow some or all of the energy released by CCO formation to be dissipated into the bulk.

Given the potential for the ice to promote association reactions, we wanted to examine the potential influence of the ice surface on $N_2$ reactivity with a particular focus on its reaction with atomic carbon. In the gas-phase, the reaction of $N_2$ with C-atoms leads to the formation of the diazocarbene radical, CNN. Recent MRCI calculations of the $^3A''$ potential energy surface of this process based on CASSCF calculations employing large basis set functions confirm that CNN formation is barrierless, with CNN predicted to be 107 kJ/mol more stable than the separated reagents C and $N_2$.[10, 11] As the exit channel of this reaction, leading to N + CN radicals as products, is endothermic with respect to the C + $N_2$ entrance channel by more than 200 kJ/mol, the only possible outcome for this process at low temperature is association to form CNN. Consequently, under the low pressure and low temperature conditions of the ISM, the energized CNN radical will simply redissociate to C + $N_2$ in a similar manner to gas-phase CCO described above. Nevertheless, if this reaction takes place on the ice surface, the CNN radical could be effectively stabilized by transferring a significant fraction of the excess energy to the bulk. As a result, the CNN radical would be available to react with other species present on the surface or arriving from the gas-phase such as atomic hydrogen. In particular, CNN could be a potential precursor for the formation of cyanamide, $NH_2CN$, and its tautomer carbodiimide, HNCNH, although the formation of these species would require an initial reorganization of the C-N-N backbone into an N-C-N one. $NH_2CN$ was first detected in the ISM by Turner et al. in dark cloud Sgr B2.[12] More recently, Coutens et al detected $NH_2CN$ towards the solar-type protostars IRAS 16293–2422 B and NGC 1333 IRAS2A, deriving abundances of $2 \times 10^{-10}$ and $5 \times 10^{-11}$ respectively.[13] HNCNH was first detected by McGuire et al. via its maser emission features towards Sgr B2(N).[14]

For CNN to form on the surface of interstellar ice from C and $N_2$, three different scenarios need to be considered. The first two situations involve one of the reagents already bound to the surface with the other reagent accreting from the gas-phase in an Eley-Rideal type mechanism, colliding directly with the absorbed species. The final situation that should be considered is one where the two species are already present on the surface and can then react by diffusion through a Langmuir-Hinshelwood type process or because they are already present in adjacent absorption sites. $N_2$ has been shown to be only weakly bound to the surface of amorphous solid water (ASW) with experimentally[15] and theoretically[16-18] determined binding energies of around 1200 K. Despite this, the binding energy is thought to be large enough to prohibit $N_2$ diffusion at typical dense cloud temperatures (10 K), with only atomic hydrogen, atomic nitrogen and $H_2$ displaying significant mobility at these temperatures. In contrast, atomic carbon has been shown to form a relatively strong covalent type bond with water on ASW with typical binding energies[18-21] of around 10000 K, prohibiting the diffusion of atomic carbon on ASW (a recent experimental and theoretical study has provided some evidence however, regarding the possible diffusion of atomic carbon on interstellar ice analogues which would mean that some of these atoms could be available to react even if they are already present on the surface).[22] Furthermore, recent theoretical studies by Ferrero et al. of the reactivity of atomic carbon already bound to water (C-OH$_2$) indicate that reaction with closed shell molecules such as $NH_3$, CO, $CO_2$ and $H_2$ does not occur.[23] These findings suggest that when atomic carbon is already present on the ice surface, it is unlikely to be available for reaction with either $N_2$ coming from the gas-phase or when $N_2$ is in an adjacent surface site. Indeed, the only seemingly credible scenario for the in situ formation of CNN on the ice surface is the one where C arrives from the gas-phase to react in an Eley-Rideal type mechanism with $N_2$ pre-adsorbed on the surface. A similar methodology was used by Molpeceres et al.[24] who studied the reactivity of C with $NH_3$ already bound to a $(H_2O)_{14}$ water cluster. In contrast to the results of Ferrero et al.,[23] Molpeceres et al.[24] showed that C reacts with $NH_3$ in the absence of a barrier to form a $CNH_3$ intermediate with a strong covalent C-N bond.

In this study, we studied the various pathways for $NH_2CN$ and HNCNH formation, starting from $N_2$ bound to a water ice cluster in different adsorption sites, followed by reaction with C-atoms accreting from the gas-phase. We then examined the various isomerization and hydrogenation steps required to form the final products. These pathways are then included in a state of the art three-phase gas-grain astrochemical model alongside an updated reaction network to examine whether $NH_2CN$ and HNCNH can be produced efficiently in the interstellar medium.

## 2 Theoretical Methods

The calculations described in the present paper were performed using an ASW ice cluster of 18 water molecules as described in the work of Rimola et al.[5] and in later studies by these authors.[25, 26] This cluster size was preferred over other ones available due to the good diversity of adsorption sites for a comparatively low computational cost. The cluster was initially optimized at the present level of theory prior to use.

All of the calculations reported here were performed using ORCA version 6.0.1.[27, 28] Specifically, density functional theory (DFT) was used to obtain the structures, employing the $\omega$B97X functional with the D3 dispersion correction[29] coupled with the Karlsruhe group triple-$\zeta$ def2-TZVP basis set. This functional-basis set combination has been seen in previous work to exhibit good performance for similar systems.[26] An unrestricted formalism was employed to treat open-shell systems. Geometry optimizations were performed to determine the stationary points on the potential energy surface which were subsequently verified through analytical harmonic frequency calculations. Transition state (TS) structures, were either located through the standard eigenvector following TS optimization procedure as called by the keyword OptTS, or through climbing image nudged elastic band methods (CI-NEB) followed by TS optimization using the keyword NEB-TS.[30] In this way, TS structures were confirmed to possess a single imaginary frequency, while stable intermediate structures were verified to display only positive values for the harmonic frequencies. Additional settings that were used for all calculations were TightSCF coupled with the defgrid3 integration grid. Furthermore, to account for artificial overbinding effects brought about by basis set superposition errors, due to the finite size of the basis sets used, a semi-empirical correction was applied which approximates the Boys and Bernardi geometrical counterpoise correction (gCP) in the intermolecular case[31] using the ORCA keyword gCP(DFT/TZ). In this way, the optimized energy $E_{OPT}$ of any particular cluster was composed of the sum of three terms, the final single point energy, $E$, the gCP correction and the D3 dispersion correction.

As harmonic frequency calculations were performed for all optimized structures, the zero-point energy (ZPE) contributions were also determined. In this way, the relative energies $E_{REL}$, such as those for a reaction or isomerization process going from reactants A to products B, could be calculated by the formula

$$E_{REL} = E_{OPT}(B) + ZPE_B - (E_{OPT}(A) + ZPE_A) = E_{OPT}(B) - E_{OPT}(A) + \Delta ZPE \quad (1)$$

So, for the reaction of atomic hydrogen with a CNN radical already absorbed on the ASW cluster to form HNNC, this becomes

$$E_{REL} = E_{(H_2O)_{18}\cdots HNNC} + ZPE_{(H_2O)_{18}\cdots HNNC} - (E_{(H_2O)_{18}\cdots CNN} + ZPE_{(H_2O)_{18}\cdots CNN} + E_{H(g)})$$

where $E_{(H_2O)_{18}\cdots HNNC}$ and $E_{(H_2O)_{18}\cdots CNN}$ are the optimized energies of the $(H_2O)_{18}\cdots HNNC$ and $(H_2O)_{18}\cdots CNN$ complexes, $ZPE_{(H_2O)_{18}\cdots HNNC}$ and $ZPE_{(H_2O)_{18}\cdots CNN}$ are the zero-point energies of these optimized structures and $E_{H(g)}$ is the single-point energy of an isolated (gas-phase) hydrogen atom. The changes in energy brought about by deformation of the water cluster and/or the adsorbed molecule were not considered in the present work as these differences were found to be very small compared with both the overall energy change and the likely error in the calculated reaction energy.

Reactions involving the association of two radical species were studied using the broken symmetry formalism as integrated in ORCA using the keyword flipspin to converge to the high-spin state initially, with the orbitals generated then used to converge to the low-spin state solution. All structures were manipulated and viewed using the molecule editing and visualization software Avogadro.[32]

## 3 Results and Discussion

### 3.1 N₂ binding sites and reaction with atomic carbon

Three different binding sites were examined during the course of this work, denoted a-c as shown in Figure 1.

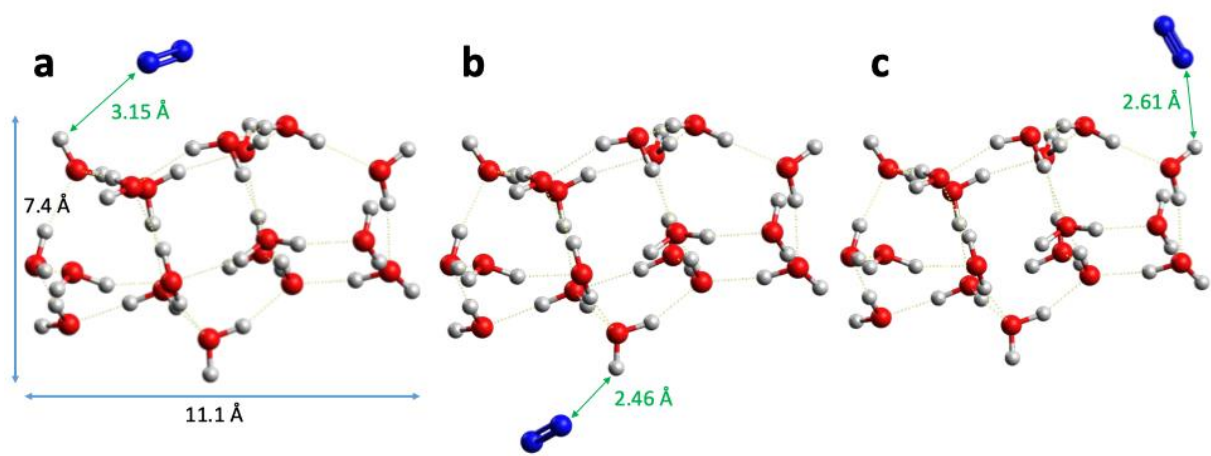

**Figure 1** The (H₂O)₁₈ cluster used in the present work, optimized at the $\omega$B97X-D3/def2-TZVP level. Panels a, b, and c represent the three different initial N₂ adsorption sites.

In site a, the $N_2$ molecule is positioned above the centre of a ring of 6 H-bonded $H_2O$ molecules with the nearest (dangling) H-atom at a distance of 3.15 Å. For site b, the $N_2$ molecule is positioned directly above two $H_2O$ molecules in the cluster forming part of a ring of 6 H-bonded $H_2O$ molecules with the nearest (dangling) H-atom at a distance of 2.46 Å. For site c, the $N_2$ molecule is positioned directly above a different two $H_2O$ molecules in the cluster with the nearest (dangling) H-atom at a distance of 2.61 Å. The calculated binding energies of the three sites are 1115, 1024 and 856 K for sites a, b and c respectively (relative to the bare cluster and gas-phase $N_2$); values that are in good agreement with those already reported in the literature for $N_2$. [15-18]

In order to examine the reactivity of C with $N_2$ on the ice surface, atomic carbon was placed at a distance of 2.5 Å from $N_2$ and the whole system was optimized over the triplet potential energy surface. In all three cases, the CNN radical was readily formed indicating the absence of a barrier for CNN formation, a result that was also observed for the formation of CNN in the gas-phase from C + $N_2$.[10, 11] The ZPE corrected energies of the $(H_2O)_{18}$...CNN system relative to $(H_2O)_{18}$...$N_2$ + $C_{(g)}$, corresponding to the reaction energy for CNN formation were calculated to be -177, -177 and -180 kJ/mol for sites a, b and c respectively. One of the $(H_2O)_{18}$...CNN clusters formed is shown in Figure 2.

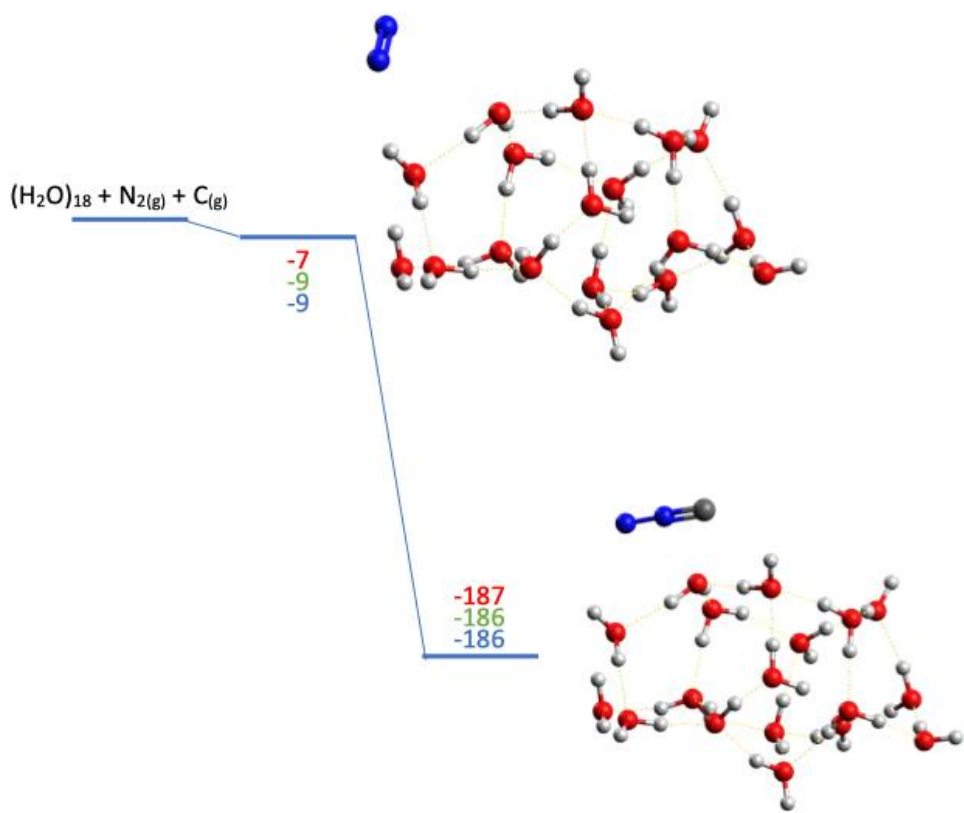

**Figure 2** Binding energies for $N_2$ and reaction energies for the formation of $(H_2O)_{18}...CNN$ at the three different adsorption sites. The clusters shown here are those corresponding to site c. All energies are corrected for ZPE differences in kJ/mol with values in blue, green and red representing the calculated energies at sites a, b and c respectively, relative to the bare cluster and isolated $N_2$ and C.

In order to verify that the opposite process (($H_2O)_{18}...C + N_{2(g)}$) does not take place (that is, the reaction of $N_2$ coming from the gas-phase with C atoms already present on the water ice cluster), we first placed atomic carbon at 2.5 Å from the water cluster at the same three sites. For sites a and c, atomic carbon was seen to form a strong bond with the O atom of one of the nearest $H_2O$ molecules (C-O = 1.545 Å and 1.505 Å for sites a and c respectively) with calculated binding energies of 12442 and 13118 K for sites a and c respectively, in good agreement with previous work.[18-21] For site b, the carbon atom was seen to spontaneously insert into an O-H bond of one of the $H_2O$ molecules to form HCOH. Now, when we place $N_2$ at 2.5 Å from the C-atom for sites a and c, the $N_2$ molecule drifts away from the strongly bound C-atom eventually forming a weak bond with the water cluster with no reaction occurring with the carbon atom.

**3.2 Gas-phase reaction pathways to guide the surface reactivity studies**

As several pathways are available for the formation of $NH_2CN$ from CNN radicals present on the surface, there are potentially a large number of computationally expensive calculations to perform to derive the most energetically favorable ones. To examine which pathways are likely to be the most efficient ones as a guide for the cluster calculations, we conducted preliminary studies at the same level of theory ($\omega$B97X-D3/def2-TZVP) in the absence of the water ice cluster. Two intermediate steps must be accomplished for all pathways transforming the CNN radical (or hydrogenated equivalent) on the surface into NCN (or hydrogenated equivalent). (I) CNN (or its hydrogenated equivalent) must first be able to isomerize into a cyclic intermediate species c-CNN (or its hydrogenated equivalent). (II) N-N bond fission must occur in the intermediate c-CNN species (or its hydrogenated equivalent) to form NCN (or its hydrogenated equivalent). As hydrogenation of the CNN and NCN radicals can occur at various stages of the process, we therefore need to consider several different scenarios as discussed below:

### 3.2.1 CNN → NCN

The isomerization of CNN into NCN occurs in the absence of the initial hydrogenation of the CNN backbone. Here the energy available to overcome the various activation barriers for isomerization can be considered as equal to the exothermic energy release of the previous reaction, that is the energy released by the C + N$_2$ association step in this case. The calculated triplet potential energy surface for the CNN → NCN gas-phase isomerization is shown in Figure 3.

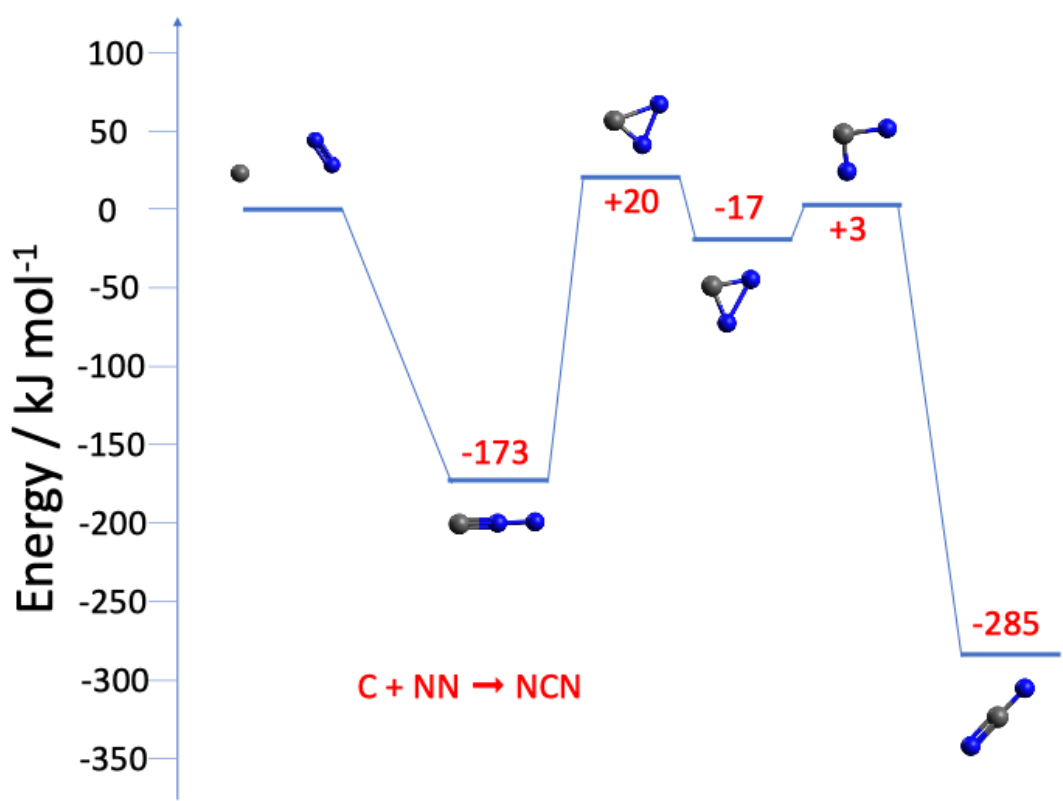

**Figure 3** Potential energy surface for the gas-phase formation of NCN by the C + N$_2$ reaction calculated at the $\omega$B97X-D3/def2-TZVP level. Energies are in kJ/mol and have been corrected for ZPE differences.

The C + N$_2$ → CNN reaction is barrierless, with an overall exothermicity of 173 kJ/mol at the $\omega$B97X-D3/def2-TZVP level. Previous work[10, 11] using a multireference method (MRCI-F12+Q) based on CASSCF wavefunctions and a larger set of basis functions (aug-cc-pVTZ) derived a smaller overall exothermicity of 107 kJ/mol for this process. In the current work, both the cyclization step and the N-N bond breaking step are predicted to be endothermic, with energies of +20 kJ/mol and +3 kJ/mol, although the energies calculated here are likely to underestimate the barriers considering the values derived in earlier studies (+73 kJ/mol and +37 kJ/mol from Lu et al.[8] for the cyclization and N-N bond breaking steps respectively). The gas-phase energies for this isomerization indicate that this process is unlikely to be important for the formation of NCN (and $NH_2CN$ by subsequent hydrogenation steps) due the high energy of the TSs connecting the various intermediates, unless the water cluster plays an important role in stabilizing these species. This possibility is examined below in section 3.3.1.

**3.2.2 H + CNN → HCNN/HNNC → HNCN**

The CNN radical formed by the C + $N_2$ reaction is hydrogenated at either the terminal N atom or at the C atom end. Here, the energy available to overcome the various activation barriers for isomerization is equal to the exothermic energy release of the previous reaction, that is, the energy released by the H + CNN association step in this case, where H can react at either the terminal C end or the terminal N end of the CNN molecule. The calculated doublet potential energy surface for the HCNN/HNNC → HNCN gas-phase isomerization is shown in Figure 4.

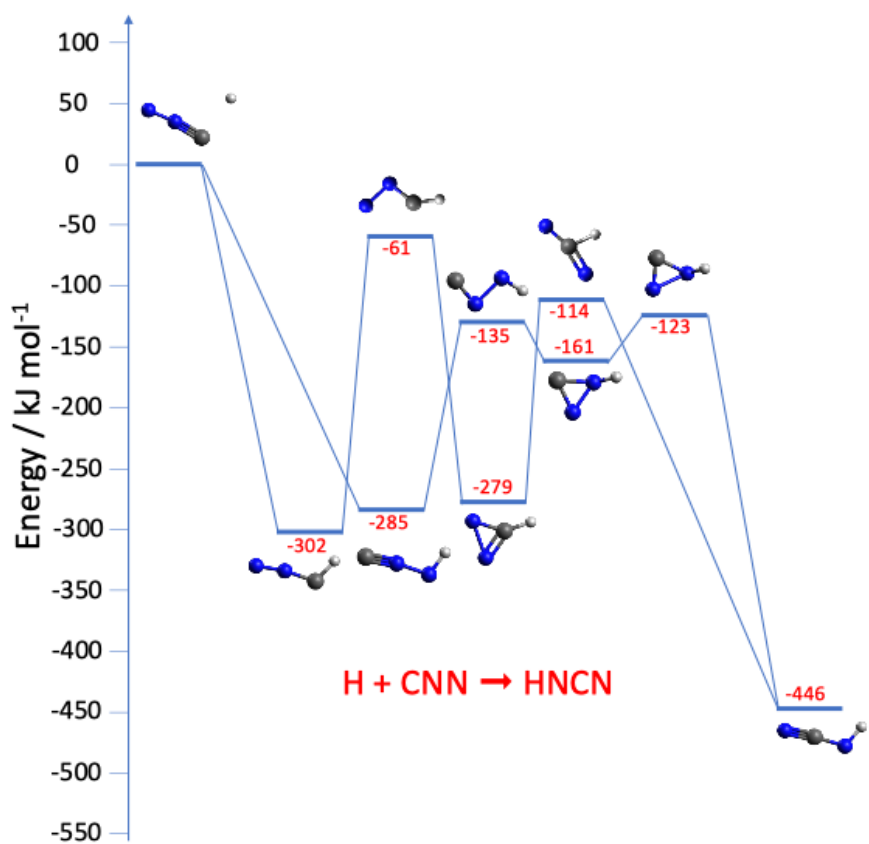

**Figure 4** Potential energy surface for the gas-phase formation of HNCN by the H + CNN reaction calculated at the $\omega$B97X-D3/def2-TZVP level. Energies are in kJ/mol and have been corrected for ZPE differences.

The H + CNN hydrogenation steps are calculated to be considerably more exothermic than the equivalent C + N$_2$ one described earlier (-302 kJ/mol and -285 kJ/mol for HCNN and HNNC formation respectively). Nevertheless, the formation of c-HCNN from HCNN involves a rather high barrier (-61 kJ/mol) potentially making this process a less favorable one. Interestingly, the N-N bond breaking step here is accompanied by H-atom transfer from the C-atom to one of the N-atoms leading to HNCN formation rather than the formation of a NCHN type species. For HNNC, the barriers for formation of the cyclic intermediate and N-N bond breaking steps are somewhat lower (-135 kJ/mol and -123 kJ/mol respectively). Consequently, this could represent a potentially important pathway for NH$_2$CN and NHCHCN formation on the ASW surface as the final hydrogenation steps over the singlet potential energy surface are barrierless and exothermic.

### 3.2.3 H + HCNN/HNNC → HCNNH/NH$_2$NC → NH$_2$CN

As with the previous example, the energy available to overcome the various isomerization steps comes from the exothermicity of the last hydrogenation step, that is, the H + HCNN/HNNC reaction. The calculated singlet potential energy surface for the HCNNH/NH$_2$NC → NH$_2$CN gas-phase isomerization is shown in Figure 5.

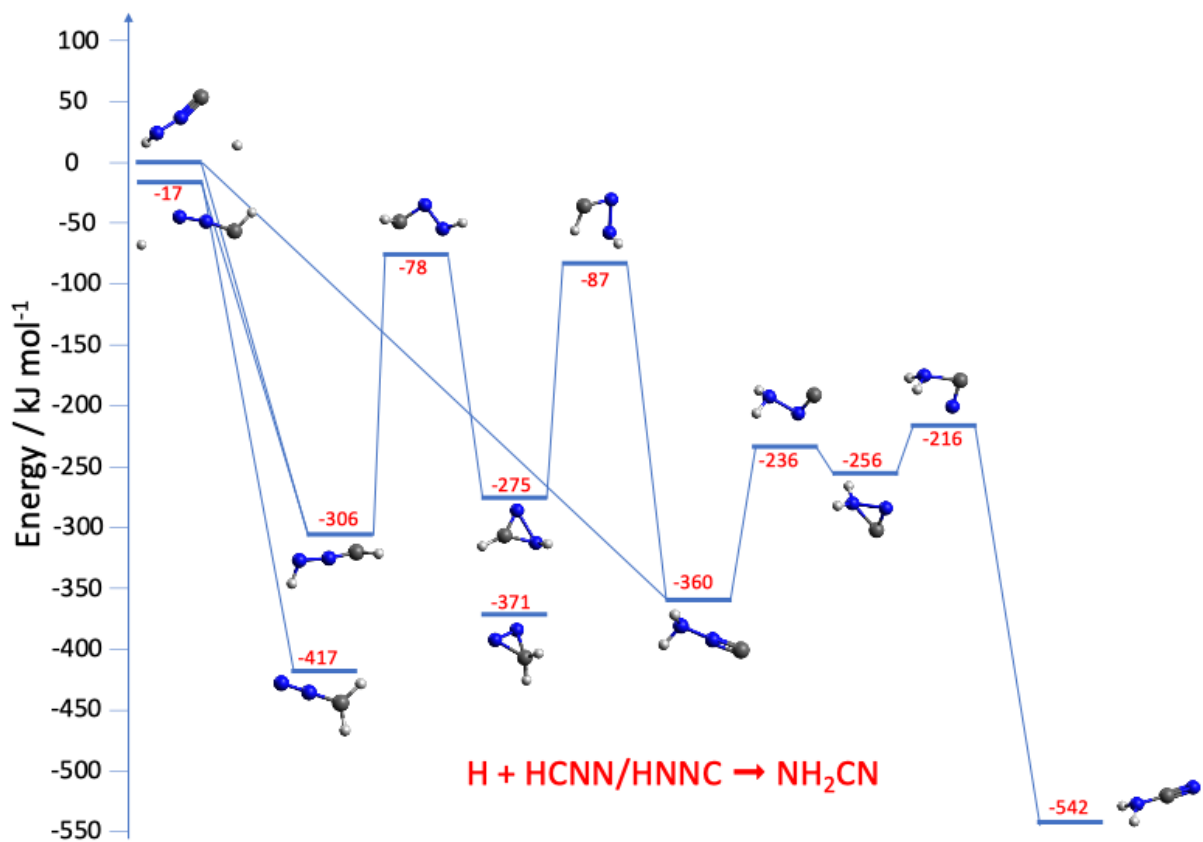

**Figure 5** Potential energy surface for the gas-phase formation of NH$_2$CN by the H + HNNC and H + HCNN reactions calculated at the $\omega$B97X-D3/def2-TZVP level. Energies are in kJ/mol and have been corrected for ZPE differences.

It can be seen from Figure 5 that when hydrogenation of HCNN occurs at the terminal N to form HCNNH, the TSs for cyclization and H-atom transfer are rather high (-78 and -87 kJ/mol respectively), leading to NH$_2$NC formation initially. In contrast, hydrogenation of HNNC at the nitrogen end leads directly to the barrierless formation of NH$_2$NC which can subsequently undergo cyclization and bond breaking to form NH$_2$CN over low barriers (-236 and -216 kJ/mol respectively). For the H + HCNN reaction, there is a second barrierless pathway (not shown in Figure 5) that produces CH$_2$NN. However, CH$_2$NN rapidly evolves to form CH$_2$ + N$_2$ and thus does not contribute to the production of NH$_2$CN or HNCNH. This reaction is included in our global model but is not discussed in further detail.

The calculations on these gas-phase isomerization processes indicate two potentially important pathways for NH$_2$CN formation that we should consider for further calculations on the ASW cluster. These are H + CNN → HNNC → HNCN followed by hydrogenation of HNCN to form $NH_2CN$ and HNCNH and H + HNNC → $NH_2NC$ → $NH_2CN$. In addition, there is also the possibility of the interconversion of HNCNH and $NH_2CN$ via a concerted type mechanism on the ice surface according to the earlier work of Tordini et al.,[33] representing an alternative pathway that is inaccessible in the gas-phase. Here, HNCN formed through the equivalent ice surface mechanism to the one shown in Figure 4 can undergo hydrogenation at either end of molecule as discussed above and then exchange one of its H-atoms with water ice. With the help of the H-bonding network of $H_2O$ molecules, a H-atom can then be donated back to the N-C-N bearing molecule at the other terminal N-atom. This effect will be discussed below in section 3.3.4. Now we will discuss the likely main $NH_2CN$/HNCNH formation routes on the ASW cluster based on the gas-phase results.

**3.3 CNN reactivity on ASW ice clusters**

**3.3.1 $(H_2O)_{18}$...$N_2$ + $C_{(g)}$ → $(H_2O)_{18}$...CNN → $(H_2O)_{18}$...c-CNN → $(H_2O)_{18}$...NCN**

In these simulations, atomic carbon was placed at a distance of 2.5 Å from the nitrogen molecule and the whole system was optimized. Reaction between C and $N_2$ was seen to occur spontaneously, indicating the barrierless nature of the reaction in a similar manner to the isolated gas-phase process. The calculated structures and energies for the reaction of C with $(H_2O)_{18}$...$N_2$ and the subsequent isomerization of $(H_2O)_{18}$...CNN to $(H_2O)_{18}$...NCN are shown in Figure 6.

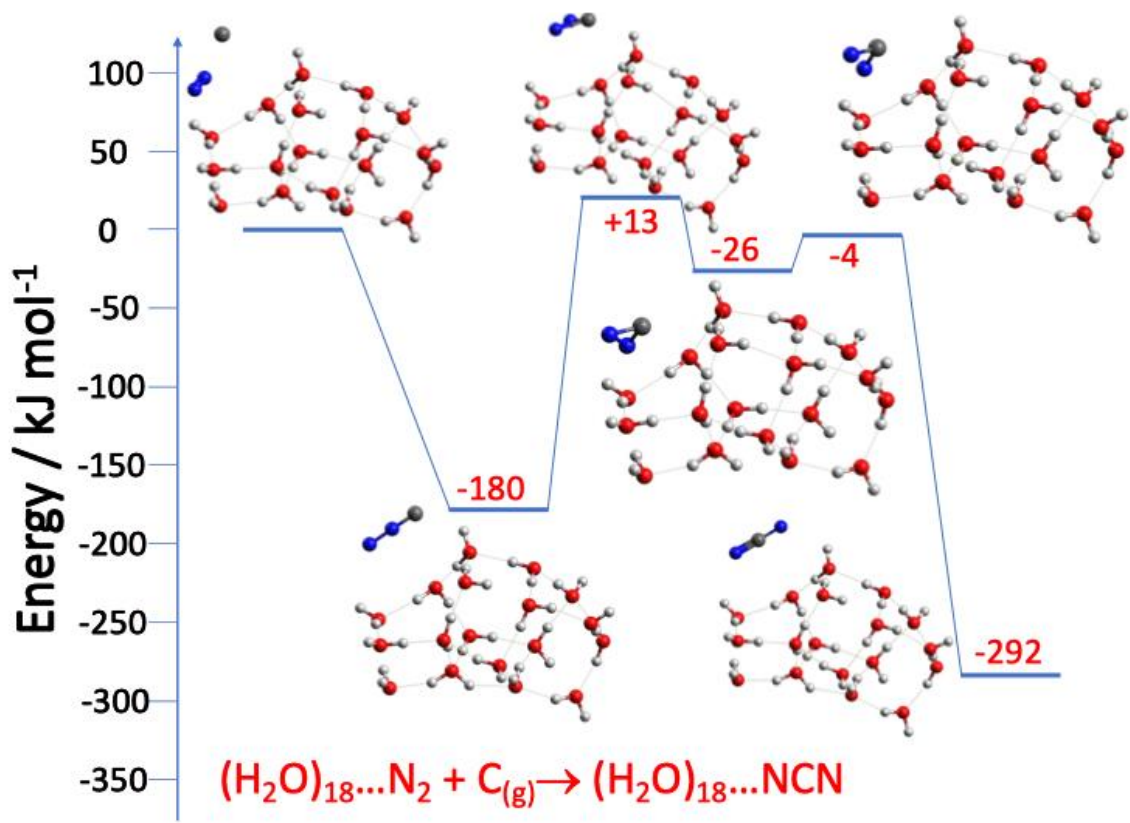

**Figure 6** Potential energy surface for the formation of NCN by the C + N$_2$ reaction on the (H$_2$O)$_{18}$ ASW cluster calculated at the $\omega$B97X-D3/def2-TZVP level. The clusters shown here correspond to those where reaction is occurring at site c. All energies are corrected for ZPE differences.

CNN formation by the C + N$_2$ reaction on the ASW cluster is slightly more exothermic than the equivalent gas-phase process shown in Figure 3, presumably due to additional stabilizing interactions with the surface. A similar weak stabilization effect is also found for all of the subsequent isomerization steps towards NCN formation. Although the isomerization barriers CNN → c-CNN and c-CNN → NCN are lower than those calculated in the gas-phase, the present results indicate that the level of stabilization is too low for the isomerization of CNN to NCN on the ASW surface to play an important role in the eventual formation of cyanamide/carbodiimide. Moreover, as a non-negligible fraction of the 180 kJ/mol available for isomerization is likely to be dissipated into the bulk, this isomerization process is assumed to contribute negligibly to NH$_2$CN and/or HNCNH formation in the present model.

**3.3.2 (H$_2$O)$_{18}$...CNN + H$_{(g)}$ → (H$_2$O)$_{18}$...HNCN then (H$_2$O)$_{18}$...HNCN + H$_{(g)}$ → (H$_2$O)$_{18}$...NH$_2$CN/HNCNH**

In this scenario and in all those to follow, H atoms were placed at a distance of 2.5 Å from the molecule (in this case H was placed at 2.5 Å from both ends of the CNN molecule) at the three sites and these systems were optimized over the doublet potential energy surface. Unsurprisingly, as a barrierless radical-radical association reaction, HNNC and/or HCNN were readily formed from H + CNN in all cases as shown in Figure 7.

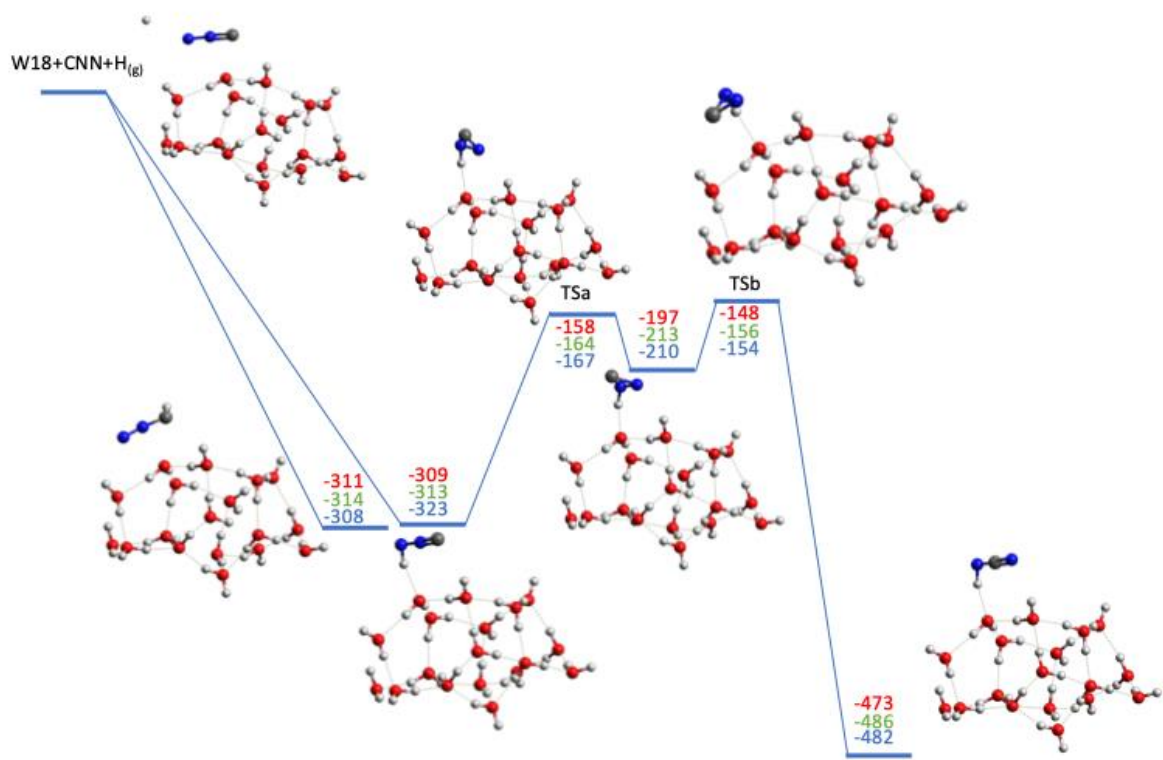

**Figure 7** HNCN formation pathways following the reaction of CNN with atomic hydrogen on the $(H_2O)_{18}$ amorphous solid water cluster. The clusters shown here correspond to those where reaction is occurring at site c. All energies are corrected for ZPE differences in kJ/mol with values in blue, green and red representing the calculated energies at sites a, b and c respectively.

Given that the gas-phase calculations indicate that the pathway for the isomerization of HCNN into HNCN is characterized by at least one high TS at the $\omega$B97X-D3/def2-TZVP level (see Figure 4), no further calculations were performed on this system. By comparison with the gas-phase values shown in Figure 4, it appears that the presence of the ASW cluster does not lead to a substantial stabilization of the HCNN intermediate (-308 to -314 kJ/mol on $(H_2O)_{18}$ versus -302 kJ/mol in the gas-phase). In contrast, HNNC displays a more marked stabilization of around 30 kJ/mol with respect to the gas-phase value (-309 to -323 kJ/mol on $(H_2O)_{18}$ versus -285 kJ/mol in the gas-phase). A similar level of stabilization is also observed for all the other stationary point structures along the HNNC → c-HNNC → HNCN isomerization pathway. Once HNCN is present on the surface, both $NH_2CN$ and HNCNH can be formed readily when a H-atom diffusing across the ASW surface or arriving from the gas-phase encounters the HNCN radical. As before, this process is simulated by placing the H-atom at a distance of 2.5 Å from either end of the HNCN molecule and the whole system is optimized over the singlet potential energy surface. These calculations confirm the barrierless, exothermic nature of this radical association step with calculated energies for $NH_2CN$ formation of -387, -399 and -388 kJ/mol for sites a, b and c respectively, and slightly lower calculated energies for HNCNH formation of -374, -380 and -371 kJ/mol for the sites a, b and c respectively. Quoted energies are relative to the $(H_2O)_{18}$..HNCN + $H_{(g)}$ level.

### 3.3.3 $(H_2O)_{18}$...CNN + $H_{(g)}$ → $(H_2O)_{18}$...HNNC then $(H_2O)_{18}$...HNNC + $H_{(g)}$ → $(H_2O)_{18}$...$NH_2NC$ → $(H_2O)_{18}$...$NH_2CN$

In this scenario, CNN hydrogenation initially occurs as described in the previous section to form HNNC on the $(H_2O)_{18}$ cluster. Then a second H-atom reacts with HNNC in a radical association step to form $NH_2NC$. $NH_2NC$ isomerization then occurs through a cyclic intermediate followed by N-N bond fission to form $NH_2CN$. Our earlier calculations on the equivalent gas-phase pathway (se Figure 5) indicated that this process could represent the most energetically accessible route for $NH_2CN$ formation. The singlet potential energy surface for this isomerization process for the three different adsorption sites is shown in Figure 8.

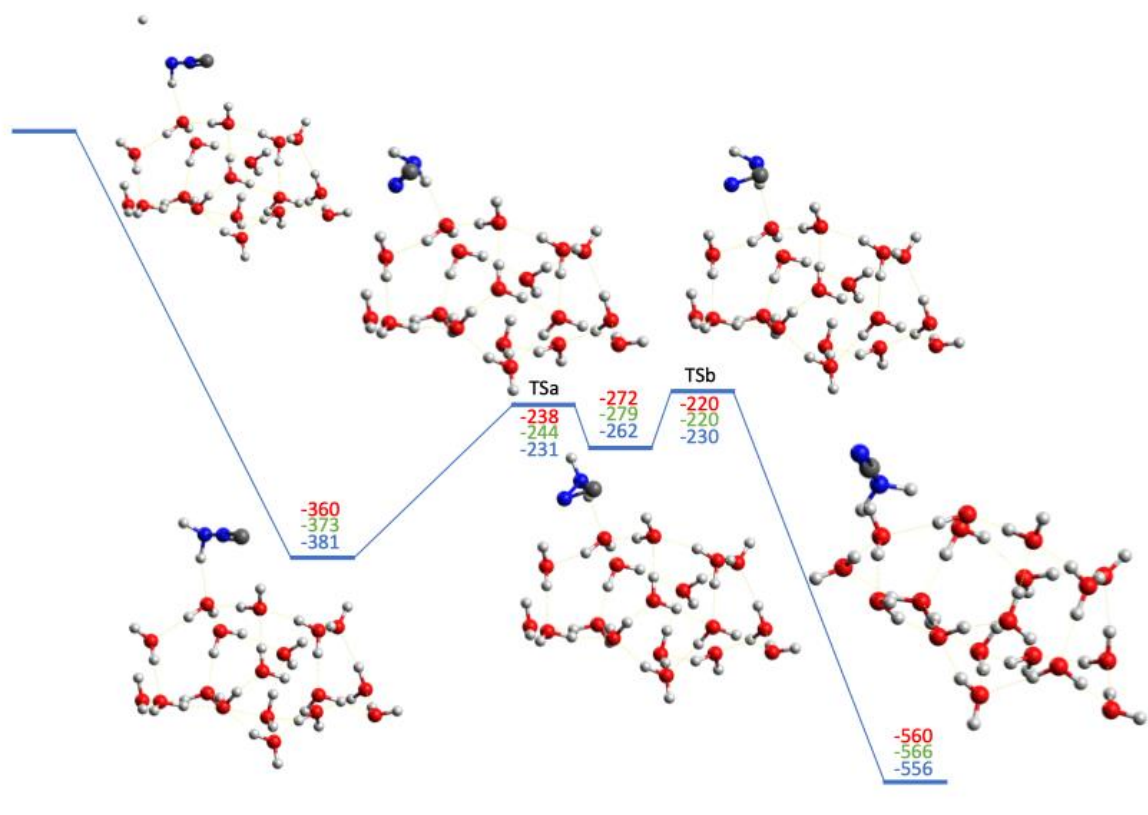

**Figure 8** NH$_2$CN formation pathways following the reaction of HNNC with atomic hydrogen on a (H$_2$O)$_{18}$ amorphous solid water cluster. The clusters shown here are those where reaction is occurring at site c. All energies are corrected for ZPE differences in kJ/mol with values in blue, green and red representing the calculated energies at sites a, b and c respectively.

NH$_2$NC formation on the ASW cluster is barrierless and exothermic for all three adsorption sites with exothermicity values that are similar to or slightly larger than those obtained by the equivalent gas-phase process. The isomerization step for NH$_2$NC to form c-NH$_2$NC is characterized by a barrier (TSa) with values very similar to those obtained in the gas-phase, while c-NH$_2$NC is found to be slightly more stable (10-15 kJ/mol) on the ASW cluster than in the gas-phase. The second isomerization step involving N-N bond breaking between c-NH$_2$NC and NH$_2$CN is also characterized by a barrier (TSb) that is slightly more stable than the gas-phase equivalent considering the three different adsorption sites. For at least one of the calculated adsorption sites (W18-c in particular, with a lesser effect for W18-a), the formation of NH$_2$CN is accompanied by a slight reorganization of the water cluster structure (see figure 8), almost certainly due to the large dipole moment of NH$_2$CN and its subsequent formation of additional hydrogen bonds with the water cluster. This is reflected by the additional stabilization energy for NH$_2$CN on the cluster with (H$_2$O)$_{18}$...NH$_2$CN calculated to be approximately 20 kJ/mol more stable than gas-phase NH$_2$CN when compared with (H$_2$O)$_{18}$...HNNC + H$_{(g)}$ or HNNC + H respectively.

### 3.3.4 (H$_2$O)$_{18}$...NH$_2$CN ↔ (H$_2$O)$_{18}$...HNCNH concerted isomerization

To test the possibility for the interconversion of NH$_2$CN and HNCNH through a concerted type mechanism on the ASW cluster we performed a series of scans whereby an N-H bond, where the H-atom is also hydrogen bonded to an O atom of a H$_2$O molecule in the cluster, was stretched from its equilibrium value. In Figure 9, we show a couple of snapshots of one of these scans (HNCNH is positioned at site a of the (H$_2$O)$_{18}$ cluster in this instance).

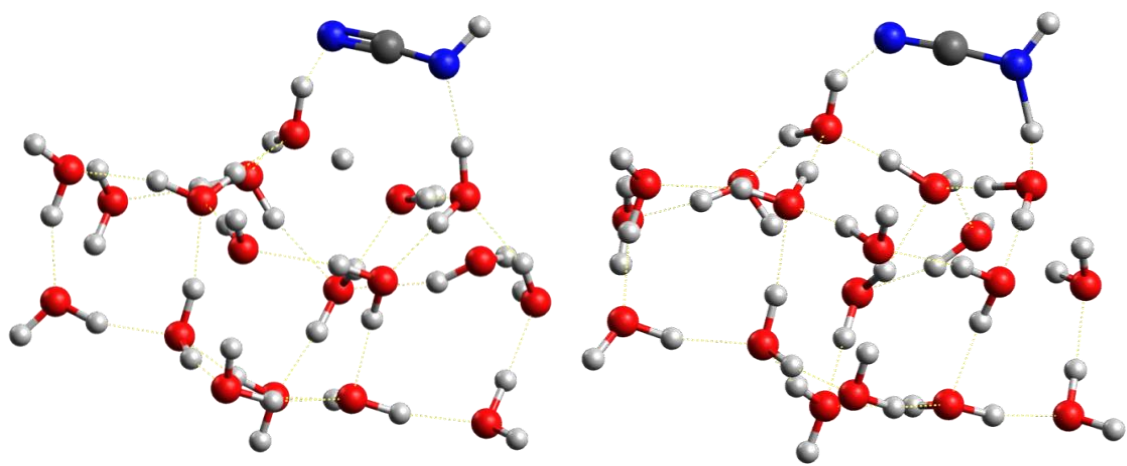

**Figure 9** (H$_2$O)$_{18}$...HNCNH → (H$_2$O)$_{18}$...NH$_2$CN concerted isomerization process at site a. (Left panel) One of the H-atoms of HNCNH (the one bound to a cluster H$_2$O molecules through hydrogen bonding) is transferred to H$_2$O which itself liberates a H-atom. (Right panel) The 'free' H-atom in the left panel binds to an adjacent H$_2$O molecule which donates a H-atom to a neighbouring H$_2$O molecule which finally donates a H-atom back to the adsorbed molecule to form NH$_2$CN.

Here, the H-atom from HNCNH is transferred to one of the cluster H$_2$O molecules (H-atom now bound to the water molecule just below the left-hand N-atom in the left panel), which liberates a H-atom (left panel) to be transferred to an adjacent H$_2$O molecule. This concerted mechanism continues in the right panel following an intermediate H-transfer, with a final H-atom transfer to the other terminal N atom of the HNCNH molecule to finally form NH$_2$CN. While it is difficult to isolate an individual TS for this process due the quasi-simultaneous nature of the four bond formation and fission processes, the energy given by the constrained scan itself indicates that this concerted isomerization process requires approximately 100 kJ/mol of additional energy above the $(H_2O)_{18}$...HNCNH well. For site b, a similar result is found, with three $H_2O$ molecules involved in the H-atom transfer with an energy of less than 40 kJ/mol required for the overall concerted isomerization. For site c, where the two ends of the N and H atoms of the HNCNH molecule are connected by a cycle of six hydrogen bonded $H_2O$ molecules, the H-atom transfer process does not occur indicating the likelihood that this mechanism does not proceed over longer distances. Interestingly, when we try to run the scan in the opposite direction (that is by stretching the newly formed N-H bond of $NH_2CN$), the reverse process leading to HNCNH formation does not occur in this instance. If we examine more closely the reason for this failure, it becomes clear that the nitrile group nitrogen atom $NH_2C\equiv N$ is characterized by a significantly smaller negative charge than the corresponding amine group nitrogen atom of H-NCNH as shown by the calculated Mulliken atomic charges for these systems. Consequently, as the hydrogen bond formed by the nitrile type nitrogen atom is generally weaker and longer (> 2 Å) than the corresponding amine type one (< 2 Å), H-atom transfer through the hydrogen bonding network appears to be less favourable for the reverse process. As a result, we would expect the preferential isomerization of HNCNH to $NH_2CN$ following HNCNH formation on the ice surface. While this does not rule out the presence of HNCNH on interstellar ice surfaces due to its possible stabilization by transferring its excess energy to the bulk, it does indicate that $NH_2CN$ could be present at higher abundance levels than HNCNH.

## 4 Astrophysical Model and Implications

To obtain a view as comprehensive as possible of the formation pathways of $NH_2CN$ and HNCNH in astrophysics—and to assess the role of the C + $N_2$ reaction—we conducted a search for other potential formation routes for these species by drawing in particular on the work of Ramal-Olmedo et al. [34] To do so, we examined possible efficient pathways at low temperature (with no or low barriers) leading to species containing one carbon atom and two nitrogen atoms, starting from those assumed (or detected) to be present in interstellar gas and ices. We identified three families of reactions (followed by hydrogenation) leading to $NH_2CN$ and HNCNH, mostly on grains as the gas phase production of $NH_2CN$ or HNCNH is inefficient with our gas phase network:

(I) reaction with s-$N_2$: s-C + s-$N_2$ (mostly through the Eley-Rideal mechanism C + s-$N_2$) and s-CH + s-$N_2$

(II) reaction of s-N with s-CN and s-HCNH (derived from s-HCN) and reaction of s-NH with s-CN and s-HCNH (derived from s-HCN)

(III) reaction of s-$NH_2$ with s-CN

It should be noted that species that are unstable and easily isomerize (such as HNNC or $NH_2NC$), and species that are stable but produced in small quantities and do not lead to $NH_2CN$ or HNCNH (such as HCNNH produced by s-H + s-HCNN, and $H_2CNNH_2$) are not considered in the network. In our network, according to the theoretical calculations described above, we consider for the following processes:

(1) the C + s-$N_2$ reaction leads to s-CNN formation alone.

s-CNN is formed rather than s-NCN as the isomerization process s-CNN $\rightarrow$ s-NCN is characterized by a real barrier at the present level of theory (see Figure 6).

(2) the s-H + s-CNN reaction leads to s-HCNN (30%) and s-HNCN (70%).

Although the s-H + s-CNN reaction should result in the formation of equivalent amounts of s-HCNN and s-HNNC, given their similar exothermic energies (see Figure 7), we consider that all of the s-HNNC isomerizes to s-HNCN given the low submerged barriers for this process. Indeed, as the HNNC molecule must reorient following its formation, with the H-atom binding to an oxygen atom of one of the surface $H_2O$ molecules, it is likely that isomerization will occur before the excess energy can be dissipated to the cluster. In contrast, we consider that only a fraction of s-HCNN isomerizes to s-HNCN. This is due to the higher submerged barriers predicted by the gas-phase calculations for the latter isomerization (see Figure 4).

(3) the s-H + s-HNCN reaction leads to s-$NH_2CN$ (80%) and s-HNCNH (20%).

Although s-$NH_2CN$ and s-HNCNH should be formed in similar quantities due to the barrierless attack of s-H at both ends of the s-HNCN molecule, $NH_2CN$ formation is favoured due to the concerted isomerization of s-HNCNH to s-$NH_2CN$ as shown in Figure 9, while the reverse process is inhibited.

(4) the s-H + s-HCNN reaction leads to s-$CH_2$ + s-$N_2$ (49%), s-$H_2CNN$ (1%), s-HCN + NH (30%) and s-$NH_2CN$ + H (20%).

All of these four reactions are barrierless (see Table S1 for the references).

The complete list of reactions included in the chemical network is presented in Table S1.

It should be noted that our results are quite different from the ones of Ramal-Olmedo et al.[34] where a non-negligible amount of NH$_2$CN was produced in the gas phase. The main difference is for the CN + CH$_3$NH$_2$ reaction which does not lead to NH$_2$CN + CH$_3$ in our network, in contrast to the studies by Sleiman et al.[35] and Ramal-Olmedo et al.,[34] but to CH$_3$NH + HCN instead. We have reexamined this reaction and found a previously unexplored pathway, CN + CH$_3$NH$_2$ → CH$_3$NH$_2$...CN → HCN + CH$_3$NH, significantly lower in energy as shown in Figure 10, with a mechanism similar to the CN + NH$_3$ reaction. Consequently, it is no longer necessary to artificially lower the calculated barriers to reproduce the experimental kinetic results as in Sleiman et al.,[35] and more importantly, this reaction will not lead to NH$_2$CN formation.

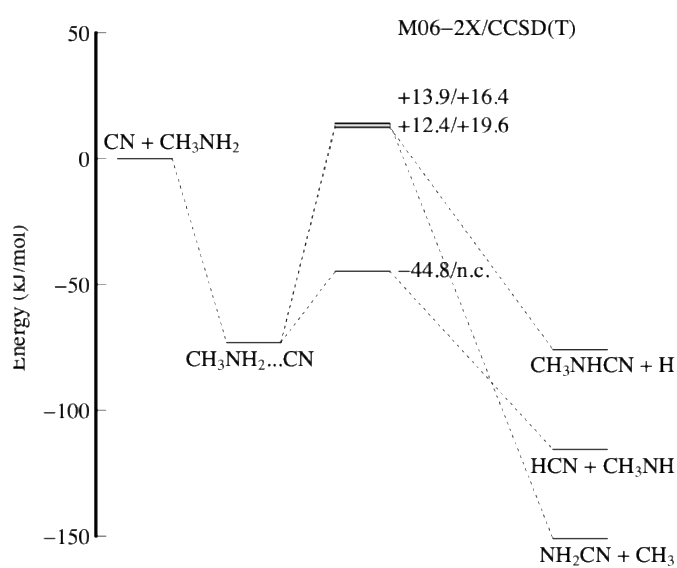

**Figure 10** Diagram of potential energy for the CN + CH$_3$NH$_2$ reaction calculated at the M06-2X/aug-cc-pVTZ level. The energy of the TS calculated at the CCSD(T)/aug-cc-pVTZ level from Sleiman et al.[35] is also reported when it was calculated ('n.c.' means ''not calculated'').

Despite this result, we conducted a test assuming that the CN + CH$_3$NH$_2$ reaction produced NH$_2$CN to see if this process could represent an important gas-phase source of NH$_2$CN in the simulations. Our test results showed that only low abundances of NH$_2$CN are produced by the CN + CH$_3$NH$_2$ reaction. This can be explained by the fact that although CN is abundant in cold regions, CH$_3$NH$_2$ is only present at low abundance levels in the gas phase. Conversely, in warm regions CN is only present at low abundance levels while CH$_3$NH$_2$ is relatively abundant as a result of its production on grains followed by desorption into the gas-phase.

To understand the effect of the C + s-$N_2$ reaction on the astrophysical chemistry of $NH_2CN$ and HNCNH, we use the Nautilus code[36] to calculate the abundances of these species in a low mass protostar such as IRAS16293-B where $NH_2CN$ was unambiguously detected.[13] The Nautilus code[36] is a 3-phase gas, dust-grain ice-surface and dust-grain ice-mantle time dependent chemical model employing kida.uva.2024[37] as the basic reaction network updated recently for a better description of COM chemistry on interstellar dust grains and in the gas-phase,[38-41] including the reactions reviewed in this work and presented in Table S1. There are 800 individual species included in the network that are involved in approximately 9000 separate reactions. The physical model to describe the evolution of the protostar comprises two successive evolutionary stages of a low-mass protostar: a uniform stage, corresponding to the pre-stellar phase, or the cold-core (homogenous cloud with a total density (nH + 2nH$_2$) equal to $5.0 \times 10^4$ cm$^{-3}$, a temperature equal to 10 K for both the gas and dust, a visual extinction of 10 mag, a cosmic-ray ionization rate of $1.3 \times 10^{-17}$ s$^{-1}$, a standard external UV field of 1 $G_0$ with the initial abundances used given in Table 1. This is followed (after $1 \times 10^6$ years) by the collapse phase as described in Manigand et al.[38] The C/O gas phase elemental ratio is equal to 1 in this work. This ratio, which is higher than the cosmic abundance ratio (0.6), allows for a much better agreement between the model and the entire set of observations for dense molecular clouds such as TMC-1. Otherwise, there is insufficient carbon to consume atomic oxygen leading to CO formation, while the remaining oxygen atoms limit the radical abundance levels, thereby inhibiting the chemistry. This choice was already implemented by Hincelin et al.[42] in their study of $O_2$ formation and also by Byrne et al.[43] and Mallo et al.[44] in their studies of aromatic species. Physically, this oxygen depletion is explained by the formation of water ice during the evolution of the dense cloud; water ice that does not desorb in the warmer, less-dense phases of this evolution, unlike CO[45-46] which dissociates into C and O. We also consider a slight depletion in nitrogen stored in the form of $NH_3$ on ice formed concurrently with water during the evolution of the dense cloud.

**Table 1:** Elemental abundances

| Element | Abundance[a] |
|---|---|
| $H_2$ | 0.5 |
| He | 0.09 |
| $C^+$ | $1.4 \times 10^{-4}$ |
| N | $3.0 \times 10^{-5}$ |
| O | $1.4 \times 10^{-4}$ |
| s-$H_2O$[b] | $1.0 \times 10^{-4}$ |
| $S^+$ | $4.0 \times 10^{-6}$ |
| $Fe^+$ | $2.0 \times 10^{-8}$ |
| $Cl^+$ | $1.0 \times 10^{-7}$ |
| F | $6.7 \times 10^{-9}$ |

[a]Relative to total hydrogen (nH + 2nH$_2$)

[b]s-$H_2O$ means $H_2O$ on dust grains

The grain surface and the mantle are both chemically active for these simulations, while accretion and desorption are only allowed between the surface and the gas-phase. The dust-to-gas ratio (in terms of mass) is 0.01. A sticking probability of 1 is assumed for all neutral species while desorption occurs by both thermal and non-thermal processes (cosmic rays, chemical desorption) including sputtering of ices by cosmic-ray collisions.[47] A full description of the surface reaction formalism and simulations is available in Ruaud et al.[36]

The evolution of the abundances of $NH_2CN$, HNCNH and $NH_2CHO$ (to compare with the observations of IRAS16293-B by Coutens et al.[13]) during the cold-core phase and the collapse phase is shown in Figure 11 (panel b) for our nominal model.

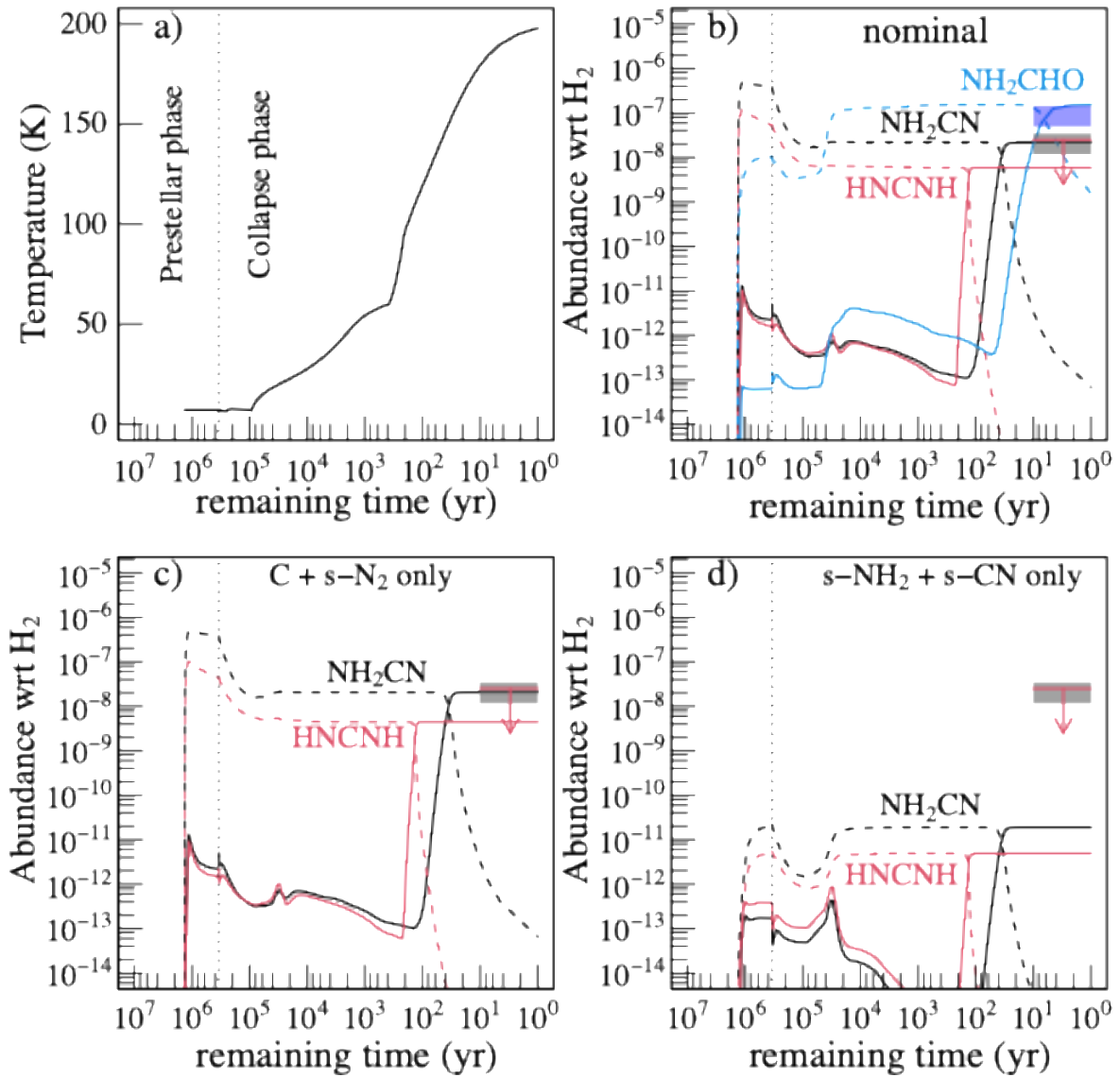

**Figure 11** Time evolution of the temperature (panel a) and the abundances (panels b-d) in the gas phase (solid lines) and on the grain surface (dashed lines) of the species NH$_2$CN, HNCNH and NH$_2$CHO during the cold-core phase and the collapse phase. The time axis is inverted to better visualize the evolution of abundances. Observations in IRAS16293-B are taken from Coutens et al.[13, 48] (NH2CHO and NH$_2$CN are represented by the blue and grey blocks respectively, while HNCNH is not detected and only the upper limit is shown as a red bar).

The calculated abundances are compared with observations by dividing the column densities by a total H$_2$ column density of 1.5×10$^{23}$ cm$^{-2}$, and display good agreement. While this H$_2$ column density value is below the lower limit derived by Jørgensen et al.,[49] it successfully reproduces the observations of major species such as CH$_3$OH, CH$_3$CHO, etc. The species

$NH_2CN$, HNCNH and $NH_2CHO$ are mostly produced on grain surfaces and released into the gas-phase when the temperature reaches the desorption temperature. The first species to desorb is HNCNH. The binding energy of HNCNH has not been measured to the best of our knowledge but it is calculated to be 4500 K by Wakelam et al.[18] $NH_2CN$ and $NH_2CHO$ desorb at much higher temperature. The binding energy of $NH_2CN$ has not been measured either to the best of our knowledge but is calculated to be 7000 K by Wakelam et al.,[18] and the binding energy of $NH_2CHO$ has been measured between 7500 K and 9600 K,[50-51] with the study by Wakelam et al.[18] deriving a value of 8500 K. The high binding energy of $NH_2CN$ results in desorption only in the hottest regions of the protostar. This may limit its overall desorption and could make comparisons with other species challenging, except for those desorbing at similar temperatures, such as $NH_2CHO$.

To assess the importance of individual reactions to $NH_2CN$ or HNCNH production, we performed runs considering only one production reaction (leading to $NH_2CN$ or HNCNH) at a time. By far the most efficient reaction for $NH_2CN$ and HNCNH production is C + s-$N_2$ (Eley-Rideal mechanism) as shown in Figure 11, panel c. Indeed, when only this reaction is included in the simulations, it leads to 95% of the $NH_2CN$ produced with the full network and 75 % of the HNCNH. A similar outcome would be expected with mixed $H_2O$/CO ices as incoming C-atoms react with s-$N_2$ by the Eley-Rideal mechanism, so the nature of the surface is unlikely to have a large influence on the barrierless nature of the reaction. The s-$NH_2$ + s-CN reaction is very inefficient, as shown in Figure 11, panel d. This is due to the low abundance of s-CN, which is highly reactive and undergoes tunneling reactions with s-$H_2$ and reacts without a barrier with s-$NH_3$, and probably also with s-$H_2O$, which according to a recent theoretical study is possible by a concerted effect of water molecules in the ice.[52] For the production of s-HNCNH, the direct pathway s-N + s-HCNH is not negligible but may be overestimated. Indeed, s-HCNH is produced by the hydrogenation of s-HCN (and s-HNC), which is abundant in the model (about 2% of the ice), yet has never been detected.

It is worth noting that, while $NH_2CN$ and $NH_2CHO$ are similar molecules, their formation pathways are very different even though they are both formed on grain surfaces. $NH_2CN$ is mainly produced by the Eley-Rideal type reaction C + s-$N_2$ followed by hydrogenation, which remains efficient even at low temperatures since neither species needs to diffuse on the ice for reaction to occur. In contrast, $NH_2CHO$ is primarily formed by the reaction s-$NH_2$ + s-HCO, which is only efficient if $NH_2$ and/or HCO can move on the ice surface, and thus only occurs during the warm phase of the protostar.

It should be noted that the abundance of $NH_2CN$ in dense molecular clouds is not negligible, with values around $1 \times 10^{-11}$ / $H_2$. This is due to the fact that $NH_2CN$ is efficiently produced on grains during the cold phase reaching a value as high as $5 \times 10^{-7}$. Then desorption through the interaction of cosmic rays with the ice becomes an efficient way to deliver some $NH_2CN$ to the gas phase as described in Wakelam et al.[47]; a mechanism which is implemented in this version of Nautilus.

Our model, incorporating the reaction of atomic carbon with s-$N_2$ present on ice through an Eley-Rideal type mechanism, yields simulated $NH_2CN$ abundances in good agreement with those observed in IRAS16293-B.[13] While our model produces a lower but non-negligible amount of HNCNH, it remains consistent with the upper limit derived from observations.[13] Thus, $NH_2CN$ could be an interesting proxy for probing $N_2$ in protostars, with $N_2$ being undetectable by rotational spectroscopy due to its lack of a dipole moment. Moreover, the $^{14}N/^{15}N$ fractionation of $NH_2CN$ is potentially interesting because it should reproduce the observed fractionation of $N_2$, ($^{14}N/^{15}N > 441$ based on observations of $N_2H^+$[53,54] and models of nitrogen fractionation[55,56]) with the same fractionation for both nitrogen atoms in $NH_2CN$. Quantification of the $^{14}N/^{15}N$ ratio in $NH_2CN$ would allow us to place constraints on the fractionation of $N_2$ in protostars and to validate the proposed formation pathways of $NH_2CN$. Indeed, if $NH_2CN$ were not produced from $N_2$, the amine group and nitrile group nitrogen atoms could exhibit different isotope ratios following the fractionation of their precursor (N, NH, CN, …), and could even be enriched in $^{15}N$ if it were formed from CN and $NH_3$ as these species are observed to be enriched in dense molecular clouds.[53, 57, 58] The availability of spectroscopic data for both $^{15}N$ isotopologues[59] makes it possible to investigate the $^{14}N/^{15}N$ ratios in protostellar environments.

## 5 Conclusions

The present work reports the results of a quantum chemical study of the C + $N_2$ → CNN reaction in the gas-phase and on the surface of an amorphous solid water cluster. While the gas-phase C + $N_2$ association reaction is shown to be unimportant at low interstellar densities, the equivalent ice surface reaction occurs readily when $N_2$ is already adsorbed on the surface and C-atoms are depleting from the gas phase via an Eley-Rideal type mechanism. Despite the large predicted exothermicity of the reaction, direct isomerization to NCN is unlikely to occur due to the high barriers for this process. Instead, exothermic reactions of atomic hydrogen with CNN are shown to promote the isomerization of the various HCNN and $H_2$CNN intermediate species, allowing HNCN and $H_2$NCN type species to form readily over low predicted barriers. An extensive series of gas-phase calculations of these hydrogenation reactions allowed us to identify the likely dominant pathways on the ice surface, which were later confirmed by additional calculations of these dominant pathways at various sites of an 18 molecule ASW cluster, leading to the efficient formation of both $NH_2$CN and HNCNH on the ASW surface. Additional calculations were performed, highlighting the potential importance of the transformation of HNCNH to $NH_2$CN through a concerted type mechanism whereby a hydrogen atom of HNCNH is transferred to a water molecule of the cluster, with the H-atom given back at the other end of the molecule to yield $NH_2$CN. In contrast, these calculations indicate that this concerted mechanism does not proceed in the opposite direction, suggesting the preferential formation of $NH_2$CN on the ASW surface.

The effects of these reactions on the simulated protostellar abundances of $NH_2$CN and HNCNH were tested by including them in a three-phase astrochemical model. Furthermore, the network of formation and destruction reactions of other small nitrogen bearing species was updated to provide a better description of the overall chemistry of these species in interstellar environments. The model results clearly demonstrate the importance of the C + $N_2$ reaction on $NH_2$CN and HNCNH formation with this process generating $NH_2$CN and HNCNH abundances in good agreement with the observed values and several orders of magnitude greater than those produced by the corresponding model without it.

**Author Information**

**Corresponding Author**

*Email: kevin.hickson@u-bordeaux.fr.

**Notes**

The authors declare no competing financial interest.

**Supporting Information**

Table of reactions added to the network (Table S1) and species geometries at the $\omega$B97X-D3/def2-TZVP level of theory.

**Acknowledgements**

K. M. H. acknowledges support by the thematic action "Physique et Chimie du Milieu Interstellaire" (PCMI) of the INSU National Programme "Astro", with contributions from CNRS Physique & CNRS Chimie, CEA, and CNES. A. C. acknowledges financial support from the European Research Council (ERC) under the European Union's Horizon 2020 research and innovation programme (ERC Starting Grant "Chemtrip", grant agreement No 949278).

**References**

(1) Daranlot, J.; Hincelin, U.; Bergeat, A.; Costes, M.; Loison, J. C.; Wakelam, V.; Hickson, K. M. Elemental Nitrogen Partitioning in Dense Interstellar Clouds. *Proc. Natl. Acad. Sci. USA* **2012**, *109*, 10233-10238. DOI: 10.1073/pnas.1200017109.

(2) Daranlot, J.; Hu, X.; Xie, C.; Loison, J. C.; Caubet, P.; Costes, M.; Wakelam, V.; Xie, D.; Guo, H.; Hickson, K. M. Low Temperature Rate Constants for the N($^4$S) + CH(X$^2\Pi_r$) Reaction. Implications for N$_2$ Formation Cycles in Dense Interstellar Clouds. *Phys. Chem. Chem. Phys.* **2013**, *15*, 13888-13896. DOI: 10.1039/c3cp52535j.

(3) Daranlot, J.; Jorfi, M.; Xie, C.; Bergeat, A.; Costes, M.; Caubet, P.; Xie, D.; Guo, H.; Honvault, P.; Hickson, K. M. Revealing Atom-Radical Reactivity at Low Temperature through the N + OH Reaction. *Science* **2011**, *334*, 1538-1541. DOI: 10.1126/science.1213789.

(4) Hily-Blant, P.; Walmsley, M.; Pineau des Forêts, G.; Flower, D. Nitrogen Chemistry and Depletion in Starless Cores. *Astron. Astrophys.* **2010**, *513*, A41. DOI: 10.1051/0004-6361/200913200.

(5) Rimola, A.; Taquet, V.; Ugliengo, P.; Balucani, N.; Ceccarelli, C. Combined Quantum Chemical and Modeling Study of CO Hydrogenation on Water Ice. *Astron. Astrophys.* **2014**, *572*, 10.1051/0004-6361/201424046.

(6) Fuchs, G., W.; Cuppen, H., M.; Ioppolo, S.; Romanzin, C.; Bisschop, S., E.; Andersson, S.; van Dishoeck, E., F.; Linnartz, H. Hydrogenation Reactions in Interstellar CO Ice Analogues. *Astron. Astrophys.* **2009**, *505*, 629-639.

(7) Fedoseev, G.; Qasim, D.; Chuang, K.-J.; Ioppolo, S.; Lamberts, T.; van Dishoeck, E. F.; Linnartz, H. Hydrogenation of Accreting C Atoms and CO Molecules–Simulating Ketene and Acetaldehyde Formation Under Dark and Translucent Cloud Conditions. *Astrophys. J.* **2022**, *924*, 110. DOI: 10.3847/1538-4357/ac3834.

(8) Ferrero, S.; Ceccarelli, C.; Ugliengo, P.; Sodupe, M.; Rimola, A. Formation of Complex Organic Molecules on Interstellar CO Ices? Insights from Computational Chemistry Simulations. *Astrophys. J.* **2023**, *951*, 150. DOI: 10.3847/1538-4357/acd192.

(9) Choi, H.; Mordaunt, D. H.; Bise, R. T.; Taylor, T. R.; Neumark, D. M. Photodissociation of Triplet and Singlet States of the CCO Radical. *J. Chem. Phys.* **1998**, *108*, 4070-4078. DOI: 10.1063/1.475839.

(10) Lu, D.; Urzúa-Leiva, R.; Denis-Alpizar, O.; Guo, H. Hyperthermal Dynamics and Kinetics of the C($^3$P) + N$_2$(X$^1\Sigma_g^+$) → CN(X$^2\Sigma^+$) + N($^4$S) Reaction. *J. Phys. Chem. A* **2023**, *127*, 2839-2845. DOI: 10.1021/acs.jpca.3c00210.

(11) Urzúa-Leiva, R.; Denis-Alpizar, O. Study of the CN(X$^2\Sigma^+$) + N($^4$S) Reaction at High Temperatures: Potential Energy Surface and Thermal Rate Coefficients. *J. Phys. Chem. A* **2021**, *125*, 8168-8174. DOI: 10.1021/acs.jpca.1c04903.

(12) Turner, B. E.; Liszt, H. S.; Kaifu, N.; Kisliakov, A. G. Microwave Detection of Interstellar Cyanamide. *Astrophys. J.* **1975**, *201*, L149-L152. DOI: 10.1086/181963.

(13) Coutens, A.; Willis, E. R.; Garrod, R. T.; Müller, H. S. P.; Bourke, T. L.; Calcutt, H.; Drozdovskaya, M. N.; Jørgensen, J. K.; Ligterink, N. F. W.; Persson, M. V.; et al. First Detection of Cyanamide (NH$_2$CN) towards Solar-type Protostars. *Astron. Astrophys.* **2018**, *612*, 10.1051/0004-6361/201732346.

(14) McGuire, B. A.; Loomis, R. A.; Charness, C. M.; Corby, J. F.; Blake, G. A.; Hollis, J. M.; Lovas, F. J.; Jewell, P. R.; Remijan, A. J. Interstellar Carbodiimide (HNCNH): a New Astronomical Detection from the GBT Primos Survey via Maser Emission Features. *Astrophys. J. Lett.* **2012**, *758*, L33. DOI: 10.1088/2041-8205/758/2/L33.

(15) Smith, R. S.; May, R. A.; Kay, B. D. Desorption Kinetics of Ar, Kr, Xe, N$_2$, O$_2$, CO, Methane, Ethane, and Propane from Graphene and Amorphous Solid Water Surfaces. *J. Phys. Chem. B* **2016**, *120*, 1979-1987. DOI: 10.1021/acs.jpcb.5b10033.

(16) Ferrero, S.; Zamirri, L.; Ceccarelli, C.; Witzel, A.; Rimola, A.; Ugliengo, P. Binding Energies of Interstellar Molecules on Crystalline and Amorphous Models of Water Ice by Ab Initio Calculations. *Astrophys. J.* **2020**, *904*, 11. DOI: 10.3847/1538-4357/abb953.

(17) Das, A.; Sil, M.; Gorai, P.; Chakrabarti, S. K.; Loison, J. C. An Approach to Estimate the Binding Energy of Interstellar Species. *Astrophys. J., Suppl. Ser.* **2018**, *237*, 9. DOI: 10.3847/1538-4365/aac886.

(18) Wakelam, V.; Loison, J. C.; Mereau, R.; Ruaud, M. Binding energies: New Values and Impact on the Efficiency of Chemical Desorption. *Mol. Astrophys.* **2017**, *6*, 22-35. DOI: https://doi.org/10.1016/j.molap.2017.01.002.

(19) Shimonishi, T.; Nakatani, N.; Furuya, K.; Hama, T. Adsorption Energies of Carbon, Nitrogen, and Oxygen Atoms on the Low-temperature Amorphous Water Ice: A Systematic Estimation from Quantum Chemistry Calculations. *Astrophys. J.* **2018**, *855*, 27. DOI: 10.3847/1538-4357/aaaa6a.

(20) Duflot, D.; Toubin, C.; Monnerville, M. Theoretical Determination of Binding Energies of Small Molecules on Interstellar Ice Surfaces. *Front. Astron. Space Sci.* **2021**, *8*, 645243.

(21) Molpeceres, G.; Kästner, J.; Fedoseev, G.; Qasim, D.; Schömig, R.; Linnartz, H.; Lamberts, T. Carbon Atom Reactivity with Amorphous Solid Water: $H_2O$-Catalyzed Formation of $H_2CO$. *J. Phys. Chem. Lett.* **2021**, *12*, 10854-10860. DOI: 10.1021/acs.jpclett.1c02760.

(22) Tsuge, M.; Molpeceres, G.; Aikawa, Y.; Watanabe, N. Surface Diffusion of Carbon Atoms as a Driver of Interstellar Organic Chemistry. *Nature Astron.* **2023**, *7*, 1351-1358.

(23) Ferrero, S.; Ceccarelli, C.; Ugliengo, P.; Sodupe, M.; Rimola, A. Formation of Interstellar Complex Organic Molecules on Water-rich Ices Triggered by Atomic Carbon Freezing. *Astrophys. J.* **2024**, *960*, 22. DOI: 10.3847/1538-4357/ad0547.

(24) Molpeceres, G.; Tsuge, M.; Furuya, K.; Watanabe, N.; San Andrés, D.; Rivilla, V. M.; Colzi, L.; Aikawa, Y. Carbon Atom Condensation on $NH_3$–$H_2O$ Ices. An Alternative Pathway to Interstellar Methanimine and Methylamine. *J. Phys. Chem. A* **2024**, *128*, 3874-3889. DOI: 10.1021/acs.jpca.3c08286.

(25) Enrique-Romero, J.; Rimola, A.; Ceccarelli, C.; Ugliengo, P.; Balucani, N.; Skouteris, D. Reactivity of HCO with $CH_3$ and $NH_2$ on Water Ice Surfaces. A Comprehensive Accurate Quantum Chemistry Study. *ACS Earth Space Chem.* **2019**, *3*, 2158-2170. DOI: 10.1021/acsearthspacechem.9b00156.

(26) Perrero, J.; Enrique-Romero, J.; Martínez-Bachs, B.; Ceccarelli, C.; Balucani, N.; Ugliengo, P.; Rimola, A. Non-energetic Formation of Ethanol via CCH Reaction with Interstellar $H_2O$ Ices. A Computational Chemistry Study. *ACS Earth Space Chem.* **2022**, *6*, 496-511. DOI: 10.1021/acsearthspacechem.1c00369.

(27) Neese, F. The ORCA Program System. *WIREs Comput. Mol. Sci.* **2012**, *2*, 73-78. DOI: 10.1002/wcms.81.

(28) Neese, F. Software update: The ORCA Program System—Version 5.0. *WIREs Comput. Mol. Sci.* **2022**, *12*, e1606. DOI: 10.1002/wcms.1606.

(29) Grimme, S.; Ehrlich, S.; Goerigk, L. Effect of the Damping Function in Dispersion Corrected Density Functional Theory. *J. Comput. Chem.* **2011**, *32*, 1456-1465. DOI: https://doi.org/10.1002/jcc.21759 (acccessed 2025/10/27).

(30) Ásgeirsson, V.; Birgisson, B. O.; Bjornsson, R.; Becker, U.; Neese, F.; Riplinger, C.; Jónsson, H. Nudged Elastic Band Method for Molecular Reactions Using Energy-Weighted Springs Combined with Eigenvector Following. *J. Chem. Theory Comput.* **2021**, *17*, 4929-4945. DOI: 10.1021/acs.jctc.1c00462.

(31) Kruse, H.; Grimme, S. A Geometrical Correction for the Inter- and Intra-Molecular Basis Set Superposition Error in Hartree-Fock and Density Functional Theory Calculations for Large Systems. *J. Chem. Phys.* **2012**, *136*, 154101. DOI: 10.1063/1.3700154.

(32) Hanwell, M. D.; Curtis, D. E.; Lonie, D. C.; Vandermeersch, T.; Zurek, E.; Hutchison, G. R. Avogadro: An Advanced Semantic Chemical Editor, Visualization, and Analysis Platform. *J. Cheminf.* **2012**, *4*, 17. DOI: 10.1186/1758-2946-4-17.

(33) Tordini, F.; Bencini, A.; Bruschi, M.; De Gioia, L.; Zampella, G.; Fantucci, P. Theoretical Study of Hydration of Cyanamide and Carbodiimide. *J. Phys. Chem. A* **2003**, *107*, 1188-1196. DOI: 10.1021/jp026535r.

(34) Ramal-Olmedo, J. C.; Menor-Salván, C. A.; Miyoshi, A.; Fortenberry, R. C. Gas-Phase Molecular Formation Mechanisms of Cyanamide ($NH_2CN$) and Its Tautomer Carbodiimide (HNCNH) under Sgr B2(N) Astrophysical Conditions. *Astron. Astrophys.* **2023**, *672*, A49. https://doi.org/10.1051/0004-6361/202245811

(35) Sleiman, C.; El Dib, G.; Rosi, M.; Skouteris, D.; Balucani, N.; Canosa, A. Low Temperature Kinetics and Theoretical Studies of the Reaction CN + $CH_3NH_2$: A Potential Source of Cyanamide and Methyl Cyanamide in the Interstellar Medium. *Phys. Chem. Chem. Phys.* **2018**, *20* (8), 5478-5489, 10. http://dx.doi.org/10.1039/C7CP05746F

(36) Ruaud, M.; Wakelam, V.; Hersant, F. Gas and Grain Chemical Composition in Cold Cores as Predicted by the Nautilus 3-Phase Model. *Mon. Not. R. Astron. Soc.* **2016**, *459*, 3756-3767. DOI: 10.1093/mnras/stw887.

(37) Wakelam, V.; Gratier, P.; Loison, J.-C.; Hickson, K. M.; Penguen, J.; Mechineau, A. The 2024 KIDA Network for Interstellar Chemistry. *Astron. Astrophys.* **2024**, *689*, A63.

(38) Manigand, S.; Coutens, A.; Loison, J.-C.; Wakelam, V.; Calcutt, H.; Müller, H. S. P.; Jørgensen, J. K.; Taquet, V.; Wampfler, S. F.; Bourke, T. L.; et al. The ALMA-PILS Survey: First Detection of the Unsaturated 3-Carbon Molecules Propenal ($C_2H_3CHO$) and Propylene ($C_3H_6$) Towards IRAS 16293–2422 B. *Astron. Astrophys.* **2021**, *645*, A53. DOI: 10.1051/0004-6361/202038113.

(39) Hickson, K. M.; Loison, J.-C.; Wakelam, V. Kinetic Study of the Gas-Phase $C(^3P)$ + $CH_3CN$ Reaction at Low Temperatures: Rate Constants, H-Atom Product Yields, and Astrochemical Implications. *ACS Earth Space Chem.* **2021**, *5*, 824-833. DOI: 10.1021/acsearthspacechem.0c00347

(40) Coutens, A.; Loison, J.-C.; Boulanger, A.; Caux, E.; Müller, H. S. P.; Wakelam, V.; Manigand, S.; Jørgensen, J. K. The Alma-Pils Survey: First Tentative Detection of 3-Hydroxypropenal (HOCHCHCHO) in the Interstellar Medium and Chemical Modeling of the $C_3H_4O_2$ Isomers. *Astron. Astrophys.* **2022**, *660*, L6. DOI: 10.1051/0004-6361/202243038.

(41) Hickson, K. M.; Loison, J.-C.; Wakelam, V. A Low-Temperature Kinetic Study of the $C(^3P)$ + $CH_3OCH_3$ Reaction: Rate Constants, H Atom Product Yields, and Astrochemical Implications. *ACS Earth Space Chem.* **2024**, *8*, 1087-1100. DOI: 10.1021/acsearthspacechem.4c00014.

(42). Hincelin, U.; Wakelam, V.; Hersant, F.; Guilloteau, S.; Loison, J. C.; Honvault, P.; Troe, J. Oxygen Depletion in Dense Molecular Clouds: A Clue to a Low $O_2$ Abundance? *Astron. Astrophys.* **2011**, *530*, 61.

(43) Byrne, A. N.; Xue, C.; Van Voorhis, T.; McGuire, B. A. Sensitivity Analysis of Aromatic Chemistry to Gas-Phase Kinetics in a Dark Molecular Cloud Model. *Phys. Chem. Chem. Phys.* **2024**, *26*, 26734-26747, 10.1039/D4CP03229B. DOI: 10.1039/D4CP03229B.

(44) Mallo, M.; Agúndez, M.; Cabezas, C.; Roncero, O.; Cernicharo, J.; Molpeceres, G. Ion-Molecule Routes Towards Cycles in TMC-1. *Astron. Astrophys.* **2025**, *704*, A249. https://doi.org/10.1051/0004-6361/202557647

(45) Hincelin, U.; Commerçon, B.; Wakelam, V.; Hersant, F.; Guilloteau, S.; Herbst, E. Chemical and Physical Characterization of Collapsing Low-Mass Prestellar Dense Cores. *Astrophys. J.* **2016**, *822*, 12. DOI: 10.3847/0004-637X/822/1/12.

(46) Wakelam, V.; Ruaud, M.; Gratier, P.; Bonnell, I. A. Influence of Galactic Arm Scale Dynamics on the Molecular Composition of the Cold and Dense ISM – II. Molecular Oxygen Abundance. *Mon. Not. R. Astron. Soc.* **2019**, *486*, 4198-4202. DOI: 10.1093/mnras/stz1122

(47) Wakelam, V.; Dartois, E.; Chabot, M.; Spezzano, S.; Navarro-Almaida, D.; Loison, J.-C.; Fuente, A. Efficiency of Non-Thermal Desorptions in Cold-Core Conditions. *Astron. Astrophys.* **2021**, *652*, A63. https://doi.org/10.1051/0004-6361/202039855

(48) Coutens, A.; Jørgensen, J. K.; van der Wiel, M. H. D.; Müller, H. S. P.; Lykke, J. M.; Bjerkeli, P.; Bourke, T. L.; Calcutt, H.; Drozdovskaya, M. N.; Favre, C.; et al. The ALMA-PILS Survey: First Detections of Deuterated Formamide and Deuterated Isocyanic Acid in the Interstellar Medium. *Astron. Astrophys.* **2016**, *590*.

(49) Jørgensen, J. K.; van der Wiel, M. H. D.; Coutens, A.; Lykke, J. M.; Müller, H. S. P.; van Dishoeck, E. F.; Calcutt, H.; Bjerkeli, P.; Bourke, T. L.; Drozdovskaya, M. N.; et al. The ALMA Protostellar Interferometric Line Survey (PILS). *Astron. Astrophys.* **2016**, *595*, A117. https://doi.org/10.1051/0004-6361/201628648

(50) Chaabouni, H.; Diana, S.; Nguyen, T.; Dulieu, F. Thermal Desorption of Formamide and Methylamine from Graphite and Amorphous Water Ice Surfaces. *Astron. Astrophys.* **2018**, *612*, A47. https://doi.org/10.1051/0004-6361/201731006

(51) Minissale, M.; Aikawa, Y.; Bergin, E.; Bertin, M.; Brown, W. A.; Cazaux, S.; Charnley, S. B.; Coutens, A.; Cuppen, H. M.; Guzman, V.; et al. Thermal Desorption of Interstellar Ices: A Review on the Controlling Parameters and Their Implications from Snowlines to Chemical Complexity. *ACS Earth Space Chem.* **2022**, *6*, 597-630. DOI: 10.1021/acsearthspacechem.1c00357.

(52) Rimola, A.; Skouteris, D.; Balucani, N.; Ceccarelli, C.; Enrique-Romero, J.; Taquet, V.; Ugliengo, P. Can Formamide Be Formed on Interstellar Ice? An Atomistic Perspective. *ACS Earth Space Chem.* **2018**, *2*, 720-734. DOI: 10.1021/acsearthspacechem.7b00156.

(53) Daniel, F.; Gérin, M.; Roueff, E.; Cernicharo, J.; Marcelino, N.; Lique, F.; Lis, D. C.; Teyssier, D.; Biver, N.; Bockelé-Morvan, D. Nitrogen Isotopic Ratios in Barnard 1: A Consistent Study of the $N_2H^+$, $NH_3$, CN, HCN, and HNC Isotopologues. *Astron. Astrophys.* **2013**, *560*, A3. http://dx.doi.org/10.1051/0004-6361/201321939

(54) Redaelli, E.; Bizzocchi, L.; Caselli, P.; Harju, J.; Chacon-Tanarro, A.; Leonardo, E.; Dore, L. $^{14}N/^{15}N$ Ratio Measurements in Prestellar Cores with $N_2H^+$ : New Evidence of $^{15}$N-

Antifractionation. *Astron. Astrophys.* **2018**, *617*, A7. https://doi.org/10.1051/0004-6361/201833065

(55) Furuya, K.; Aikawa, Y. Depletion of Heavy Nitrogen in the Cold Gas of Star-Forming Regions. *Astrophys. J.* **2018**, *857*, 105. DOI: 10.3847/1538-4357/aab768

(56) Loison, J.-C.; Wakelam, V.; Gratier, P.; Hickson, K. M. Chemical Nitrogen Fractionation in Dense Molecular Clouds. *Mon. Not. R. Astron. Soc.* **2018**, *484*, 2747-2756. DOI: 10.1093/mnras/sty3293

(57) Lis, D. C.; Wootten, A.; Gerin, M.; Roueff, E. Nitrogen Isotopic Fractionation in Interstellar Ammonia. *Astrophys. J. Lett.* **2010**, *710*, L49. DOI: 10.1088/2041-8205/710/1/L49

(58) Kahane, C.; Al-Edhari, A. J.; Ceccarelli, C.; López-Sepulcre, A.; Fontani, F.; Kama, M. First Measurement of the $^{14}$N/$^{15}$N Ratio in the Analog of the Sun Progenitor OMC-2 FIR4. *Astrophys. J.* **2018**, *852*, 130. DOI: 10.3847/1538-4357/aa9e88

(59) Coutens, A.; Zakharenko, O.; Lewen, F.; Jørgensen, J. K.; Schlemmer, S.; Müller, H. S. P. Laboratory Spectroscopic Study of the $^{15}$N Isotopomers of Cyanamide, H$_2$NCN, and a Search for them toward IRAS 16293−2422 B. *Astron. Astrophys.* **2019**, *623*, 10.1051/0004-6361/201834605.

# The Reaction between Atomic Carbon and Molecular Nitrogen as a Source of Cyanamide and Carbodiimide on Interstellar Ices

Kevin M. Hickson,*,[1] Jean-Christophe Loison[1] and Audrey Coutens[2]

[1] Univ. Bordeaux, CNRS, Bordeaux INP, ISM, UMR 5255, F-33400 Talence, France

[2] Institut de Recherche en Astrophysique et Planétologie, Université de Toulouse, CNRS, CNES, 9 av. du Colonel Roche, 31028 Toulouse Cedex 4, France

Email: kevin.hickson@u-bordeaux.fr

**Table S1** Summary of reaction review.

Species or reactions highlighted in gray such as HNNC are not included in the network

**Gas phase reactions:**
$\Delta E$ in kJ/mol, k = $\alpha \times (T/300)^{\beta} \times \exp(-\gamma/T)$ cm$^3$ molecule$^{-1}$ s$^{-1}$

|   | Reaction | ΔE | α | β | γ (K) | Comments |
|---|---|---|---|---|---|---|
| 1. | NH + H$_2$CN → NH$_2$ + HCN<br>→ H$_2$CNN + H | -273<br>-116 | 3.0e-11<br>3.0e-11 | 0<br>0 | 0<br>0 | Radical-radical reaction |
| 2. | NH + H$_2$NC → NH$_2$ + HCN<br>→ HNCNH + H | -398<br>-241 | 3.0e-11<br>3.0e-11 | 0<br>0 | 0<br>0 | Radical-radical reaction |
| 3. | NH$_2$ + H$_2$CN → NH$_3$ + HCN<br>→ H + HCNNH$_2$ | -335<br>+161 | 6.0e-11<br>0 | 0 | 0 | Deduced from (Yelle *et al.* 2010) |
| 4. | NH$_2$ + H$_2$NC → NH$_3$ + HNC<br>→ H + H$_2$NCNH | -408<br>+42 | 6.0e-11<br>0 | 0 | 0 | Radical-radical reaction |
| 5. | CN + NH$_2$ → NH + HCN<br>→ H + HNCN | -145<br>-101 | 1.0e-10<br>1.0e-10 | 0<br>0 | 0<br>0 | Radical-radical reaction |
| 6. | CN + NH$_3$ → HCN + NH$_2$ | -83 | 2.77e-11 | -1.14 | 0 | (Blitz *et al.* 2009, Meads *et al.* 1993, Talbi & Smith 2009, Sims *et al.* 1994) |
| 7. | CN + CH$_3$NH$_2$ → HCN + CH$_3$NH<br>→ HCN + CH$_2$NH$_2$<br>→ NH$_2$CN + CH$_3$<br>→ CH$_3$NHCN + H | -116<br>-146<br>-151<br>-76 | 4.0e-10<br>0<br>0<br>0 | 0 | 0 | Global rate from (Sleiman *et al.* 2018). Their branching ratios are derived from theoretical calculations for pathways with a barrier. They missed the most favorable pathway: CN + CH$_3$NH$_2$ → CH$_3$NH$_2$...CN → HCN + CH$_3$NH with a mechanism similar to CN + NH$_3$ reaction. |

**Grain surface reactions** (s- is omitted):
$\Delta E$ in kJ/mol, **BR**=branching ratio, **γ (K)** = activation barrier

|   | Reaction | ΔE | BR | γ (K) | Comments |
|---|---|---|---|---|---|
| 8. | H + NCN → HNCN | -344 | 1 | 0 | Radical-radical reaction |
| 9. | H + CNN → HCNN<br>→ HNNC<br>→ c-HNNC<br>→ c-HCNN<br>→ HNCN<br>→ CN + NH<br>→ CH + N$_2$ | -308<br>-295<br>-171<br>-283<br>-452<br>+27<br>-171 | 0.3<br>0<br>0<br>0<br>0.7<br>0<br>0 | 0<br><br><br><br>0<br><br> | CNN is a triplet state with •C=N=N• configuration. Both HCNN and HNNC should be produced. Some HCNN should isomerize into HNCN and most of the HNNC (this work). |
| 10. | H + HNCN → NH$_2$CN<br>→ HNCNH<br>→ NH + HNC | -399<br>-394<br>+8 | 0.8<br>0.2 | 0<br>0 | The lonely electron of HNCN is located on nitrogen atoms H-N•-C≡N n H-N=C=N• so both NH$_2$CN and HNCNH should be produced. However, NH$_2$CN seems to be favored in presence of water cluster (this work). |
| 11. | H + HCNN → H$_2$CNN<br>→ CH$_2$(a$^1$A$_1$) + N$_2$ | -410<br>-237 | 0.0<br>1 | 0<br>0 | No exit TS on the singlet surface leading to CH$_2$(a$^1$A$_1$) + N$_2$. CH$_2$(a$^1$A$_1$) + N$_2$ can give back to |

| | | | | | | |
|---|---|---|---|---|---|---|
| | | → $CH_2(X^3B_1) + N_2$ | -291 | 0.4 | 0 | $H_2CNN$ or relaxed to $CH_2(X^3B_1) + N_2$ or $CH_2(a^1A_1)$ can react with $H_2O$ species nearby. $CH_2(X^3B_1) + N_2$ can lead to $H_2CNN$ in its triplet excited state (+140 kJ/mol above the ground singlet state) through a TS localized 75 kJ/mol above the $CH_2(X^3B_1) + N_2$ energy. The TS for $HCNNH \rightarrow H_2CNN$ is located +96 kJ/mol above the H + HCNN energy) but HCNNH can evolve toward dissociation into HNC + NH (through a TS located -58 kJ/mol below the entrance H + HCNN) or toward $NH_2CN$. |
| | | → HCNNH | -294 | 9 | | |
| | | → $HCN + NH(X^3\Sigma^-)$ | -188 | 0 | | |
| | | → $NH_2CN$ | -543 | 0 | | |
| | | → HNCNH | -538 | 0.3 | | |
| | | | | 0.2 | | |
| | | | | 0 | | |
| 12. | $H + NH_2CN$ | → $NH_2CHN$ | -111 | 0 | 3550 | M06-2X/AVTZ calculations (this work). We mix $NH_2CHN$ and $NH_2CNH$ to avoid to introduce too many species ending into $NH_2CHNH$ |
| | | → $NH_2CNH$ | -86 | 1 | 3800 | |
| | | → $HNCN + H_2$ | -27 | 0 | 6440 | |
| | | → $HCN + NH_2$ | -24 | 0 | | |
| 13. | $H + HNCNH$ | → $NH_2CNH$ | -91 | 1 | 4940 | M06-2X/AVTZ calculations (this work). We mix HNCHNH and $NH_2CNH$ to avoid to introduce too many species ending into $NH_2CHNH$ |
| | | → HNCHNH | -88 | 0 | 4930 | |
| | | → $HNCN + H_2$ | -32 | 0 | 5200 | |
| | | → $HNC + NH_2$ | +25 | 0 | | |
| 14. | $H + H_2CNN$ | → $H_2CNNH$ | -150 | 0 | 3020 | M06-2X/AVTZ calculations (this work). We neglect the $H_2CNNH$ formation considering the large barrier. We also neglect the abstraction channel considering the low barrier for the $N_2$ formation. |
| | | → $CH_3 + N_2$ | -325 | 1 | 700 | |
| | | → $HCNN + H_2$ | -16 | 0 | - | |
| 15. | $H + NH_2CNH$ | → $NH_2CHNH$ | -423 | 0.2 | 0 | The TS for $NH_2CHNH \rightarrow NH_3 + HNC$ is located 343 kJ/mol above the $NH_2CHNH$ energy. So, some of the $NH_2CHNH$ should dissociate. |
| | | → $NH_3 + HNC$ | -325 | 0.3 | 0 | |
| | | → $NH_3 + HCN$ | -377 | 0 | 0 | |
| | | → $H_2 + NH_2CN$ | -339 | 0.5 | 0 | |
| 16. | $H + NH_2CHNH$ | → $NH_2CH_2NH$ | -66 | 0 | 4820 | M06-2X/AVTZ calculations (this work). We mix $NH_2CH_2NH$ and $NH_2CHNH_2$ and $NH_2CHN$ and $NH_2CNH$. |
| | | → $NH_2CHNH_2$ | -102 | 0.5 | 4800 | |
| | | → $H_2 + NH_2CNH$ | -3 | 0 | | |
| | | → $H_2 + NH_2CHN$ | -28 | 0.5 | 4150 | |
| | | → $H_2 + NHCHNH$ | 0 | 0 | 7800 | |
| 17. | $H + NH_2CHNH_2$ | → $NH_2CH_2NH_2$ | -355 | 0.0 | 0 | The TS for $NH_2CH_2NH_2 \rightarrow NH_3 + CH_2NH$ is located 141 kJ/mol above the $NH_2CH_2NH_2$ energy. So, most of the $NH_2CH_2NH_2$ dissociate. |
| | | → $NH_3 + CH_2NH$ | -317 | 1 | 0 | |
| | | | | 0.9 | | |
| | | | | 9 | | |
| 18. | $H + NH_2CH_2NH_2$ | → $H_2 + NH_2CHNH_2$ | -71 | 1 | 3400 | M06-2X/AVTZ calculations (this work). We mix $NH_2CH_2NH$ and $NH_2CHNH_2$. |
| | | → $H_2 + NH_2CH_2NH$ | -35 | 0 | 3400 | |
| | | → $NH_3 + CH_2NH_2$ | -120 | 0 | large | |
| 19. | $C + N_2$ | → CNN | -124 | 1 | 0 | No barrier (Hickson *et al.* 2016, Lu *et al.* 2023, Urzúa-Leiva & Denis-Alpizar 2021). |
| | | → N + CN | +196 | | | |
| 20. | $CH + N_2$ | → HCNN | -137 | 1 | 0 | (Brownsword *et al.* 1996, Le Picard & Canosa 1998, Berman *et al.* 2007) |
| 21. | $CH_3 + NH_2CNH$ | → $NH_2CN + CH_4$ | -342 | 1 | 0 | Radical-radical reaction, we neglect the adduct ($H_2N-C(CH_3)=NH$) formation. |
| 22. | $CH_3 + NH_2CHNH_2$ | → $NH_2CHNH + CH_4$ | -326 | 1 | 0 | Radical-radical reaction, we neglect the adduct ($H_2N-CH(CH_3)-NH_2$) formation. The TS for $NH_2CHNH \rightarrow NH_3 + HNC$ is located 343 kJ/mol above the $NH_2CHNH$ energy. |
| | | → $NH_3 + HNC + CH_4$ | -280 | 0 | | |
| 23. | N + CN | → NCN | -464 | 1 | 0 | The unpaired electron of CN is primarily localized on the carbon atom, so the main product formed will be NCN. Given its greater stability, its isomerization to CNN is unlikely. |
| | | → $^3CNN$ | -356 | 0 | 0 | |
| 24. | $N + H_2CN$ | → $H_2CNN$ | -310 | 0 | | The unpaired electron of $H_2CN$ is primarily localized on the nitrogen atom, so the main product formed will be $H_2CNN$ in its excited triplet state. $H_2CNN$ will mostly dissociate into $CH_2(X^3B_1) + N_2$ through a TS localized -251 kJ/mol below the $N + H_2CN$. Very few, if any, $H_2CNN$ should be stabilized. |
| | | → $H_2CNN$ | -444 | 0 | | |
| | | → $CH_2(X^3B_1) + N_2$ | -326 | 1 | 0 | |
| | | → $CH_2(a^1A_1) + N_2$ | -272 | | | |
| 25. | $N + {^2HCNH}$ | → HNCHN | -329 | 0 | 0 | The TS from HNCHN toward HNCHN is located -196 kJ/mol below the N + HNCHN level on the singlet surface. We cannot find the exit TS from |
| | | → HNCHN | -238 | 0 | 0 | |
| | | → HNCNH | -605 | 0.5 | 0 | |

| # | Reactants | Products | ΔH (kJ/mol) | branching | barrier | Comments |
|---|---|---|---|---|---|---|
| | | → HCN + NH | -255 | 0.5 | 0 | HNCHN toward HCN + NH but considering the reactivity of NH, it should be around 50 kJ/mol above the HCN + NH level so around -200 kJ/mol. We consider only HNCNH formation but this reaction is obviously complex. We consider also NH + HCN production. |
| 26. | N + CH$_3$NH | → CH$_3$ + N$_2$ + H | -282 | 1 | 0 | Radical-radical reaction. |
| 27. | N + CH$_2$NH$_2$ | → NCH$_2$NH$_2$ | -282 | 0 | 0 | Calculations at M06-2X/AVTZ level. We favor HCN + NH$_3$, the TS for NH$_2$CHNH → HCN + NH$_3$ being localized -306 kJ/mol below the N + CH$_2$NH$_2$ energy. |
| | | → H$_2$CN + NH$_2$ | -189 | 0 | 0 | |
| | | → NH$_2$CHN + H | -171 | 0 | 0 | |
| | | → NH$_2$CNH + H | -146 | 0 | 0 | |
| | | → NH$_2$CHNH | -569 | 0 | 0 | |
| | | → HCN + NH$_3$ | -523 | 1 | 0 | |
| | | → NH$_2$CN + H$_2$ | -486 | 0 | 0 | |
| 28. | N + CNN | → CN + N$_2$ | -574 | 1 | 0 | CNN is a triplet state with •C=N=N• configuration. The attack of the nitrogen atom on C or terminal N leads in both cases to CN + N$_2$ |
| 29. | N + HCNN | → HCN + N$_2$ | -790 | 1 | 0 | The attack of the nitrogen atom on C or terminal N leads in both cases to HCN + N$_2$ |
| 30. | N + H$_2$CNN | → H$_2$CN + N$_2$ | -485 | 0.2 | 0 | Radical-radical reaction. Most of the H$_2$CN should dissociate into H + HCN considering the exothermicity of the reaction. |
| | | → H + HCN + N$_2$ | -380 | 0.8 | | |
| 31. | NH + CN | → HNCN | -479 | 1 | 0 | |
| | | → NCN + H | -136 | 0 | 0 | |
| 32. | NH + H$_2$CN | → NH$_2$ + HCN | -273 | 0.5 | 0 | H$_2$CNNH is likely produced but is not considered in the network. |
| | | → H$_2$CNNH | -265 | 0.5 | 0 | |
| | | → H$_2$CNN + H | -116 | 0 | 0 | |
| | | → HCNNH + H | 0 | 0 | 0 | |
| | | → CH$_2$ + N$_2$ + H | +2 | 0 | 0 | |
| 33. | NH + HCNH | → NH$_2$ + HCN | -305 | 0.5 | 0 | |
| | | → HNCNH + H | -277 | 0.5 | 0 | |
| 34. | NH + CH$_2$NH$_2$ | → NH$_2$ + CH$_2$NH | -221 | 0.5 | 0 | |
| | | → HNC + NH$_3$ + H | -143 | 0.5 | 0 | |
| 35. | NH + CH$_3$NH | → NH$_2$ + CH$_2$NH | -251 | 0.5 | 0 | |
| | | → N$_2$ + H$_2$ + CH$_3$ | -380 | 0.5 | 0 | |
| 36. | NH$_2$ + CN | → NH$_2$CN | -500 | 0.8 | 0 | Some NH$_2$CN may isomerize into HNCNH (this work). |
| | | → HNCNH | -496 | 0.2 | 0 | |
| | | → HNCN + H | -101 | 0 | 0 | |
| 37. | NH$_2$ + H$_2$CN | → H$_2$CNNH$_2$ | -258 | 0.5 | 0 | H$_2$CNNH$_2$ is likely produced but is not considered in the network. The TS (H$_2$CNNH$_2$→ HCN + NH$_3$) = -10 kJ/mol below the NH$_2$ + H$_2$CN energy. |
| | | → NH$_3$ + HCN | -335 | 0.5 | 0 | |
| 38. | NH$_2$ + HCNH | → NH$_2$CHNH | -413 | 0.5 | 0 | |
| | | → NH$_3$ + HCN | -367 | 0.25 | 0 | |
| | | → NH$_3$ + HNC | -314 | 0.25 | | |
| 39. | CN + NH$_3$ | → HCN + NH$_2$ | -83 | 1 | 0 | (Blitz et al. 2009, Meads et al. 1993, Talbi & Smith 2009, Sims et al. 1994) |
| 40. | HCO + CNN | → HCNN + CO | -248 | 0.5 | 0 | CNN is a triplet state with •C=N=N• configuration |
| | | → HNNC + CO | -235 | 0 | | |
| | | → HNCN + CO | -392 | 0.5 | 0 | |
| | | → CNNCHO | -230 | 0 | | |
| | | → NNCCHO | -240 | 0 | | |
| 41. | HCO + NCN | → HNCN + CO | -283 | 1 | 0 | |
| | | → NCNCHO | -268 | 0 | | |
| 42. | HCO + HCNN | → H$_2$CNN + CO | -350 | 0.0 | 0 | |
| | | → CH$_2$ + N$_2$ + CO | -232 | 1 | 0 | |
| | | → HCNNCHO | -271 | 0.49 | | |
| | | → HCN + NCHO | -138 | | | |
| | | → HCN + HNCO | -496 | 0 | 0 | |
| | | → HCOCHNN | -374 | 0 | | |
| | | | | 0.5 | | |

S3

| # | Reaction | | ΔH | branching | Ea | Notes |
|---|---|---|---|---|---|---|
| | | | 0 | | | |
| 43. | HCO + HNCN | → NH$_2$CN + CO | -339 | 0.8 | 0 | |
| | | → HNCNH + CO | -334 | 0.2 | 0 | |
| | | → NCN + H$_2$CO | -18 | 0 | | |
| | | → HCONHCN | -359 | 0 | | |
| 44. | HCO + NH$_2$CN | → HNCN + H$_2$CO | +37 | 0 | | Endothermic reaction. |
| 45. | HCO + HNCNH | → HNCN + H$_2$CO | +33 | 0 | | Endothermic reaction. |
| 46. | HCO + NH$_2$CNH | → NH$_2$CN + H$_2$CO | -275 | 0.5 | 0 | |
| | | → NH$_2$CHNH + CO | -363 | 0.5 | 0 | |
| 47. | HCO + NH$_2$CHNH$_2$ | → NH$_2$CHNH + H$_2$CO | -259 | 0.5 | 0 | |
| | | → NH$_2$CH$_2$NH$_2$ + CO | -295 | 0.5 | 0 | |
| 48. | CH$_2$OH + CNN | → HCNN + H$_2$CO | -181 | 0.5 | | CNN is a triplet state with •C=N=N• configuration |
| | | → HNNC + H$_2$CO | -168 | 0 | | |
| | | → HNCN + H$_2$CO | -325 | 0.5 | | |
| | | → C$_2$H$_2$ + OH + N$_2$ | -264 | 0.5 | | |
| 49. | CH$_2$OH + NCN | → HNCN + H$_2$CO | -217 | 1 | 0 | |
| 50. | CH$_2$OH + HCNN | → H$_2$CNN + H$_2$CO | -283 | 0.2 | 0 | |
| | | → CH$_2$ + N$_2$ + H$_2$CO | -165 | 0.8 | 0 | |
| 51. | CH$_2$OH + HNCN | → NH$_2$CN + H$_2$CO | -272 | 0.8 | 0 | |
| | | → HNCNH + H$_2$CO | -268 | 0.2 | 0 | |
| 52. | CH$_2$OH + NH$_2$CN | → HNCN + CH$_3$O | +6 | 0 | | Endothermic reaction. |
| 53. | CH$_3$O + CNN | → HCNN + H$_2$CO | -215 | 0 | | CNN is a triplet state with •C=N=N• configuration |
| | | → HNNC + H$_2$CO | -202 | 0 | | |
| | | → HNCN + H$_2$CO | -359 | 0.5 | | |
| | | → N$_2$ + CO + CH$_3$ | -529 | 0.5 | | |
| 54. | CH$_3$O + NCN | → HNCN + H$_2$CO | -251 | 1 | 0 | |
| 55. | CH$_3$O + HCNN | → H$_2$CNN + H$_2$CO | -317 | 0.1 | 0 | |
| | | → CH$_2$ + N$_2$ + H$_2$CO | -199 | 0.9 | 0 | |
| 56. | CH$_3$O + HNCN | → NH$_2$CN + H$_2$CO | -306 | 0.8 | 0 | |
| | | → HNCNH + H$_2$CO | -302 | 0.2 | 0 | |
| 57. | CH$_3$O + NH$_2$CN | → HNCN + CH$_3$OH | -28 | 1 | 1760 | M06-2X/AVTZ calculations (this work) |
| 58. | CH$_3$O + HNCNH | → HNCN + CH$_3$OH | -32 | 1 | 1580 | M06-2X/AVTZ calculations (this work) |

Berman M.R., Tsuchiya T., Gregusova A., Perera S.A., Bartlett R.J., 2007, J. Phys. Chem A*,* 111**,** 6894

Blitz M.A., Seakins P.W., Smith I.W.M., 2009, Physical Chemistry Chemical Physics*,* 11**,** 10824

Brownsword R.A., Herbert L.B., Smith I.W.M., Stewart D.W.A., 1996, J. Chem. Soc. Faraday Trans.**,** 1087

Hickson K.M., Loison J.-C., Lique F., Klos J., 2016, The Journal of Physical Chemistry A*,*

Le Picard S.D., Canosa A., 1998, Geophys. Res. Lett. *,* 25**,** 485

Lu D., Urzúa-Leiva R., Denis-Alpizar O., Guo H., 2023, The Journal of Physical Chemistry A*,*

Meads R.F., Maclagan R.G.A.R., Phillips L.F., 1993, J. Phys. Chem.*,* 97**,** 3257

Sims I.R., Queffelec J.L., Defrance A., Rebrion-Rowe C., Travers D., Bocherel P., Rowe B.R., Smith I.W.M., 1994, J. Chem. Phys.*,* 100**,** 4229

Sleiman C., El Dib G., Rosi M., Skouteris D., Balucani N., Canosa A., 2018, Physical Chemistry Chemical Physics*,* 20**,** 5478

Talbi D., Smith I.W.M., 2009, Physical Chemistry Chemical Physics*,* 11**,** 8477

Urzúa-Leiva R., Denis-Alpizar O., 2021, The Journal of Physical Chemistry A*,* 125**,** 8168

Yelle R.V., Vuitton V., Lavvas P., Klippenstein S.J., Smith M.A., Horst S.M., Cui J., 2010, Faraday Discuss.*,* 147**,** 31

**Calculated species geometries at the ωB97X-D3/def2-TZVP level of theory**

**Gas-phase calculations**
CNN
  C     -2.31779114895971    0.57636525845194    0.00000010907878
  N     -1.29024761746858    1.24566184388544   -0.00000021880282
  N     -0.29181668247170    1.90415536316261    0.00000010972404

c-CNN
  C     -2.32134159571596    0.61049855034587    0.00548478521397
  N     -1.17657997867200    1.18961320279913   -0.01957676493300
  N     -2.63938495071204    1.83491126835500    0.22091950931902

NCN
  C     -2.04610872550570    1.20985027620061    0.06868514794465
  N     -0.94782246774331    0.70687959425692   -0.11462051949083
  N     -3.14003741615336    1.72194567481965    0.25298534854314

HCNN
  C     -2.15062971058207    0.15516621232626   -0.00006569145700
  N     -1.37590188271708    1.16480210347108   -0.00001294026948
  N     -0.54827393623514    1.94987352532115    0.00002996813248
  H     -3.20937382966570    0.41298843008151    0.00005505909401

c-HCNN
  N     -1.83138978821954    2.02256398699062   -0.00587001624877
  C     -1.99400424931131    0.73819801435450    0.00042348901620
  N     -0.72005026041812    0.96942842950096    0.00322920030735
  H     -2.73873570205099   -0.04736043084611    0.00222332692522

HNNC
  N     -2.17796899997937    0.10948801847026   -0.00007152049272
  N     -1.39846632098774    1.13091922606884   -0.00001023418547
  C     -0.56841927779981    1.97468185313242    0.00003225194642
  H     -3.13932476043307    0.46774117352848    0.00005589823178

c-HNNC
  N     -1.98362718047858    0.55343011050641   -0.20816839805288
  N     -1.74990427768135    1.90778299363218   -0.25395742236517
  C     -0.77847022401593    1.11918822289147    0.21830166787034
  H     -2.77217767702410    0.10242894416992    0.24383054804772

HNCN
  N     -2.18475578072506    0.14505158773198   -0.00006628074310
  C     -1.38239285328330    1.12847053967396   -0.00001231353144
  N     -0.55497838509777    1.97592995198540    0.00003148571640
  H     -3.16205234009387    0.43337819180866    0.00005350395814

$H_2CNN$
  C     -2.11527508441566    0.15377158840048   -0.00825935383783
  N     -1.29716016953304    1.15106014819625   -0.00201584949867

| | | | |
|---|---|---|---|
| N | -0.58335252296212 | 2.02143052213143 | 0.00322810334391 |
| H | -3.16984131559379 | 0.37281799909610 | 0.00289879807700 |
| H | -1.69421791039539 | -0.83752305852425 | -0.02401405138441 |

c-$H_2CNN$

| | | | |
|---|---|---|---|
| C | -1.99990434919548 | 0.35595707353203 | -0.01527572443429 |
| N | -0.85891925832271 | 1.06813286241654 | 0.56798950875356 |
| N | -1.20554274885431 | 1.48108222185759 | -0.51710051343408 |
| H | -2.98228330583509 | 0.57558078712379 | 0.38221204315938 |
| H | -1.81319734069235 | -0.61919574562995 | -0.44598766734458 |

$H_2NNC$

| | | | |
|---|---|---|---|
| N | -2.11597734529375 | -0.15388137527357 | 0.30374267719545 |
| N | -1.28214406587503 | 0.86530575243463 | 0.00850315624844 |
| C | -0.54944160508133 | 1.75702771934030 | -0.14771452436864 |
| H | -2.99154686029921 | -0.02202393656499 | -0.18717818471531 |
| H | -1.68780297285060 | -1.02692179523637 | 0.02220305564007 |

c-$H_2NNC$

| | | | |
|---|---|---|---|
| N | -1.98122522965234 | 0.05037082476811 | 0.09438181463713 |
| N | -0.90216329988499 | 0.73153477723045 | -0.66959250672920 |
| C | -1.12188586163716 | 1.34585168017427 | 0.39200857226427 |
| H | -2.89938554750896 | 0.12842034513814 | -0.30966363973257 |
| H | -1.72225291071640 | -0.83667126261098 | 0.49242193956037 |

$NH_2CN$

| | | | |
|---|---|---|---|
| N | -2.16226620392981 | -0.13353184886104 | -0.04372731243790 |
| N | -0.31970755515364 | 1.49928813480884 | -0.41802807025555 |
| C | -1.17147594183166 | 0.74997866500020 | -0.22058022687849 |
| H | -3.05564564705838 | 0.26196576680688 | 0.20138968699445 |
| H | -1.91781750142636 | -0.95819435305488 | 0.48050210257749 |

$NH_2NC$

| | | | |
|---|---|---|---|
| N | -1.22847930637496 | 0.74817934380684 | 0.15349613070032 |
| N | -1.25681428077829 | 2.09728875670068 | 0.16476520149773 |
| C | -1.31968339252014 | 3.25594852842100 | 0.26445553975960 |
| H | -1.79591737515277 | 0.40884632409632 | -0.61318509884340 |
| H | -0.27184725661046 | 0.43550411499296 | 0.04386447929377 |

HNNCH

| | | | |
|---|---|---|---|
| N | -1.88856879922941 | 0.34526378361275 | -0.09980861831936 |
| N | -1.15139627043216 | 1.30860050400033 | 0.11155360024148 |
| C | -0.41716608857459 | 2.15496558392254 | 0.49037386790105 |
| H | -2.76199464895240 | 0.67504902778608 | -0.50984654810836 |
| H | 0.39722830338860 | 2.66679405257825 | 0.00778353308518 |

c-HNNCH

| | | | |
|---|---|---|---|
| N | -1.58122469820845 | 0.66774820692139 | 0.23737029516512 |
| N | -1.82123499169525 | 2.26144413458627 | 0.34818346447214 |
| C | -0.71812570003227 | 1.77070797045401 | 0.11121941142123 |
| H | -1.90970991341835 | 0.41235439804434 | -0.69615956192479 |
| H | 0.32262005115432 | 2.01681679929399 | -0.05025755973369 |

HNCNH(1)

| | | | |
|---|---|---|---|
| N | -3.32414904626964 | 0.22611983111838 | 4.76884326198633 |
| N | -0.91585408804725 | -0.07106556245615 | 4.78210383373135 |
| C | -2.13035512753682 | -0.00708235210377 | 4.79330496712164 |
| H | -0.46322164938557 | -0.81333821673751 | 4.27005367046501 |
| H | -3.92320502253694 | -0.17643796408850 | 5.47392594368551 |

HNCNH(2)

| | | | |
|---|---|---|---|
| N | -3.23476455502869 | 0.32454359447514 | 5.16766352862813 |
| N | -0.89550035280482 | 0.00107344353470 | 4.60921143459158 |
| C | -2.06648628896129 | 0.08533237839154 | 4.92785969337627 |
| H | -3.92869592479538 | -0.39344242811139 | 5.02288508444296 |
| H | -0.22068813480982 | -0.32500385028999 | 5.28493576836105 |

**Ice cluster calculations**

W18

| | | | |
|---|---|---|---|
| H | 1.51130970927166 | 0.98923813734242 | -0.98621451797292 |
| O | 1.38075973008310 | 2.41652474169391 | 0.06237494676342 |
| O | 1.32517569177874 | 1.39395101587019 | 2.63545565896749 |
| O | 2.64342953464707 | -0.85087887599910 | 2.37184886127335 |
| O | 1.08339857160992 | -2.98496815456198 | 1.63109517301428 |
| O | 4.14276832183148 | 1.59298600118772 | -1.07679723472606 |
| O | 4.72606440686168 | -0.67497496596701 | 0.30741157886565 |
| O | 4.19903920532740 | -2.68691410533017 | -1.63358748542464 |
| O | 1.56333399460962 | -2.29274301714194 | -1.08441055390778 |
| O | 1.67943512812236 | 0.38584748527367 | -1.74238780296344 |
| O | -0.82503248162015 | 0.41158000247452 | -2.65311621642241 |
| O | -1.22435726969604 | -2.17299662332698 | -1.62635190310662 |
| O | -3.76909107261714 | -3.05247866939762 | -1.03201571727585 |
| O | -1.46614275420964 | 2.26658320528490 | -0.56648615040274 |
| O | -1.45120874914766 | 1.36641661215493 | 2.07970940051691 |
| O | -3.11705987175577 | -0.62163324840653 | 2.39914678612808 |
| O | -1.74767144907506 | -2.62054363773251 | 1.10291056881427 |
| O | -4.16511421272346 | 2.07387129604580 | -1.03462762211652 |
| O | -4.62175058757741 | -0.44174347476109 | -0.02831166352068 |
| H | -4.92941142669456 | 2.58700236175868 | -0.77167642568861 |
| H | -4.34178780511937 | 1.15393472005879 | -0.74288146648507 |
| H | -4.22635451356888 | -0.49118545533864 | 0.85647712170022 |
| H | -4.45481807626900 | -1.30274474573835 | -0.43973584620632 |
| H | 0.11638040670262 | -2.96311868897530 | 1.66791176998791 |
| H | 1.28870110130792 | -2.88781975497929 | 0.68695899439665 |
| H | 0.48883771290834 | 2.74526332355073 | -0.12432221619601 |
| H | 1.38273759485583 | 2.17062614732035 | 1.00760706238734 |
| H | -2.42223347370284 | 2.44697810720116 | -0.66521871360768 |
| H | -1.29007690080966 | 1.62146801973942 | -1.27231700107545 |
| H | -2.44400181999655 | -3.20016371275337 | 0.77182646445735 |
| H | -2.20975550981203 | -1.95341545797285 | 1.64974824333994 |
| H | -2.49672480186346 | 0.15704744579791 | 2.33728696969766 |
| H | -3.46486905288072 | -0.64179849939753 | 3.29144880564716 |
| H | -2.94050401521144 | -2.83818215862135 | -1.50383842329226 |
| H | -4.26598976177078 | -3.65638653797153 | -1.58507563618322 |
| H | -1.22038816767315 | -1.27232359738070 | -1.99836196415254 |
| H | -1.26017214778350 | -2.07810451819982 | -0.65587585630624 |
| H | -0.52524796368633 | 1.30473476994312 | 2.36608027418860 |
| H | -1.43789105147258 | 1.65787253055873 | 1.14751953655652 |
| H | 0.15330264730530 | 0.43811694136223 | -2.52293038554668 |
| H | -1.02015747574438 | 0.71924989454418 | -3.53947701448619 |
| H | 3.32530537179835 | -0.86309505675999 | 1.68400120441755 |
| H | 2.06997346117902 | -1.63200427329083 | 2.21533576862738 |
| H | 3.65464324449927 | 2.27281182711373 | -0.59815104521259 |
| H | 3.45487341932840 | 1.17363825681404 | -1.62138301342215 |
| H | 4.59855009605135 | 0.19851376564312 | -0.13208196805093 |
| H | 5.58712247281040 | -0.64959063677600 | 0.72948343680962 |
| H | 2.38009645215362 | -2.65341772525180 | -1.47853746532373 |
| H | 0.77666935806666 | -2.57622756023391 | -1.57334496157042 |
| H | 4.66277269375149 | -2.43798490120399 | -2.43406343903623 |
| H | 4.46271646813018 | -2.04002434901867 | -0.95377334010876 |
| H | 1.82564231719600 | 0.53253068393837 | 2.55828514658744 |
| H | 1.63231299532794 | 1.82620868977801 | 3.43354720356908 |
| H | 1.69672830496585 | -0.53252157996175 | -1.40067292692309 |

W18-a..N$_2$

| | | | |
|---|---|---|---|
| H | 1.63892868272035 | 1.19266129995821 | -0.95772294796604 |
| O | 1.04512166276709 | 2.74376021080469 | -0.60590051724168 |
| O | 0.52812188351280 | 2.59920441621950 | 2.14402047855025 |
| O | 1.75928984003069 | 0.42539693814653 | 2.95926228056319 |
| O | 0.26811121283771 | -1.86184436302386 | 2.63776068827236 |
| O | 4.58951701873352 | -0.04840984426872 | -0.92100995475097 |
| O | 4.31216179314748 | -0.41085788752187 | 1.82718793194405 |
| O | 3.98516389857524 | -2.75978013936288 | 0.04453777605294 |
| O | 1.34349190821174 | -2.03415823422996 | 0.01746856376640 |
| O | 1.94149058355416 | 0.31544950051225 | -1.30604887564909 |
| O | -0.26940732258276 | -0.13327683533919 | -2.80998789062388 |
| O | -1.22482014841877 | -2.21405265109484 | -1.15273847513918 |
| O | -3.87021961007161 | -2.96426678816092 | -0.88480776292677 |
| O | -1.58953622681221 | 2.10480239402837 | -1.60312757216227 |
| O | -2.09806969005699 | 2.24461993112432 | 1.13833937411182 |
| O | -3.89554074221777 | 0.43453321798561 | 1.71509447235501 |
| O | -2.33700945825652 | -1.79553973755306 | 1.38640445192474 |
| O | -4.12745167409331 | 1.71830987921668 | -2.54594613178732 |
| O | -4.95317256986684 | -0.26212906568676 | -0.85175239516391 |
| H | -4.86333677785798 | 2.32412821368196 | -2.63601638260765 |
| H | -4.44810952230772 | 0.97663243103909 | -1.98658612991411 |
| H | -4.72555152587149 | -0.00937936329364 | 0.05698171398293 |
| H | -4.68148564816178 | -1.18787272282818 | -0.94561648404074 |
| H | -0.68277543052268 | -1.86921651499810 | 2.45635744242696 |
| H | 0.67192254133172 | -2.05326715120683 | 1.77493437066568 |
| H | 0.16907682448573 | 2.85103831469309 | -1.00544878586151 |
| H | 0.91762068311088 | 2.84755210545758 | 0.35405986609591 |
| H | -2.49485129034169 | 2.16503662950049 | -1.97226242950455 |
| H | -1.20877115675109 | 1.31702244470147 | -2.02370112206391 |
| H | -2.96360627457195 | -2.48417580168599 | 1.13561368595492 |
| H | -2.88315113518565 | -1.00635562248849 | 1.58300263439593 |
| H | -3.23362604673125 | 1.16530936019315 | 1.56217726288922 |
| H | -4.35242744097883 | 0.62893388031740 | 2.53462224716439 |
| H | -2.95782258666951 | -2.86572542253127 | -1.21907219564529 |
| H | -4.24922884996110 | -3.73515046346042 | -1.30861304336845 |
| H | -1.02739069633764 | -1.49397622138322 | -1.77697036159247 |
| H | -1.47686296394222 | -1.80297258686260 | -0.30387163795326 |
| H | -1.24067854325789 | 2.32602396009070 | 1.58855875572428 |
| H | -1.90315075045895 | 2.17306525035030 | 0.18404571020430 |
| H | 0.63784096665701 | 0.01635288832736 | -2.46011436135946 |
| H | -0.22277938495848 | -0.12790650231911 | -3.76721759403442 |
| H | 2.60452138729361 | 0.15399460871023 | 2.57092481795923 |
| H | 1.18523773315067 | -0.36959963177495 | 2.95577599662277 |
| H | 5.17375422945397 | 0.47189729351828 | -1.47371478781632 |
| H | 3.66468306162818 | 0.17444248318164 | -1.18131343507973 |
| H | 4.51956623674409 | -0.08186950764656 | 0.93255504094673 |
| H | 5.04766609586842 | -0.16812411546429 | 2.39219921062530 |
| H | 2.19338851671439 | -2.51016238830111 | -0.07803229714364 |
| H | 0.65365743004495 | -2.46068624984954 | -0.51412187255405 |
| H | 4.37840634684171 | -2.11617569395621 | -0.56019542369406 |
| H | 4.14275815169733 | -2.36635165391268 | 0.91339162558714 |
| H | 1.00871896752730 | 1.78257360031982 | 2.45494981120653 |
| H | 0.66544161133083 | 3.27018635663236 | 2.81445973347672 |
| H | 1.67032061828911 | -0.38836395670111 | -0.67956266604413 |
| N | -1.12591386125951 | 1.38761164614390 | 4.91632626863973 |
| N | -2.00818256100135 | 0.81894787449604 | 4.62998131173723 |

W18-a..C

| | | | |
|---|---|---|---|
| H | 1.48681913911458 | 1.29714868255240 | -0.90720187211160 |
| O | 1.10776025378548 | 2.82890882646650 | -0.32869550979897 |
| O | 1.13903093323835 | 2.10872152082843 | 2.33218553270735 |
| O | 2.61075833314359 | -0.04472787241110 | 2.38926137935432 |
| O | 0.41631486666516 | -3.09974714430216 | 1.85758236480269 |
| O | 4.45133508921791 | 1.12482409982861 | -1.33316233231929 |
| O | 4.82880875386070 | -0.76357131116139 | 0.55731508245443 |
| O | 3.88037470435061 | -2.81950599348567 | -1.05906139353641 |
| O | 1.33109179256960 | -2.20073402533748 | -0.45446560393577 |
| O | 1.75145974952320 | 0.47639648853270 | -1.39999442362481 |
| O | -0.50089583941314 | 0.31144333386370 | -2.85837973193796 |
| O | -1.29564947590764 | -2.00739502531647 | -1.52376067104852 |
| O | -3.84560478440773 | -2.92814894064586 | -0.93884408827143 |
| O | -1.61793775186251 | 2.22881676572244 | -1.02665068001440 |
| O | -1.56852599475130 | 1.57466441345725 | 1.70522215961006 |
| O | -3.25564687071702 | -0.28781208615199 | 2.38418184176176 |
| O | -2.02642302618035 | -2.36425477797909 | 1.20089080847754 |
| O | -4.32330996203437 | 2.01083804044503 | -1.34715968113010 |
| O | -4.85387087766031 | -0.31947951612520 | -0.00148986104547 |
| H | -5.01773435469868 | 2.62959456228513 | -1.11991016311312 |
| H | -4.54727440305115 | 1.16642323812805 | -0.90003327610091 |
| H | -4.44732493857811 | -0.30520203517306 | 0.87945218025677 |
| H | -4.64144076324996 | -1.18359274510454 | -0.38183319686086 |
| H | -0.55245514679520 | -2.90503380065975 | 1.73384360743321 |
| H | 0.86203742906970 | -2.79972623488138 | 1.01476484137409 |
| H | 0.17735568106390 | 2.98327589754957 | -0.55222990265421 |
| H | 1.14787813830182 | 2.77241811103931 | 0.64479164171245 |
| H | -2.58481432443817 | 2.33011176645281 | -1.15001790064101 |
| H | -1.34343549322511 | 1.59469789797334 | -1.70755415206860 |
| H | -2.74138193506528 | -2.88820181203694 | 0.81465881345692 |
| H | -2.45010234612443 | -1.66437735071217 | 1.74842135322680 |
| H | -2.60588565396543 | 0.44828279558434 | 2.19787007886645 |
| H | -3.49334390840972 | -0.23482240147866 | 3.31093234721775 |
| H | -3.03241161595836 | -2.73272804275746 | -1.44119250768622 |
| H | -4.32545019187897 | -3.60699870709105 | -1.41513447633635 |
| H | -1.16423339296126 | -1.19409700991273 | -2.04540330974378 |
| H | -1.43492786632562 | -1.74040201721872 | -0.59976753196984 |
| H | -0.65358953175135 | 1.66369048014084 | 2.02204890628890 |
| H | -1.56091014768621 | 1.76388704334359 | 0.74889855760518 |
| H | 0.41545621906683 | 0.37606804057246 | -2.49925254129323 |
| H | -0.46530992659606 | 0.48659018149676 | -3.79981458763494 |
| H | 3.31225063677423 | -0.34407783099526 | 1.79669809751258 |
| H | 2.07535187198514 | -0.84257398641734 | 2.64699862599286 |
| H | 4.43281531814049 | 2.06589441853199 | -1.14753033014704 |
| H | 3.51936938029485 | 0.87662730905973 | -1.49989789811412 |
| H | 4.84023803086700 | 0.00336281589523 | -0.05991637748514 |
| H | 5.65678965123204 | -0.76328508926427 | 1.04061975483147 |
| H | 2.18829516532569 | -2.55060047489227 | -0.78704948401055 |
| H | 0.62133722155765 | -2.40177108321575 | -1.08539444103761 |
| H | 4.28667556877788 | -2.72034286424864 | -1.92114514556147 |
| H | 4.32483366492591 | -2.17844834879068 | -0.47071353323247 |
| H | 1.72340012790816 | 1.29886840444975 | 2.34657608765163 |
| H | 1.32629077393475 | 2.59772504065552 | 3.13453442879670 |
| H | 1.63274136794348 | -0.29208271297070 | -0.81469276290573 |
| C | 0.96762552104022 | -2.28589684775767 | 3.04922570569867 |

W18-a..CNN from W18-a..N$_2$ + C

| | | | |
|---|---|---|---|
| H | 1.62829576343364 | 1.22268205759396 | -0.91986772856701 |
| O | 1.03705930428136 | 2.77606847120024 | -0.58377644994373 |
| O | 0.54004452892621 | 2.61723010202308 | 2.16466378585513 |
| O | 1.68196060014995 | 0.40564244103962 | 3.02745562682472 |
| O | 0.23282756477821 | -1.89039281578992 | 2.64784105154172 |
| O | 4.57677995247852 | 0.00846841992899 | -0.78130123527761 |
| O | 4.26843596122407 | -0.39209858344987 | 1.96166620295426 |
| O | 3.97798195175385 | -2.71898011104565 | 0.12964834786166 |
| O | 1.33274499430249 | -2.02187105567088 | 0.04290340543895 |
| O | 1.93645506978121 | 0.34376649183755 | -1.26009542993309 |
| O | -0.26381957632723 | -0.10825109077213 | -2.78250360510083 |
| O | -1.22527496059065 | -2.21459410770452 | -1.15933900483783 |
| O | -3.88323657423011 | -2.94083838194580 | -0.91668096559217 |
| O | -1.59217708514823 | 2.11269378563442 | -1.56165736655925 |
| O | -2.08479753157572 | 2.21426878755177 | 1.18202935000747 |
| O | -3.89565935727365 | 0.41453910958540 | 1.75078180148685 |
| O | -2.36042577862614 | -1.83262163672063 | 1.38027395270890 |
| O | -4.12889766775617 | 1.76596011176831 | -2.51605802171211 |
| O | -4.94572479213931 | -0.23184115645665 | -0.83870933324781 |
| H | -4.86344766966793 | 2.37502309270109 | -2.59424758304239 |
| H | -4.44682191049209 | 1.01916140591183 | -1.96192505876015 |
| H | -4.71244697327634 | 0.00443020812315 | 0.07299796346586 |
| H | -4.68211693405941 | -1.15833107454888 | -0.94738307950976 |
| H | -0.71631000167843 | -1.92900480044707 | 2.46053293787040 |
| H | 0.64914540999284 | -2.06235266402899 | 1.78646316218474 |
| H | 0.15762141997372 | 2.87441376376880 | -0.97901578542183 |
| H | 0.91458986632831 | 2.88612407830216 | 0.37603482179820 |
| H | -2.49213743412955 | 2.19255613712742 | -1.94034864883165 |
| H | -1.21537674241511 | 1.32694142209389 | -1.98978858481360 |
| H | -2.99638257399772 | -2.50466556486318 | 1.10816708003725 |
| H | -2.89530952369109 | -1.04391755458875 | 1.60346462466091 |
| H | -3.22926995333435 | 1.14169168710326 | 1.60703836076014 |
| H | -4.38557919122120 | 0.62090942132107 | 2.54767147073548 |
| H | -2.96792750824362 | -2.84476550365212 | -1.24359827486772 |
| H | -4.26535092927776 | -3.70277237792022 | -1.35377269421351 |
| H | -1.02220659606712 | -1.48607225365799 | -1.77178883148774 |
| H | -1.47888726692102 | -1.81531119038721 | -0.30562575474635 |
| H | -1.22207790384922 | 2.31954298444756 | 1.61698556817753 |
| H | -1.90346821377388 | 2.15268168932517 | 0.22402959719969 |
| H | 0.64075853285041 | 0.04302196056738 | -2.42661976669395 |
| H | -0.21099181615443 | -0.09752740416636 | -3.73941846467311 |
| H | 2.55077902576248 | 0.14259249063914 | 2.68700300971944 |
| H | 1.11738382486640 | -0.39571330975031 | 2.98635101516621 |
| H | 5.17178418284068 | 0.53936356155961 | -1.31218437153439 |
| H | 3.65870058758320 | 0.21996586447348 | -1.07258213235111 |
| H | 4.47733847281945 | -0.05464638527243 | 1.07043162831801 |
| H | 5.00157226403386 | -0.15070501313454 | 2.53038612048068 |
| H | 2.19190474172402 | -2.48425243867644 | -0.03733606765337 |
| H | 0.65824276986827 | -2.45536407338239 | -0.50182911430346 |
| H | 4.38062704043710 | -2.05975357642664 | -0.45180745240102 |
| H | 4.11815692417011 | -2.34819789765062 | 1.01114161654750 |
| H | 0.99668000417969 | 1.78899867416159 | 2.47727441130923 |
| H | 0.66042475178673 | 3.26987740734266 | 2.85803094355211 |
| H | 1.65481663023409 | -0.35785443239365 | -0.63661321443697 |
| N | -1.15668388140021 | 1.23459936978475 | 4.91133901503520 |
| N | -1.83698607914137 | 0.28040043531821 | 4.68554925761230 |
| C | -0.46498988900180 | 2.22172111016829 | 5.13767390830289 |

W18-a..CNNH from W18-a..CNN + H

| | | | |
|---|---|---|---|
| H | 1.60402655816074 | 1.88074484081295 | -0.82336271677207 |
| O | 0.80209136944917 | 3.40049751444090 | -0.67264561544802 |
| O | 0.19779476152427 | 2.68719078740030 | 2.08963648603781 |
| O | 1.37139380752784 | 0.20660198088370 | 1.73954872395120 |
| O | -0.08932877469828 | -1.90720703680893 | 2.68381256621078 |
| O | 4.42068103727099 | 0.04226152888530 | -0.56759532651353 |
| O | 4.10857239666498 | -0.24118209019889 | 2.15914286874425 |
| O | 3.54839983785901 | -2.61187713720700 | 0.45514192372172 |
| O | 0.88070518910628 | -3.52326276056780 | 0.74482908621837 |
| O | 1.88512449748875 | 0.93154261598373 | -0.85328034593261 |
| O | -0.01482841395477 | -0.08800429300544 | -2.46027059308919 |
| O | -1.10321431363087 | -2.08126476337279 | -0.84177809587432 |
| O | -3.74265358506364 | -2.84497206668176 | -0.84005002104412 |
| O | -1.57659585201414 | 2.22704664401148 | -1.77929490745523 |
| O | -2.34808766192152 | 2.41409037312644 | 0.89541586819687 |
| O | -4.16056519920268 | 0.64262692118881 | 1.51450095069148 |
| O | -2.60552188529895 | -1.60866243539024 | 1.53254679472132 |
| O | -4.00683859174839 | 1.67569635746997 | -2.88127590605677 |
| O | -4.93289224031679 | -0.19769854160082 | -1.11801822363966 |
| H | -4.74753848781160 | 2.24402341909517 | -3.09345145417358 |
| H | -4.35713109786364 | 0.97234984862432 | -2.29118617031708 |
| H | -4.79420933267460 | 0.10546125484429 | -0.20634063984992 |
| H | -4.63110930825829 | -1.11811433733664 | -1.13592514494462 |
| H | -1.02649162824691 | -1.84356915353821 | 2.42095971433536 |
| H | 0.29070045311007 | -2.61077383785297 | 2.10351767014812 |
| H | -0.05253298858027 | 3.29810111811074 | -1.12165404284986 |
| H | 0.60150527274703 | 3.49302877926014 | 0.27051561493902 |
| H | -2.44581950061398 | 2.21511716585192 | -2.23316552496208 |
| H | -1.13285510706465 | 1.41595781849454 | -2.07520632124134 |
| H | -3.16850848038753 | -2.26414600236082 | 1.09925085827109 |
| H | -3.16585428512568 | -0.81529672441345 | 1.65423439480244 |
| H | -3.49129907694193 | 1.36123717084147 | 1.33239182999653 |
| H | -4.75943831234425 | 0.97021563996045 | 2.18685085010221 |
| H | -2.80377273765567 | -2.73462549699673 | -1.08963877166121 |
| H | -4.03735109953033 | -3.68558382000657 | -1.19279230690809 |
| H | -0.76660244244758 | -1.40716626857496 | -1.46078649050098 |
| H | -1.42208651476448 | -1.60876939252839 | -0.05452359159541 |
| H | -1.52329028065571 | 2.51155574831661 | 1.40172308537254 |
| H | -2.08498181954721 | 2.32359088889345 | -0.04061245738610 |
| H | 0.78354798021805 | 0.23986967875113 | -1.98278444817460 |
| H | 0.22471879611587 | -0.21980827313873 | -3.37900110622637 |
| H | 2.27663763977115 | 0.06226363435493 | 2.06903028961502 |
| H | 0.82318416376348 | -0.54555019310265 | 2.06041204841063 |
| H | 5.09497165128693 | 0.36169430997290 | -1.16834171944184 |
| H | 3.57302722259679 | 0.47911883200971 | -0.81035596372144 |
| H | 4.43492445548015 | 0.02453125651506 | 1.27861121264001 |
| H | 4.69136735957887 | 0.14379854934320 | 2.81507354501819 |
| H | 1.79832598545856 | -3.27437427921429 | 0.53541421381228 |
| H | 0.32874765254006 | -3.18241600085336 | 0.02514704828716 |
| H | 3.84016974977989 | -1.95091784719477 | -0.18529070806380 |
| H | 3.78351015655771 | -2.22146935635226 | 1.30477044789938 |
| H | 0.62673035067747 | 1.80829663120209 | 2.04240112337785 |
| H | 0.19958622993916 | 2.90560663784789 | 3.02809017596181 |
| H | 1.62085076068249 | 0.57334590905803 | 0.02275683160569 |
| N | 0.29514559440282 | 0.20578829951502 | 5.12248965736008 |
| N | 0.30886216412587 | -1.04019347996567 | 5.42521404727223 |
| C | 0.30088346875330 | 1.37135544654081 | 4.93905953242899 |
| H | 0.18202336702683 | -1.55549604534334 | 4.53181322139310 |

W18-a..HCNN from W18-a..CNN + H

| | | | |
|---|---|---|---|
| H | 1.52903507004246 | 1.28147360611322 | -0.77929234417866 |
| O | 0.88223671939324 | 2.79157189909107 | -0.39913271932555 |
| O | 0.19260973834653 | 2.41700091817002 | 2.26215270441759 |
| O | 1.44014968298541 | 0.23061396458706 | 3.17489353371027 |
| O | 0.06619086883913 | -2.08296323273138 | 2.59763461391032 |
| O | 4.50585647169545 | 0.12556527298801 | -0.57068695035278 |
| O | 4.09621102086882 | -0.42994327447403 | 2.13250112569669 |
| O | 3.94490369541044 | -2.66148744959382 | 0.18199515906911 |
| O | 1.28811494758952 | -2.00221044633740 | 0.03313622544526 |
| O | 1.88678408098692 | 0.42808688823383 | -1.13881692317275 |
| O | -0.20965370545858 | -0.02077546928541 | -2.79912574165434 |
| O | -1.20563177271201 | -2.18850995190178 | -1.28263938690197 |
| O | -3.82695710393724 | -3.04607793083110 | -1.19131680432198 |
| O | -1.67369804492651 | 2.14485658752702 | -1.61808537883969 |
| O | -2.34530836456486 | 2.12338850018498 | 1.10073400694412 |
| O | -4.09936014565605 | 0.24754078375307 | 1.52994417398728 |
| O | -2.47759141164875 | -1.94484803703271 | 1.20968571364577 |
| O | -4.14547474264543 | 1.68655801587799 | -2.70998691153447 |
| O | -4.98742013877002 | -0.37278043161036 | -1.11776213428211 |
| H | -4.89654406743895 | 2.27062785410738 | -2.81710166701276 |
| H | -4.46640434328801 | 0.91904893121086 | -2.18710714512181 |
| H | -4.79953459883007 | -0.14937957177659 | -0.19170981447230 |
| H | -4.68986695105716 | -1.28748659159693 | -1.23524506725803 |
| H | -0.87442995264899 | -2.08580087579568 | 2.36928990692077 |
| H | 0.51401598101865 | -2.18958290778648 | 1.74212000079317 |
| H | 0.03165886500243 | 2.91164706356185 | -0.84600087788677 |
| H | 0.69057275575246 | 2.82651204988039 | 0.55730938919861 |
| H | -2.55832690783177 | 2.18216324126009 | -2.03597156392575 |
| H | -1.24177085897460 | 1.38215618402647 | -2.03502534141033 |
| H | -3.07673770437193 | -2.61771824782363 | 0.86483699064773 |
| H | -3.04689713055251 | -1.17947463199465 | 1.42881649341530 |
| H | -3.44856001361654 | 0.99998978851215 | 1.42583718387312 |
| H | -4.66184013494579 | 0.45414848109123 | 2.27726519255117 |
| H | -2.90271933474729 | -2.89904990940331 | -1.47136429545940 |
| H | -4.14878417376601 | -3.81832592934324 | -1.65777272525327 |
| H | -0.99603299571907 | -1.43810499422053 | -1.86590126083200 |
| H | -1.51257171623228 | -1.82199613371189 | -0.43178627112835 |
| H | -1.51055009319107 | 2.18161768596734 | 1.59943204422069 |
| H | -2.10096780774088 | 2.11399843631069 | 0.15575558678248 |
| H | 0.66841409439114 | 0.13933074795888 | -2.38501795860968 |
| H | -0.10095516936784 | 0.02357791769927 | -3.75027847907215 |
| H | 2.31802370249610 | 0.00970341560715 | 2.82953502451328 |
| H | 0.89953416087291 | -0.58282024032925 | 3.08288773554762 |
| H | 5.10735332569574 | 0.69803205739023 | -1.04823193518689 |
| H | 3.59408207370544 | 0.33163123832449 | -0.88715641921083 |
| H | 4.32672807819429 | -0.03877450897561 | 1.26889837687129 |
| H | 4.80427011616228 | -0.20333340424185 | 2.73765423177172 |
| H | 2.15643138233644 | -2.44804799919040 | -0.03788204397318 |
| H | 0.64365680287176 | -2.41708347558652 | -0.56088782010396 |
| H | 4.35018899814594 | -1.96426811661408 | -0.35129042951532 |
| H | 4.03693756735469 | -2.32674214896869 | 1.08418735918402 |
| H | 0.65548476719076 | 1.58436635922459 | 2.53152724300006 |
| H | 0.28890829077690 | 2.99701868880593 | 3.03224854896931 |
| H | 1.59797203128957 | -0.31059061969655 | -0.56285214258281 |
| N | 0.30034378412511 | 2.07951093189472 | 5.70464057478341 |
| N | -0.21077161464793 | 1.12107430979118 | 6.02936424615155 |
| C | 0.77424446432732 | 3.14650487868084 | 5.17265408987823 |
| H | 1.48380817058972 | 3.65869057088404 | 5.82284862453687 |

W18-a..c-HNNC from W18-a..HNNC

| | | | |
|---|---:|---:|---:|
| H | 1.65927386326882 | 1.89903922350182 | -0.86902599697222 |
| O | 0.82228340461738 | 3.39887866834112 | -0.59755608143084 |
| O | 0.23275937986548 | 2.58429784055166 | 2.17891084239720 |
| O | 1.46718050941032 | 0.15214459392539 | 1.63613358468217 |
| O | -0.09892678797155 | -1.87086393992043 | 2.68851379934697 |
| O | 4.46351948456754 | 0.00640268993519 | -0.64095584439285 |
| O | 4.19078713473101 | -0.30997521304778 | 2.08225282710743 |
| O | 3.57784210125608 | -2.65979829539458 | 0.41020693256698 |
| O | 0.90200049329413 | -3.49857334550725 | 0.79399555595542 |
| O | 1.94494920332779 | 0.95363214360840 | -0.92783217523069 |
| O | -0.01734826166959 | -0.07623877687515 | -2.45762616308752 |
| O | -1.07459040827505 | -2.09635077963994 | -0.83972619725195 |
| O | -3.71955014563883 | -2.86564458821271 | -0.84810004527642 |
| O | -1.56061344975789 | 2.22975051721267 | -1.69775663558430 |
| O | -2.30891874361588 | 2.36715733497558 | 0.98549486060663 |
| O | -4.13100512169610 | 0.60242240327627 | 1.57695253737263 |
| O | -2.58419664390075 | -1.63627472447471 | 1.51911870792623 |
| O | -3.99088057141603 | 1.69961714488242 | -2.81497586576511 |
| O | -4.90189685424344 | -0.20197836638889 | -1.07143026391600 |
| H | -4.72930625283901 | 2.27715057725373 | -3.00970465311746 |
| H | -4.34035213048683 | 0.98916076460521 | -2.23345231989132 |
| H | -4.77088471467794 | 0.08760316864598 | -0.15450341516881 |
| H | -4.60116565379825 | -1.12206819506670 | -1.10394052118212 |
| H | -1.03408488222285 | -1.82901633376023 | 2.40705215888287 |
| H | 0.30259960487214 | -2.59535650282263 | 2.14794586527746 |
| H | -0.04036746315871 | 3.29740346204337 | -1.03136110656839 |
| H | 0.63913902175616 | 3.47816488583940 | 0.34959792774381 |
| H | -2.42897708472962 | 2.22783967218819 | -2.15301988076286 |
| H | -1.11738916789135 | 1.42418677929652 | -2.00907329928542 |
| H | -3.15147241137091 | -2.29517292512971 | 1.09712298695635 |
| H | -3.14677433019005 | -0.84557730237019 | 1.65812623417220 |
| H | -3.45536122089573 | 1.31994102847630 | 1.40975059708035 |
| H | -4.70488857803823 | 0.90387422150877 | 2.28253493552575 |
| H | -2.78274568697215 | -2.75692164858255 | -1.10381117443168 |
| H | -4.02304012936882 | -3.69876845886880 | -1.21094026955556 |
| H | -0.74463519614564 | -1.41610615757696 | -1.45546448322515 |
| H | -1.41275762988784 | -1.62840721902526 | -0.05751871918440 |
| H | -1.47195097239162 | 2.44277911674288 | 1.47771319683135 |
| H | -2.06388553287933 | 2.29493776345070 | 0.04331917384577 |
| H | 0.80156764398192 | 0.24018091672075 | -2.00864198743484 |
| H | 0.18094386011582 | -0.18528513729293 | -3.38911437875687 |
| H | 2.36443883116535 | -0.00727330356422 | 1.98109083326425 |
| H | 0.89253061745444 | -0.57744050831570 | 1.94955113441328 |
| H | 5.14886111778221 | 0.30426932325604 | -1.24033096702910 |
| H | 3.62847121696819 | 0.46056527596192 | -0.89359850971907 |
| H | 4.50307023947750 | -0.03230969436037 | 1.19995629883262 |
| H | 4.76974289409147 | 0.08435472743296 | 2.73604868675032 |
| H | 1.82088467347615 | -3.26907477050907 | 0.56657339031270 |
| H | 0.34938585997822 | -3.17045989770015 | 0.06821244221311 |
| H | 3.85556919347275 | -2.00629095374513 | -0.24366881096064 |
| H | 3.84876083047489 | -2.26846530960107 | 1.24894277519819 |
| H | 0.70041723784465 | 1.74012681712515 | 2.02831498197798 |
| H | 0.15878579883541 | 2.63005896068957 | 3.14324910746955 |
| H | 1.69866272773874 | 0.57785197881095 | -0.05400013616986 |
| N | -0.82742306296123 | 0.33742113629291 | 5.63981627546557 |
| N | 0.18996685704652 | -0.21859473229066 | 4.88141966085506 |
| C | -0.09490402345133 | 1.13409374276369 | 4.86342449201772 |
| H | 0.12271022297187 | -0.99481383127174 | 4.19146116600283 |

W18-a..HNCN from W18-a..c-HNNC

| | | | |
|---|---|---|---|
| H | 1.67608705528394 | 1.79652393498463 | -0.95837435447279 |
| O | 0.91162110515312 | 3.32484089610455 | -0.95860153478582 |
| O | 0.52637382319086 | 2.92281642370167 | 1.89503880323023 |
| O | 1.57592517299303 | 0.39566693107766 | 1.83477672135680 |
| O | 0.06523501944308 | -1.73590613405220 | 2.62038060465510 |
| O | 4.53416045340268 | 0.05827951557260 | -0.71360343362547 |
| O | 4.31790376023197 | -0.20740800755278 | 2.01724799152479 |
| O | 3.66397714866815 | -2.57800900642895 | 0.34264899910351 |
| O | 1.00895647116308 | -3.45707769733952 | 0.75801858808236 |
| O | 1.96189690015070 | 0.84965622633726 | -0.88480054304317 |
| O | 0.07559895127976 | -0.22488658257454 | -2.50712679955640 |
| O | -1.07355459362318 | -2.23179018535791 | -0.93125124530102 |
| O | -3.78557457101746 | -2.79753657329748 | -0.71961333702908 |
| O | -1.51019416248623 | 2.06977451107060 | -1.76142674424403 |
| O | -2.14392021747971 | 2.51486271253563 | 0.93778934723138 |
| O | -3.85788281125321 | 0.70378682172084 | 1.85539798856517 |
| O | -2.38051155617868 | -1.49686164752584 | 1.42697102022687 |
| O | -4.08479736947513 | 1.67498321967399 | -2.53960464458490 |
| O | -4.98617247241149 | -0.11853997410148 | -0.68017622653761 |
| H | -4.76551493764031 | 2.34049014903951 | -2.64298934384370 |
| H | -4.43469562085590 | 1.02070476412308 | -1.89499689673949 |
| H | -4.81148328898479 | 0.16391676609444 | 0.23014869553232 |
| H | -4.66661126283044 | -1.03046763334075 | -0.74918029864989 |
| H | -0.83888356518023 | -1.67774239594843 | 2.24828407645851 |
| H | 0.48527339625461 | -2.45849343569757 | 2.09430529031730 |
| H | 0.02887471996074 | 3.16806683813496 | -1.33490590585233 |
| H | 0.77083496770805 | 3.48469377577931 | -0.01354443102685 |
| H | -2.42434919539049 | 2.06150372253973 | -2.12024555707190 |
| H | -1.09775896493553 | 1.24968408880059 | -2.07468651487318 |
| H | -3.01184749760501 | -2.17024422110981 | 1.14240237860793 |
| H | -2.90806094006088 | -0.68090205602733 | 1.60222084035684 |
| H | -3.25129933329475 | 1.43889264544764 | 1.58563453391630 |
| H | -3.95762430132336 | 0.76085884757715 | 2.81306708218921 |
| H | -2.87208786956542 | -2.76134882774781 | -1.06211025072114 |
| H | -4.18496253975384 | -3.60258251345476 | -1.05148517689455 |
| H | -0.71560068522226 | -1.59550203390989 | -1.57547960931673 |
| H | -1.36127735308256 | -1.70671259410058 | -0.16351477605981 |
| H | -1.30074895117949 | 2.68341499272382 | 1.38378439728171 |
| H | -1.92136850107438 | 2.30986133786835 | 0.01080377031246 |
| H | 0.85603435642118 | 0.08547415297310 | -1.99308361246854 |
| H | 0.34529421474556 | -0.30928610419527 | -3.42305554917930 |
| H | 2.49310414576056 | 0.19306147002109 | 2.09029366167153 |
| H | 1.01497986708446 | -0.35773277504803 | 2.13549069192090 |
| H | 5.16410448337581 | 0.41955129883726 | -1.33823161580528 |
| H | 3.65297183930621 | 0.44791777807332 | -0.91749385871835 |
| H | 4.61002362065801 | 0.04115902905929 | 1.11914728488793 |
| H | 4.99821838094006 | 0.06771939340971 | 2.63311089146587 |
| H | 1.91240471074840 | -3.22345086036869 | 0.48229447285240 |
| H | 0.41801247257896 | -3.21080746067111 | 0.03058747290764 |
| H | 3.96869281597618 | -1.94905907208118 | -0.32274625008480 |
| H | 3.91096637976566 | -2.16212373409004 | 1.17700306912723 |
| H | 0.93239378478125 | 2.03658273468597 | 2.00481743669245 |
| H | 0.71620818544143 | 3.42511484512410 | 2.68864050272395 |
| H | 1.73559130650798 | 0.59178667357171 | 0.03393879326894 |
| N | -2.89490783797990 | 0.26417674033234 | 4.71803449668286 |
| N | -0.86810258472733 | -1.01265810936808 | 5.22702880731419 |
| C | -1.91139827062307 | -0.36468389413671 | 4.91916499924040 |
| H | -0.40371734243966 | -1.43180373946910 | 4.39952886848005 |

W18-a..HNCNH from W18-a..HNCN+H

| | | | |
|---|---|---|---|
| H | 1.69273108088788 | 1.80893068545012 | -0.93559868171593 |
| O | 0.91720243535686 | 3.33932017512808 | -0.88491359602103 |
| O | 0.50606829777014 | 2.86042300611187 | 1.96087494461590 |
| O | 1.59360531817526 | 0.35579190509739 | 1.80411629286870 |
| O | 0.08337937526238 | -1.73466799703899 | 2.62935295552623 |
| O | 4.53380849914369 | 0.02659451974252 | -0.73598078082858 |
| O | 4.33703801234677 | -0.25186794238477 | 1.99810360872371 |
| O | 3.65243454164384 | -2.61330226937809 | 0.33148880536586 |
| O | 0.99803010008539 | -3.51000286268419 | 0.77632683404209 |
| O | 1.97450281918080 | 0.86037203674917 | -0.88279052175206 |
| O | 0.06010806336369 | -0.21794329328290 | -2.45865923042738 |
| O | -1.08005030393863 | -2.25123596057953 | -0.90522015209069 |
| O | -3.79592899015247 | -2.78917263900364 | -0.75385369235666 |
| O | -1.50547783031253 | 2.09281918989486 | -1.73414855375214 |
| O | -2.15133864844705 | 2.53151368561084 | 0.96316123772634 |
| O | -3.83704500136621 | 0.68219495649162 | 1.85693900457313 |
| O | -2.39571940289544 | -1.54375041222457 | 1.45519085646951 |
| O | -4.07717463420759 | 1.68872290539352 | -2.53647446825454 |
| O | -4.96957081013579 | -0.11088611423063 | -0.67927333112363 |
| H | -4.75909942378566 | 2.35424368773070 | -2.63129798104843 |
| H | -4.42269384961059 | 1.03113318861513 | -1.89254389414366 |
| H | -4.78392053005467 | 0.16718352135759 | 0.23051245842151 |
| H | -4.65300805331851 | -1.02347661066278 | -0.75732087619107 |
| H | -0.82794213894948 | -1.69184941207759 | 2.27934875269589 |
| H | 0.49194667192327 | -2.47574127192998 | 2.12547702333133 |
| H | 0.03681480714451 | 3.18784745213848 | -1.26791903232001 |
| H | 0.77368624171356 | 3.47899427356036 | 0.06284321860459 |
| H | -2.41542141237660 | 2.08392697156441 | -2.10203179487946 |
| H | -1.09238406369748 | 1.26973712540721 | -2.03913909120738 |
| H | -3.03280352202815 | -2.21312086278823 | 1.17692224162915 |
| H | -2.91506528361903 | -0.71679351103092 | 1.60570794684536 |
| H | -3.23912183524060 | 1.42568814782959 | 1.59073314044257 |
| H | -3.93253445490203 | 0.72511291570609 | 2.81983691646802 |
| H | -2.87238463161105 | -2.74586292860049 | -1.06896388143450 |
| H | -4.19110997355335 | -3.57513055143708 | -1.13283387250961 |
| H | -0.71774281056029 | -1.59698706862045 | -1.52889175680398 |
| H | -1.36984182153289 | -1.75018348457943 | -0.12196685527503 |
| H | -1.30147470784423 | 2.66476013247240 | 1.41041676823678 |
| H | -1.93270213971207 | 2.33053999770993 | 0.03438264659648 |
| H | 0.84814716691751 | 0.09700392994709 | -1.95758368597977 |
| H | 0.31903192763618 | -0.31034239943344 | -3.37693588632813 |
| H | 2.50604409838593 | 0.15100810432855 | 2.07276769096970 |
| H | 1.02493671806264 | -0.39443461940689 | 2.10457806413516 |
| H | 5.16521245891264 | 0.37970313324883 | -1.36365740614401 |
| H | 3.65818173040687 | 0.43284636340572 | -0.93178274355428 |
| H | 4.62183629824407 | 0.00113343372385 | 1.09912014267049 |
| H | 5.01567063234114 | 0.03339980165388 | 2.61106252492375 |
| H | 1.89809912649048 | -3.27075762337496 | 0.49539624484089 |
| H | 0.40472200885565 | -3.25352057292440 | 0.05454143661014 |
| H | 3.94795242515640 | -1.98096546550938 | -0.33476845869215 |
| H | 3.90623267466323 | -2.19768482924000 | 1.16402853967291 |
| H | 0.92005040235118 | 1.97415198807633 | 2.03772311257364 |
| H | 0.62231348196273 | 3.29830100174650 | 2.80539213426951 |
| H | 1.75773365234398 | 0.59192633079826 | 0.03712138919488 |
| N | -3.33867772378547 | 0.23731159820192 | 4.76151819960138 |
| N | -0.93141607351978 | -0.08208170561431 | 4.78604762873295 |
| C | -2.13419379170061 | -0.00448685277206 | 4.79106763154954 |
| H | -0.44030503458064 | -0.81360448848827 | 4.26794424602941 |
| H | -3.91219231018972 | -0.17894281559483 | 5.48165397107656 |

W18-a..NH$_2$CN from W18-a..HNCN+H

| | | | |
|---|---|---|---|
| H | 1.66426485277791 | 1.78675245744377 | -0.97047835203131 |
| O | 0.91622824518850 | 3.31633807055444 | -0.99561819662969 |
| O | 0.56448509013614 | 2.95197843514714 | 1.86434263797980 |
| O | 1.59531754086815 | 0.41239116795377 | 1.84329450341259 |
| O | 0.11792048866154 | -1.72625795204862 | 2.66155238222209 |
| O | 4.52648097674113 | 0.07374019378381 | -0.74314673271647 |
| O | 4.35139478935229 | -0.17663236800900 | 1.98554953982758 |
| O | 3.70132427592030 | -2.56913965861675 | 0.33643543450161 |
| O | 1.04629456212183 | -3.44993437446402 | 0.78521221578269 |
| O | 1.95158842114507 | 0.84052328745050 | -0.88779445287001 |
| O | 0.06558230161840 | -0.25406425748050 | -2.50047481575456 |
| O | -1.05352192142074 | -2.25371957269756 | -0.89129003239226 |
| O | -3.77546705155826 | -2.79221502934396 | -0.64731527794029 |
| O | -1.50612269096492 | 2.06679754621983 | -1.79035623573089 |
| O | -2.12144077035831 | 2.57960726279978 | 0.91162716683626 |
| O | -3.78586506135867 | 0.73696191922403 | 1.90218956408276 |
| O | -2.31390657479937 | -1.45241351012042 | 1.45397675426642 |
| O | -4.10443409671369 | 1.66522835164538 | -2.47875394883116 |
| O | -4.96881967848717 | -0.10802308015367 | -0.58750614368646 |
| H | -4.77953777479067 | 2.34053069367016 | -2.55104198247704 |
| H | -4.43709443036164 | 1.01474721413611 | -1.82035572267700 |
| H | -4.76971042413705 | 0.18322320235144 | 0.31576900304470 |
| H | -4.64789612305837 | -1.01938886010256 | -0.65700607658475 |
| H | -0.78501337364501 | -1.65724819095685 | 2.28529749356842 |
| H | 0.52973957757418 | -2.44690678700880 | 2.12754865628621 |
| H | 0.03149381168236 | 3.15817668109490 | -1.36728537549408 |
| H | 0.77984025521793 | 3.49306407194742 | -0.05301029578960 |
| H | -2.43158760481190 | 2.05092208343357 | -2.11925480796030 |
| H | -1.10510712773462 | 1.23632545921866 | -2.08965918980982 |
| H | -2.96132127786345 | -2.13173640667978 | 1.22763509683245 |
| H | -2.83321381450327 | -0.62921473369552 | 1.63296643626032 |
| H | -3.20403750323460 | 1.48028226381269 | 1.60969934551020 |
| H | -3.82041308997180 | 0.77403163893546 | 2.86914836694883 |
| H | -2.86486286401486 | -2.76569490218117 | -0.99707234498751 |
| H | -4.18482000929830 | -3.59449406383036 | -0.97324531896809 |
| H | -0.70535678026004 | -1.62179803535262 | -1.54472719268428 |
| H | -1.33988944482922 | -1.72194681094412 | -0.12639224407321 |
| H | -1.27282658707721 | 2.73196804305030 | 1.35240584498127 |
| H | -1.90563559737689 | 2.34641706774900 | -0.00939490227735 |
| H | 0.84412894458179 | 0.06345100856254 | -1.98857626465066 |
| H | 0.33737410507303 | -0.34729430208780 | -3.41485821743152 |
| H | 2.51695460549377 | 0.21694717987376 | 2.08593761125683 |
| H | 1.04269363447522 | -0.34057087224832 | 2.16294773936716 |
| H | 5.14406773394997 | 0.44496908462156 | -1.37417416320910 |
| H | 3.63731332805263 | 0.45223482531252 | -0.93504872394564 |
| H | 4.61741657920899 | 0.06373709424113 | 1.07640967219391 |
| H | 5.04512770796882 | 0.11633041329045 | 2.57763266283921 |
| H | 1.94540935549752 | -3.21542847786503 | 0.49770463319933 |
| H | 0.44598499552940 | -3.21076503970810 | 0.06249931241734 |
| H | 3.99065712088336 | -1.94369118252584 | -0.33890612489702 |
| H | 3.94832276081531 | -2.14013462883377 | 1.16406113853183 |
| H | 0.95730365608281 | 2.06235594130020 | 1.99009589361255 |
| H | 0.72468933215368 | 3.45169609490176 | 2.66588902963699 |
| H | 1.73158345052071 | 0.59051657366291 | 0.03453812742581 |
| N | -2.85213585854933 | 0.28849210900602 | 4.63670228454068 |
| N | -1.04068951235537 | -1.29420146101833 | 5.24163913485963 |
| C | -1.99591849654150 | -0.42738492791653 | 4.92848504957456 |
| H | -0.44755572313426 | -1.58830002377443 | 4.46077853278807 |
| H | -0.58865355318220 | -1.15508729643059 | 6.12953551061193 |

W18-a..CNNH$_2$ from W18-a..CNNH+H

| | | | |
|---|---|---|---|
| H | 1.61000042605157 | 1.87307884320112 | -0.78071878195139 |
| O | 0.80838536023782 | 3.39321373585204 | -0.60208707743668 |
| O | 0.16521088813273 | 2.64246788446964 | 2.12359802089196 |
| O | 1.31162831613031 | 0.15007019233215 | 1.73240542024120 |
| O | -0.13356767822512 | -1.97265691764827 | 2.70005652381544 |
| O | 4.41803652443761 | 0.04537470083989 | -0.45191920681323 |
| O | 4.02305545930364 | -0.27872105812747 | 2.26098238048279 |
| O | 3.51286561061006 | -2.62071696733049 | 0.51891050337390 |
| O | 0.84520138146080 | -3.55093488050215 | 0.73244966390650 |
| O | 1.88745261656269 | 0.92345684101328 | -0.82171727479306 |
| O | -0.00992721397849 | -0.08179832114809 | -2.43356169551200 |
| O | -1.12196760784295 | -2.08061439564743 | -0.83849070909889 |
| O | -3.75446651425938 | -2.84988329963910 | -0.88321207435359 |
| O | -1.57232441571209 | 2.23567005321072 | -1.74186331766365 |
| O | -2.36994488387990 | 2.38621384075284 | 0.93282140073011 |
| O | -4.18005905576697 | 0.60855104983552 | 1.51983773134768 |
| O | -2.64770470448712 | -1.65726347109069 | 1.53005407024673 |
| O | -3.99942402706998 | 1.69127669123630 | -2.86679329977057 |
| O | -4.93842262618789 | -0.19754784316912 | -1.12624650711816 |
| H | -4.73490987459480 | 2.27140463971838 | -3.06478946581870 |
| H | -4.35418681219202 | 0.98404749661199 | -2.28408202206229 |
| H | -4.80058351959910 | 0.09362477327069 | -0.21048966420104 |
| H | -4.63791236666313 | -1.11799762254768 | -1.15706533108617 |
| H | -1.07047357351022 | -1.90622279084478 | 2.43996395037865 |
| H | 0.25298717146093 | -2.65992060176263 | 2.10514378567186 |
| H | -0.04378039288018 | 3.30050698454825 | -1.05702790783864 |
| H | 0.60124334118070 | 3.46432374503492 | 0.34224742710455 |
| H | -2.43756244431331 | 2.22630891759188 | -2.20239620062858 |
| H | -1.12526241434621 | 1.42691491598661 | -2.03860217538782 |
| H | -3.20945667179559 | -2.29730805932384 | 1.07284791713585 |
| H | -3.20238294400966 | -0.85961459231661 | 1.65140465283819 |
| H | -3.50851611945361 | 1.32882462679349 | 1.34694600071579 |
| H | -4.78100382324066 | 0.93082146672041 | 2.19271508088509 |
| H | -2.81303120305131 | -2.72857654559886 | -1.11859181181644 |
| H | -4.04064344301703 | -3.68331653012514 | -1.25894517797482 |
| H | -0.76946058863640 | -1.40048122079639 | -1.44227681331112 |
| H | -1.44472678198987 | -1.62096729447957 | -0.04535495568795 |
| H | -1.54765048656030 | 2.47598743093625 | 1.44635185663482 |
| H | -2.09995428502465 | 2.31306400297979 | -0.00242790007356 |
| H | 0.79046183796956 | 0.23743938091981 | -1.95250208502839 |
| H | 0.23125973384432 | -0.21438113360039 | -3.35167751537972 |
| H | 2.20537945289129 | 0.01204153537838 | 2.09595554178007 |
| H | 0.75548982916990 | -0.59416508893505 | 2.05422031727131 |
| H | 5.10949713755002 | 0.36857250187337 | -1.03060575450337 |
| H | 3.57617153295843 | 0.47700598281274 | -0.72254247012873 |
| H | 4.38773749293054 | 0.01163573447566 | 1.40438455089144 |
| H | 4.54849821125041 | 0.11696723693065 | 2.95761382194510 |
| H | 1.76498484342069 | -3.29495124498465 | 0.54103753287703 |
| H | 0.30020075089074 | -3.18898508633636 | 0.01756808421290 |
| H | 3.82604951260014 | -1.94919634602890 | -0.10018259923024 |
| H | 3.71409966662695 | -2.24123144524579 | 1.38231907512396 |
| H | 0.58148137496444 | 1.75988057538034 | 2.04760916328367 |
| H | 0.17821148509857 | 2.82299550555670 | 3.07229842843207 |
| H | 1.60559681903839 | 0.55061153494905 | 0.04340383577625 |
| N | 0.43617780775106 | 0.27422113124877 | 5.13074389682901 |
| N | 0.61221405752920 | -1.04420953362166 | 5.33190621353554 |
| C | 0.38477792860998 | 1.41466334721010 | 4.91705613680128 |
| H | 0.33563652496825 | -1.54719086078106 | 4.48079727810176 |
| H | 0.05868882135630 | -1.33595573043953 | 6.12719218130623 |

W18-a..c-CNNH$_2$ from W18-a..CNNH$_2$

| | | | |
|---|---|---|---|
| H | 1.63797536715921 | 1.87095053840358 | -0.76327436390743 |
| O | 0.83080309706691 | 3.37858597728244 | -0.49376783905580 |
| O | 0.23403495410559 | 2.51388222138204 | 2.26190931248957 |
| O | 1.49128408966128 | 0.08700175837276 | 1.71503229708516 |
| O | -0.09093828137345 | -1.93011928850327 | 2.70813375993202 |
| O | 4.43194776304816 | -0.03932443889136 | -0.64930920064102 |
| O | 4.24420722341986 | -0.38123045554819 | 2.07855739806363 |
| O | 3.59980227343406 | -2.71844489326316 | 0.37805800202434 |
| O | 0.91000244080282 | -3.54650631100301 | 0.77323244390928 |
| O | 1.92228597431948 | 0.92622980960337 | -0.84142144889887 |
| O | 0.00042976138490 | -0.07412859876468 | -2.43588615270201 |
| O | -1.06333141422329 | -2.11115020363532 | -0.84169045741568 |
| O | -3.72418033686289 | -2.83360758515334 | -0.87778694913394 |
| O | -1.52830185199971 | 2.24717810096657 | -1.69221781927017 |
| O | -2.27288243049059 | 2.34447308058713 | 0.99985240361666 |
| O | -4.08541718257992 | 0.58605915069476 | 1.63763347117119 |
| O | -2.56289738034925 | -1.66071523610440 | 1.51050814651138 |
| O | -3.99867989551916 | 1.74762276944707 | -2.73832355067687 |
| O | -4.90823473482579 | -0.16722997420774 | -1.00864404645222 |
| H | -4.73024948749154 | 2.34426223433799 | -2.89819163130609 |
| H | -4.34537635991921 | 1.03507609782425 | -2.15728433247896 |
| H | -4.77156847456287 | 0.10696890580318 | -0.08785196975062 |
| H | -4.59682273498622 | -1.08297050166422 | -1.06239770061075 |
| H | -1.01922386020661 | -1.87837722574898 | 2.40798208053784 |
| H | 0.30903889279853 | -2.64723333418447 | 2.16224539193418 |
| H | -0.02669284602561 | 3.29136747963568 | -0.94063095550774 |
| H | 0.63587722801605 | 3.45215979017642 | 0.45138889814286 |
| H | -2.41007538080064 | 2.24958694100807 | -2.12079024745391 |
| H | -1.10204566845244 | 1.43261950939610 | -2.00302916216840 |
| H | -3.14788759069833 | -2.30688355450764 | 1.09433772830456 |
| H | -3.11162871044666 | -0.86428805165731 | 1.67414395353699 |
| H | -3.40952289987397 | 1.29881535627184 | 1.45229456397784 |
| H | -4.60111224088610 | 0.86837205078585 | 2.39425820756349 |
| H | -2.78278240423743 | -2.73389245832860 | -1.12053419277054 |
| H | -4.03831624867210 | -3.64861285175747 | -1.27116258719852 |
| H | -0.72950984361401 | -1.42390649381746 | -1.44710534665488 |
| H | -1.40137590430847 | -1.65284862960603 | -0.05306825203423 |
| H | -1.43285472815469 | 2.40531739691325 | 1.49021516896418 |
| H | -2.03366594749148 | 2.28007217701729 | 0.05627095706374 |
| H | 0.80506221250062 | 0.24365059750763 | -1.96243807939346 |
| H | 0.22693180982366 | -0.18512295989351 | -3.36056602546540 |
| H | 2.39175173199079 | -0.09104872482573 | 2.03770330798052 |
| H | 0.90941194938893 | -0.63837094987655 | 2.03208905559371 |
| H | 5.10139941314005 | 0.27138553900235 | -1.25990104224491 |
| H | 3.59258787819881 | 0.42940522137139 | -0.86131175377107 |
| H | 4.51522176218771 | -0.10655346538877 | 1.18147127980418 |
| H | 4.83835359053503 | 0.04002801750744 | 2.70100884475122 |
| H | 1.82891871006967 | -3.31943473630330 | 0.54681041506433 |
| H | 0.35548038260841 | -3.20041727266244 | 0.05749830549035 |
| H | 3.85055972642899 | -2.04193603933668 | -0.26351760983584 |
| H | 3.89540695980282 | -2.35252519387187 | 1.21947836511852 |
| H | 0.71951256672536 | 1.68251391495937 | 2.10093931432489 |
| H | 0.02422853328028 | 2.46297572647878 | 3.20838461166860 |
| H | 1.68854633786563 | 0.53038602031572 | 0.02771064104279 |
| N | -1.38181530623465 | 0.28298636455129 | 5.30376122836259 |
| N | -0.05996041582325 | -0.37590712910112 | 5.10478983680185 |
| C | -0.55723403435607 | 1.06275932737755 | 4.79170445063423 |
| H | -0.02768579802614 | -1.07905323666909 | 4.36361644189918 |
| H | 0.45058320842914 | -0.54442386310545 | 5.95575508333305 |

W18-a..NH$_2$CN from W18-a..c-CNNH$_2$

| | | | |
|---|---|---|---|
| H | 1.54439781546148 | 1.93310708515098 | -0.79843664824325 |
| O | 0.68330261623640 | 3.44539713053060 | -0.81950563461187 |
| O | -0.05553473573303 | 2.93534785090351 | 1.93971075132206 |
| O | 1.04402538157534 | 0.45971636670774 | 1.87855431137372 |
| O | -0.27471368464625 | -1.80094608962054 | 2.76592834097184 |
| O | 4.47773378176309 | 0.38103505980560 | -0.09535861802709 |
| O | 3.62724055592663 | -0.16563079370721 | 2.52649103719893 |
| O | 3.43418273266103 | -2.33541461615097 | 0.69146643197759 |
| O | 0.82870118528927 | -3.32925704010516 | 0.85413896563381 |
| O | 1.85273789057610 | 0.99758155703764 | -0.71521654972310 |
| O | 0.06228622537679 | -0.11464710873071 | -2.38069548608647 |
| O | -1.17299681223634 | -2.05513644506803 | -0.81441096804235 |
| O | -3.77295952527926 | -2.90230392693216 | -0.88565861521019 |
| O | -1.63793959234595 | 2.14313958619409 | -1.88901880016039 |
| O | -2.57728011109927 | 2.49535343891457 | 0.71083612716859 |
| O | -4.34367626841331 | 0.66138001528292 | 1.31461350524903 |
| O | -2.77129716900212 | -1.58328146724177 | 1.51132577883619 |
| O | -3.96387179823722 | 1.46227152112253 | -3.13737082739037 |
| O | -4.98901073022942 | -0.29913343550320 | -1.31655105339478 |
| H | -4.68836117993561 | 2.00442361291925 | -3.45004144398025 |
| H | -4.35500985061120 | 0.80063912222320 | -2.52508218400013 |
| H | -4.88700962166801 | 0.04593239995061 | -0.41523745879878 |
| H | -4.67890638402637 | -1.21650582774992 | -1.28223368101173 |
| H | -1.21769184825679 | -1.77169050900523 | 2.52501765827612 |
| H | 0.13653961944835 | -2.44708346368090 | 2.13600300509606 |
| H | -0.14282514552495 | 3.28718593756865 | -1.30392319968658 |
| H | 0.42705977891696 | 3.54404000674531 | 0.11015008144665 |
| H | -2.46648040958317 | 2.07142389568619 | -2.40847414079896 |
| H | -1.13764810113281 | 1.33924297933581 | -2.10493955453789 |
| H | -3.29695774397988 | -2.25570595468374 | 1.05742090029754 |
| H | -3.34537304644209 | -0.79354414641911 | 1.56763455416763 |
| H | -3.69275327423849 | 1.39619530507337 | 1.13830785655890 |
| H | -4.99561200392883 | 0.99697478435979 | 1.93123894204408 |
| H | -2.83109111816372 | -2.75827043994724 | -1.10717115706969 |
| H | -4.02223389416313 | -3.76117010472841 | -1.22906336824857 |
| H | -0.78538170484510 | -1.39492791637428 | -1.41888514073473 |
| H | -1.51422559402072 | -1.57358346820341 | -0.04261285704607 |
| H | -1.79253536764071 | 2.67316623833512 | 1.25306359223513 |
| H | -2.25191910458372 | 2.35637283906192 | -0.20025242359466 |
| H | 0.83205857037339 | 0.21902353006460 | -1.85968732199818 |
| H | 0.36069988431075 | -0.27037613968264 | -3.27817182111551 |
| H | 1.95360391002407 | 0.28856465990176 | 2.20281610851741 |
| H | 0.50664742184207 | -0.30705811642651 | 2.17350961137404 |
| H | 5.15593791571846 | 0.78648810345214 | -0.63664329619428 |
| H | 3.61631040571409 | 0.70801133435920 | -0.42606572989427 |
| H | 4.26626510689911 | 0.28693386050894 | 1.95949820285337 |
| H | 3.80775438041077 | -0.01402746037872 | 3.46658807787773 |
| H | 1.74314242837844 | -3.03971291458845 | 0.67303320217232 |
| H | 0.28810520810296 | -3.05218711773708 | 0.09966461845628 |
| H | 3.83507491511242 | -1.74859362198741 | 0.04041551056353 |
| H | 3.52350726068489 | -1.83174298555360 | 1.51443406721417 |
| H | 0.36310959391185 | 2.05129609978511 | 2.04520499199363 |
| H | 0.11357223804023 | 3.42656542743070 | 2.74498924902314 |
| H | 1.50206427712940 | 0.70209979386644 | 0.15188091993382 |
| N | 2.85813024369502 | -0.19062202564585 | 5.24338723046145 |
| N | 0.91507181050103 | -1.72784256820888 | 5.34770530053276 |
| C | 1.96914369449954 | -0.92647927485571 | 5.29016706505127 |
| H | 0.37916084762910 | -1.81486413316720 | 4.47463733864407 |
| H | 1.04310356845901 | -2.58074201259403 | 5.86754729497616 |

W18-b..N$_2$

| | | | |
|---|---|---|---|
| H | 1.53653561276723 | 1.22785019045096 | -1.01541808241059 |
| O | 0.95764017755229 | 2.78995285263937 | -0.67994136995315 |
| O | 0.45994446855477 | 2.72603843204818 | 2.07745225076368 |
| O | 1.68444828042939 | 0.57472768606359 | 2.96215201188810 |
| O | 0.22659847450371 | -1.74277140691044 | 2.69669437382793 |
| O | 4.48304653142940 | 0.00894535103490 | -0.96219909336773 |
| O | 4.23031270606916 | -0.26359358753638 | 1.80080274278986 |
| O | 3.91240181058754 | -2.67329324322413 | 0.09897468478028 |
| O | 1.26857793308760 | -1.96383583104759 | 0.06430935887507 |
| O | 1.83209320561387 | 0.34309251775957 | -1.34982848943837 |
| O | -0.46041653675376 | -0.15257311930063 | -2.72215295214135 |
| O | -1.32720184472587 | -2.21708253881710 | -1.02958704748309 |
| O | -3.97185206294510 | -2.92749693265429 | -0.72070652014116 |
| O | -1.70156657118967 | 2.16188673460286 | -1.61601564709501 |
| O | -2.18538867668180 | 2.38329141881980 | 1.12451387075734 |
| O | -3.95254204597631 | 0.56667394369499 | 1.75131418812742 |
| O | -2.42332705245657 | -1.70524242002937 | 1.50850057405723 |
| O | -4.22591006359104 | 1.71341860288702 | -2.58722725567147 |
| O | -4.98604752771482 | -0.20614479684403 | -0.80554534040082 |
| H | -4.97726261963450 | 2.29910873988203 | -2.68461115694714 |
| H | -4.52109698082911 | 0.98451318901355 | -1.99662296846680 |
| H | -4.74553742442415 | 0.07515204402377 | 0.09195598588556 |
| H | -4.74837819795861 | -1.14431768252925 | -0.86070647514573 |
| H | -0.72789751209320 | -1.76597900352324 | 2.53917575436839 |
| H | 0.61294374424934 | -1.94328537163622 | 1.82800704206001 |
| H | 0.07326253078798 | 2.89653937158451 | -1.06067213484256 |
| H | 0.85099348814644 | 2.90591171322001 | 0.28139886557576 |
| H | -2.60614690658147 | 2.21911069079603 | -1.98481872896313 |
| H | -1.33703024769963 | 1.34708279028065 | -2.00070997278721 |
| H | -3.06747958848221 | -2.37771662261239 | 1.25729845619522 |
| H | -2.94829029639500 | -0.90077779321758 | 1.69448845629159 |
| H | -3.29260314309193 | 1.29617316495782 | 1.57894272323433 |
| H | -4.47609462485592 | 0.82797590627714 | 2.50987308089610 |
| H | -3.05905771269773 | -2.85045053428824 | -1.06121741414717 |
| H | -4.35210514976797 | -3.72228487254673 | -1.09649613167268 |
| H | -1.13997490655722 | -1.51650507748618 | -1.68021750268532 |
| H | -1.56173053283102 | -1.77610579019222 | -0.19138519214096 |
| H | -1.32406110407226 | 2.47718165848119 | 1.56497573724042 |
| H | -1.99896808562385 | 2.28520269552173 | 0.17094873261668 |
| H | 0.46757364049735 | 0.00667012802445 | -2.44095702957252 |
| H | -0.49943684908075 | -0.11822978502557 | -3.68059941823299 |
| H | 2.52209401490857 | 0.29390126687509 | 2.56431602644624 |
| H | 1.11135393264399 | -0.22076994219074 | 2.98423263653032 |
| H | 5.06112042527476 | 0.51095287949419 | -1.53789980014491 |
| H | 3.55598749285816 | 0.21348433610601 | -1.23026946260782 |
| H | 4.42707778290594 | 0.03729757748626 | 0.89402593696879 |
| H | 4.97300424875352 | -0.00418035108401 | 2.34870319937795 |
| H | 2.12097795341235 | -2.43813122728962 | -0.01838636475341 |
| H | 0.57449993673485 | -2.42294221935043 | -0.43330391434939 |
| H | 4.29254382441915 | -2.04608427743909 | -0.53097147086135 |
| H | 4.07309119301080 | -2.24822934526716 | 0.95229813100425 |
| H | 0.92966384236947 | 1.91373257626242 | 2.41586144276667 |
| H | 0.63476237760313 | 3.42804743350978 | 2.70584346831552 |
| H | 1.57542432372920 | -0.34454490846116 | -0.69986374243766 |
| N | -2.07088898912105 | 0.87921032851164 | -5.28614568843427 |
| N | -2.88973101426743 | 1.42905729379425 | -5.74415903924495 |

W18-b..C
| | | | |
|---|---|---|---|
| H | 1.36788003248411 | 0.78299768819758 | -0.95193753456123 |
| O | 1.45940875137376 | 2.35859495558939 | 0.01664418522043 |
| O | 1.46561148216262 | 1.55308405650088 | 2.66174853448158 |
| O | 2.92226572375512 | -0.58268527573673 | 2.35021261346375 |
| O | 1.41065747482509 | -2.81339906783734 | 1.94657174618040 |
| O | 3.82802525344703 | 1.57017864124193 | -1.50094531506731 |
| O | 4.69680160696670 | -0.51811723560963 | 0.01133888694067 |
| O | 4.08599880869643 | -2.80748513588464 | -1.60143799730726 |
| O | 1.49233577785609 | -2.43146796877734 | -0.84607088860985 |
| O | 1.31126911624829 | 0.16388059598724 | -1.70483219833812 |
| O | -1.36101618248193 | 0.14279491222191 | -1.72047718263036 |
| O | -1.26720125865315 | -2.66390060620438 | -1.02050614127123 |
| O | -4.04574812573581 | -2.59083445112945 | -2.07189437296826 |
| O | -1.42664162737784 | 2.45905422786803 | -0.26617606280580 |
| O | -1.33904900234591 | 1.43240778418400 | 2.38220029222690 |
| O | -2.87542863035273 | -0.70131160579923 | 2.48808606418883 |
| O | -1.49906007960037 | -2.97077418790611 | 1.81556894292833 |
| O | -4.37317293763856 | 1.17385036305048 | -2.36905083884506 |
| O | -4.47180138985562 | -0.71269259789375 | 0.08080878381881 |
| H | -5.26127988092092 | 1.50327490679117 | -2.55551772590838 |
| H | -4.20583221767747 | 0.06137261068969 | -0.42393605159509 |
| H | -3.93202529431790 | -0.71019053139913 | 0.88685919107945 |
| H | -4.29556153223707 | -2.11750559284970 | -1.26238069119403 |
| H | 0.45946231481611 | -2.86681412406337 | 2.11317525745265 |
| H | 1.48643026612294 | -2.79925875525793 | 0.97667738704276 |
| H | 0.56449311283957 | 2.69056393484344 | -0.14841642860994 |
| H | 1.50906787140225 | 2.19232946491602 | 0.97821227742068 |
| H | -2.05599910168904 | 3.09129328430566 | -0.61726025835328 |
| H | -1.51107035797828 | 1.64529554186694 | -0.81234746824762 |
| H | -1.95337040928392 | -3.76112702585479 | 2.11267290603661 |
| H | -2.00402015073782 | -2.20319489574805 | 2.15722728279278 |
| H | -2.26607693180647 | 0.08705777603918 | 2.47194735046007 |
| H | -3.42052730975544 | -0.60358512980070 | 3.27102292507636 |
| H | -3.14951878083454 | -2.90358669276072 | -1.89258345730722 |
| H | -3.99732171386029 | -0.76886132244826 | -3.21077309083151 |
| H | -1.38846173052227 | -1.70986708505019 | -1.16383327855765 |
| H | -1.40472332713532 | -2.80380653812097 | -0.06715168497398 |
| H | -0.38512828929904 | 1.39929779178202 | 2.57409660337525 |
| H | -1.42699834521543 | 1.87854344430246 | 1.52403352445975 |
| H | -0.39622716364761 | 0.15041405492292 | -1.93468468986498 |
| H | -1.86644694461949 | 0.16355732044291 | -2.54543739571466 |
| H | 3.52186853523725 | -0.60561079114182 | 1.58903951461570 |
| H | 2.39019569162580 | -1.40785208164541 | 2.31179977354169 |
| H | 3.27435955424218 | 2.17102547930620 | -0.98124914675535 |
| H | 3.17638154241030 | 1.08845148972661 | -2.03107904520134 |
| H | 4.44546472080778 | 0.28428673708027 | -0.50524114833048 |
| H | 5.60654797548893 | -0.39061348600205 | 0.28777709815028 |
| H | 2.25218151097943 | -2.83181857923585 | -1.30455700749818 |
| H | 0.64145456153927 | -2.79213655495630 | -1.15187697163788 |
| H | 4.51120080712141 | -2.68207810721312 | -2.45045893828487 |
| H | 4.36899433030757 | -2.06354925875303 | -1.04029112749392 |
| H | 2.02088693865913 | 0.72338404429111 | 2.56874763772586 |
| H | 1.82259694563436 | 2.04351512505325 | 3.40387020652663 |
| H | 1.46203916785189 | -0.73744202147514 | -1.34601341740143 |
| C | -3.95422374402117 | 0.32744239485399 | -3.31791667043993 |

W18-b..CNN from W18-b..N$_2$ + C

| | | | |
|---|---|---|---|
| H | 1.51997523427432 | 1.18494582172715 | -1.07857968639601 |
| O | 0.92204910121860 | 2.75371655756905 | -0.79780138981079 |
| O | 0.40577331468247 | 2.78207775324170 | 1.95316668702749 |
| O | 1.65091251020270 | 0.67624896125018 | 2.91385965217841 |
| O | 0.21687272480464 | -1.66313098480387 | 2.72001742224393 |
| O | 4.47918708670672 | -0.00634034930569 | -0.97542226283407 |
| O | 4.21001308311063 | -0.17818164800270 | 1.79611676994774 |
| O | 3.92468634496517 | -2.65187363794055 | 0.18196449984614 |
| O | 1.27461667258312 | -1.97213668363898 | 0.10136809225427 |
| O | 1.82624304666517 | 0.29303536999357 | -1.38192516067779 |
| O | -0.45356926041105 | -0.25828390436951 | -2.75221210346307 |
| O | -1.31159368981986 | -2.27634387855689 | -0.99902768144369 |
| O | -3.95035857858146 | -3.00960544543752 | -0.68495795412557 |
| O | -1.73305884374123 | 2.07472521239554 | -1.73700372344357 |
| O | -2.22940614180790 | 2.37591004836551 | 0.99911066693573 |
| O | -3.98727393223590 | 0.56273263699164 | 1.66724677457110 |
| O | -2.42596778101095 | -1.69773706818828 | 1.51699929004596 |
| O | -4.25667741713626 | 1.56404607877851 | -2.70173933506916 |
| O | -4.98668050877212 | -0.30606582791825 | -0.87020503005432 |
| H | -5.01252365528961 | 2.14195417629618 | -2.81471158072782 |
| H | -4.54332487844381 | 0.84727850064460 | -2.09093908867029 |
| H | -4.75678792661288 | 0.00786259637316 | 0.01927991826387 |
| H | -4.74093481288167 | -1.24363283816156 | -0.89353354998124 |
| H | -0.73665153375551 | -1.70281906575899 | 2.56028455623753 |
| H | 0.60996787406694 | -1.89221827956063 | 1.86161590263496 |
| H | 0.04271080399336 | 2.84543023462924 | -1.19272896488991 |
| H | 0.80462902810567 | 2.90136904528817 | 0.15814475974730 |
| H | -2.63810721028390 | 2.10552678911610 | -2.10591135772196 |
| H | -1.35668710464973 | 1.25089037109509 | -2.09045178844186 |
| H | -3.06023840320196 | -2.38963073182635 | 1.29541263109528 |
| H | -2.96307940116616 | -0.89488514206761 | 1.67166761031251 |
| H | -3.33259262002940 | 1.29211303975259 | 1.47790038303156 |
| H | -4.52266247858076 | 0.84348813065382 | 2.41045550810759 |
| H | -3.03468648168787 | -2.93125421404486 | -1.01791184486839 |
| H | -4.31737010190188 | -3.81943719517632 | -1.04134783465191 |
| H | -1.13172529599436 | -1.59389680892973 | -1.66931690145710 |
| H | -1.55695867866240 | -1.81339113525912 | -0.17467129326666 |
| H | -1.37347254022395 | 2.49232515411884 | 1.44478739285585 |
| H | -2.03251806533946 | 2.25592511850344 | 0.05065668258285 |
| H | 0.47164393636648 | -0.08452513799689 | -2.46954021810994 |
| H | -0.48225450996979 | -0.26169019364847 | -3.71184075867228 |
| H | 2.49409845275005 | 0.39409748488854 | 2.52870825575356 |
| H | 1.08751953609895 | -0.12526149171564 | 2.95780589928619 |
| H | 5.05703482711881 | 0.47945656930119 | -1.56506166435253 |
| H | 3.55189963849951 | 0.18190461331886 | -1.25389929317015 |
| H | 4.40900651782564 | 0.08983514456432 | 0.87961672227599 |
| H | 4.94517078971613 | 0.11318808968641 | 2.33817420104735 |
| H | 2.13211389981900 | -2.44075743025962 | 0.04122182578023 |
| H | 0.58929894989760 | -2.45364813110199 | -0.38677911543693 |
| H | 4.30579004756640 | -2.04464848421323 | -0.46670782867148 |
| H | 4.07469181614689 | -2.19616109287382 | 1.02126468859840 |
| H | 0.88274871936179 | 1.98730745355979 | 2.32233251969359 |
| H | 0.56941017683790 | 3.50730195280809 | 2.55785546186203 |
| H | 1.57224839612070 | -0.37578958395027 | -0.71139200434947 |
| N | -2.03090432177931 | 0.79776955571560 | -5.29120829699486 |
| N | -2.93998544837835 | 1.52912887152865 | -5.53309640125080 |
| C | -3.87916848335582 | 2.28070938365182 | -5.77405077901378 |

W18-b..HNNC from W18-b..CNN + H

| | | | |
|---|---|---|---|
| H | 1.57967982420465 | 1.24735171002859 | -1.02195906406025 |
| O | 1.03960827562920 | 2.82568304003328 | -0.61079909678082 |
| O | 0.48693322425292 | 2.71851540094471 | 2.13320579550159 |
| O | 1.74020282788678 | 0.56811993883568 | 2.96585110045563 |
| O | 0.24720432630191 | -1.72262332575820 | 2.66325174171920 |
| O | 4.49631241002800 | -0.01417281015745 | -1.01811819022213 |
| O | 4.26072271626385 | -0.25508392670050 | 1.74970640762300 |
| O | 3.90935639336846 | -2.67993723627403 | 0.06694619950957 |
| O | 1.27421498943791 | -1.94420442737510 | 0.02784453320709 |
| O | 1.85047025199701 | 0.36912509561377 | -1.38619170937110 |
| O | -0.57239418675749 | -0.01861718303452 | -2.58951978089700 |
| O | -1.33689056715074 | -2.18564226951491 | -1.01862867068233 |
| O | -4.00765176655845 | -2.80820387198037 | -0.95476258829379 |
| O | -1.68760276342681 | 2.33789831613470 | -1.54058757107363 |
| O | -2.16761419313786 | 2.46374916091158 | 1.20409482499773 |
| O | -3.94884836938780 | 0.60204260993954 | 1.60820562001909 |
| O | -2.44665748915815 | -1.70314541310420 | 1.51914043084152 |
| O | -4.12085696784042 | 1.65201050956391 | -2.88998192441121 |
| O | -4.82984088129291 | -0.08648553718223 | -1.01412779260206 |
| H | -4.90816034604353 | 2.15279748743543 | -3.11038768577124 |
| H | -4.39488958715618 | 0.98506363738637 | -2.20534784099279 |
| H | -4.61748750880947 | 0.17465585318901 | -0.10228096524176 |
| H | -4.64153512669361 | -1.03889022704892 | -1.07171713351277 |
| H | -0.70940418718480 | -1.74368141871727 | 2.52494538843485 |
| H | 0.62344241510747 | -1.91905012715140 | 1.78884640058872 |
| H | 0.17959778894277 | 3.00574466542029 | -1.01406285917175 |
| H | 0.90954594670827 | 2.91103716285248 | 0.35132673691389 |
| H | -2.53775407838906 | 2.45125243655645 | -1.99220435313966 |
| H | -1.33574334901641 | 1.50268037257224 | -1.90259236295931 |
| H | -3.09109045344481 | -2.39482633783380 | 1.33870068536611 |
| H | -2.96926833261530 | -0.88523540356262 | 1.64492323732948 |
| H | -3.28470936949521 | 1.34392886562528 | 1.52081388551573 |
| H | -4.54512579163988 | 0.83786582284954 | 2.32046933524924 |
| H | -3.05621667951560 | -2.75499917699972 | -1.17774600768229 |
| H | -4.37326318891417 | -3.55219253118143 | -1.43474305403931 |
| H | -1.15343369100608 | -1.43325478493007 | -1.61328634699529 |
| H | -1.57584021630686 | -1.80649851456105 | -0.15185508420799 |
| H | -1.30344140975359 | 2.50383549920065 | 1.64860379254259 |
| H | -1.98041903483793 | 2.44144058648210 | 0.24670129053462 |
| H | 0.38663288637888 | 0.10125473059487 | -2.42922799569648 |
| H | -0.75511320714269 | 0.00382223593686 | -3.53888334765422 |
| H | 2.56934358236385 | 0.29866886464365 | 2.54231058425196 |
| H | 1.16424571672145 | -0.22514988118270 | 2.97362651121806 |
| H | 5.08002325307135 | 0.46526356004996 | -1.60725140890199 |
| H | 3.57123542313537 | 0.20465511164605 | -1.28232003378072 |
| H | 4.44895596624414 | 0.03294132028951 | 0.83682142745682 |
| H | 5.00954064860221 | 0.01120398552219 | 2.28592895971766 |
| H | 2.12509857136150 | -2.42268494857238 | -0.05060129348577 |
| H | 0.57987929388284 | -2.40158705638401 | -0.47071621266796 |
| H | 4.29250829657024 | -2.05877816177804 | -0.56740190491507 |
| H | 4.08207224687887 | -2.25639537180396 | 0.91846027357932 |
| H | 0.96789916283515 | 1.90581556469728 | 2.45761977204933 |
| H | 0.67493895911859 | 3.41790317317372 | 2.76103740642001 |
| H | 1.59005470481033 | -0.32733750850195 | -0.74743696165720 |
| N | -2.33094231222121 | 0.46003848488539 | -4.83199838807718 |
| N | -2.75300935022577 | 0.37951051549638 | -6.03764631290483 |
| C | -3.04461504283825 | 0.25988195840836 | -7.17824573354680 |
| H | -3.07280562414287 | 0.88993862837097 | -4.24485568584600 |

W18-b..HCNN from W18-b..CNN + H

| | | | |
|---|---|---|---|
| H | 1.48579369476531 | 1.25929837606658 | -1.05363032818827 |
| O | 0.89341912725140 | 2.82356922232296 | -0.72639185478432 |
| O | 0.35954788421725 | 2.75459433871110 | 2.02072869464310 |
| O | 1.60696409116654 | 0.60735745150953 | 2.87511278252224 |
| O | 0.16371002041726 | -1.71662800649338 | 2.60195632804436 |
| O | 4.44240219641081 | 0.04662897236917 | -1.01607118167507 |
| O | 4.16946571936026 | -0.21546403831972 | 1.74440620761690 |
| O | 3.87131900958888 | -2.63334367471932 | 0.04676761026884 |
| O | 1.22054314169212 | -1.96073184450494 | -0.01483412629662 |
| O | 1.78974564342029 | 0.37796778457961 | -1.38646910140268 |
| O | -0.47574063747602 | -0.09165514739104 | -2.82261664324815 |
| O | -1.37022977897251 | -2.16587876947877 | -1.16091819399123 |
| O | -4.03249937588330 | -2.82000568699607 | -0.94731389079581 |
| O | -1.77219467134976 | 2.18096207584133 | -1.70745649774961 |
| O | -2.26585631179284 | 2.41215046099967 | 1.04159952751751 |
| O | -4.01644155205757 | 0.57681841344986 | 1.65491742646724 |
| O | -2.47727894823439 | -1.68999277836324 | 1.38631703067581 |
| O | -4.42055940558833 | 1.86533317046672 | -2.59613263710504 |
| O | -5.04757595863409 | -0.12747393768173 | -0.90368299381422 |
| H | -5.11033370937457 | 2.51676318308655 | -2.45624770208406 |
| H | -4.65383320581248 | 1.09594514057177 | -2.02016975694179 |
| H | -4.80321681903887 | 0.12178535019568 | 0.00297346050874 |
| H | -4.79779059271090 | -1.06038255758364 | -1.00227377616348 |
| H | -0.78932608710636 | -1.74492888842337 | 2.43831131818720 |
| H | 0.55709290878004 | -1.92761112864177 | 1.73865198652535 |
| H | 0.02339655986385 | 2.93523064821427 | -1.13510732425004 |
| H | 0.75941720435498 | 2.94282540707472 | 0.23124008138585 |
| H | -2.67463328221239 | 2.25593456586369 | -2.06786593362048 |
| H | -1.40593941144758 | 1.36259767560203 | -2.08490790256226 |
| H | -3.11467757939926 | -2.37153421695314 | 1.14534855095477 |
| H | -3.01135305188702 | -0.89286818040181 | 1.57631383180320 |
| H | -3.36132421652028 | 1.31345069051074 | 1.49350386226537 |
| H | -4.53603818908819 | 0.81667466773578 | 2.42336841226959 |
| H | -3.10375109631142 | -2.74802426565959 | -1.24499997479864 |
| H | -4.41018386995266 | -3.59231973272289 | -1.36979211977136 |
| H | -1.15735556087172 | -1.45249151332947 | -1.79043294218261 |
| H | -1.59530952375704 | -1.74404233390211 | -0.31101337690176 |
| H | -1.41020623084451 | 2.50037173731693 | 1.49558878175650 |
| H | -2.06428443939007 | 2.31942662883510 | 0.09143526648159 |
| H | 0.44638080749678 | 0.04187155043940 | -2.51181172170883 |
| H | -0.46619205662076 | -0.12709971212556 | -3.78562688532516 |
| H | 2.45228076924609 | 0.34044704348535 | 2.48358553550523 |
| H | 1.04507451144351 | -0.19648004147475 | 2.89046106607425 |
| H | 5.02075715610477 | 0.54378423492653 | -1.59565939997092 |
| H | 3.51480621079998 | 0.25134733292145 | -1.28126815030982 |
| H | 4.37243117977821 | 0.08082235806551 | 0.83733012527307 |
| H | 4.90609368948136 | 0.05245108832928 | 2.29648168604290 |
| H | 2.08262449278201 | -2.41896062409435 | -0.09111845824248 |
| H | 0.54538892175628 | -2.41901042590731 | -0.53698910652549 |
| H | 4.25630788321566 | -2.00547800639869 | -0.57964820912260 |
| H | 4.02459608189500 | -2.21031252165765 | 0.90230618480879 |
| H | 0.83865566706362 | 1.94328173252121 | 2.34987464740064 |
| H | 0.53127340990268 | 3.45225960679286 | 2.65491047115157 |
| H | 1.52904042229307 | -0.31301954164764 | -0.74285433353576 |
| N | -1.42768926167128 | 0.16968967154210 | -5.71475969824278 |
| N | -2.46846307487686 | 0.64300965375587 | -5.57452718701338 |
| C | -3.61323213837946 | 1.14934215394916 | -5.59397766108895 |
| H | -4.05665961168539 | 1.42050185391945 | -4.63043638713676 |

W18-b..c-HNNC from W18-b..HNNC

| | | | |
|---|---|---|---|
| H | 1.61671756251478 | 1.21562110573236 | -1.06663307876212 |
| O | 1.01839732061723 | 2.79607353560113 | -0.72200074544511 |
| O | 0.47776039244665 | 2.74637018935878 | 2.02876861620604 |
| O | 1.73784319929432 | 0.62032908386020 | 2.91088989331341 |
| O | 0.25638695514757 | -1.67962664306648 | 2.63591615830443 |
| O | 4.55076450623514 | -0.03164315183271 | -0.99580859452049 |
| O | 4.27637033307615 | -0.23429629422248 | 1.76586023768812 |
| O | 3.92380058168368 | -2.67595123406120 | 0.10575099253351 |
| O | 1.28209182862041 | -1.97595430354317 | 0.01230218674020 |
| O | 1.91739513925676 | 0.33456403428620 | -1.39589253897750 |
| O | -0.46734797911389 | -0.11369245648304 | -2.70383607946593 |
| O | -1.33402877139297 | -2.18623130589896 | -1.07719973380860 |
| O | -4.01944445955774 | -2.72035914917719 | -1.04893230894273 |
| O | -1.69008268844511 | 2.17350445203527 | -1.65694291569502 |
| O | -2.15704311396005 | 2.41232922288942 | 1.08452475483491 |
| O | -3.94567867055841 | 0.59735427134010 | 1.61803504790883 |
| O | -2.42217997701570 | -1.69331349291399 | 1.48182324074863 |
| O | -4.31896045462166 | 1.67828948949599 | -2.81117547764880 |
| O | -4.94553017312408 | -0.06006841740176 | -0.95251334740350 |
| H | -5.05236365224688 | 2.29194501444006 | -2.88531723381037 |
| H | -4.57030125410441 | 1.00989489703821 | -2.10931840127206 |
| H | -4.69083601627095 | 0.19440783833043 | -0.04884669298658 |
| H | -4.70095193967331 | -0.99842061920493 | -1.04900857498673 |
| H | -0.70016560927943 | -1.70054675209250 | 2.49641625706835 |
| H | 0.63052626960503 | -1.89842372838140 | 1.76567918652472 |
| H | 0.15465102393170 | 2.92935804418464 | -1.13509899821582 |
| H | 0.87942977297290 | 2.91514933214530 | 0.23484422591776 |
| H | -2.55202511631931 | 2.26876258039191 | -2.08791637617757 |
| H | -1.30794888444940 | 1.34749986637625 | -2.01304720058296 |
| H | -3.04705715086982 | -2.40629862860687 | 1.31910531404632 |
| H | -2.96762703262682 | -0.89316125093062 | 1.62194837811476 |
| H | -3.26666933263920 | 1.32060173579040 | 1.48956830658825 |
| H | -4.49784709723095 | 0.85135148566334 | 2.35892833826805 |
| H | -3.06166004632920 | -2.67348761838674 | -1.24994418771806 |
| H | -4.38296423258340 | -3.44484237947610 | -1.55926670978001 |
| H | -1.09644353735516 | -1.45462859298422 | -1.68081191584126 |
| H | -1.55041835731935 | -1.78782873073982 | -0.21457614110243 |
| H | -1.29589295164982 | 2.48995008126644 | 1.53088164794489 |
| H | -1.96564130449883 | 2.33048581498124 | 0.13110583063002 |
| H | 0.48254248162227 | 0.02335357072201 | -2.51946033576018 |
| H | -0.61837600698614 | -0.12206500905712 | -3.66755874667839 |
| H | 2.57862800025673 | 0.34651465428609 | 2.51381340527615 |
| H | 1.16830996389957 | -0.17767276419778 | 2.92724179987564 |
| H | 5.13814693876055 | 0.43580173596901 | -1.59074896078338 |
| H | 3.62579567385902 | 0.18270170469801 | -1.26799500957048 |
| H | 4.48217329714632 | 0.03941832330361 | 0.85204991030544 |
| H | 5.02415543246230 | 0.02114715016997 | 2.30866088071878 |
| H | 2.13934915664257 | -2.44554396848595 | -0.04975788346612 |
| H | 0.60257729243725 | -2.44269957169754 | -0.49607163932462 |
| H | 4.31871183741503 | -2.06173113726630 | -0.52837932340467 |
| H | 4.07964959849718 | -2.24197202642516 | 0.95524974469648 |
| H | 0.96505755796479 | 1.94275473290008 | 2.36627922379920 |
| H | 0.65514154632862 | 3.45599165978609 | 2.64799084941441 |
| H | 1.62910415087032 | -0.35180383747177 | -0.75999063968935 |
| N | -3.19763281478271 | 0.38929957088787 | -4.95577581644442 |
| N | -2.61333563959463 | 0.86903304491429 | -6.12083651240653 |
| C | -1.85710452299484 | 0.23562194073376 | -5.22795118208013 |
| H | -3.74182740392274 | 0.90550103829494 | -4.22734530619944 |

W18-b..HNCN from W18-b..c-HNNC

| | | | |
|---|---|---|---|
| H | 1.61404824024604 | 1.19210140904614 | -1.03179704350337 |
| O | 1.00429436196066 | 2.76690387518528 | -0.72524012176883 |
| O | 0.44996592048792 | 2.75490433841800 | 2.02758134209019 |
| O | 1.86496919124226 | 0.73113469138390 | 2.92599728008782 |
| O | 0.36702396426538 | -1.55978830373280 | 2.71212671998705 |
| O | 4.55722107962896 | -0.05944397174791 | -1.04388117141284 |
| O | 4.36167033478482 | -0.17001821645452 | 1.72518055009593 |
| O | 3.93629356351167 | -2.66002454019443 | 0.15200988959227 |
| O | 1.28353021967026 | -2.00966844356143 | 0.08249541407906 |
| O | 1.91293689588140 | 0.30338847127787 | -1.34550765804740 |
| O | -0.51793414403447 | -0.19704171664777 | -2.51758206849447 |
| O | -1.38733997570307 | -2.33190272215890 | -0.96619148097587 |
| O | -3.84351217965196 | -2.12507849623457 | -2.24026658334842 |
| O | -1.68305982759975 | 2.18297809788374 | -1.63588810932522 |
| O | -2.17310671807703 | 2.25662824302754 | 1.10280907232279 |
| O | -3.92934688824860 | 0.39954946972879 | 1.57560835583007 |
| O | -2.34078184073314 | -1.81113199721291 | 1.67621961465561 |
| O | -4.31588647524581 | 2.02647999566033 | -2.73843219580151 |
| O | -4.99755631957516 | 0.11039217389346 | -1.07090934999736 |
| H | -4.96352862810051 | 2.73323446349675 | -2.71388446786725 |
| H | -4.60775359735790 | 1.34293749334176 | -2.07110290146468 |
| H | -4.67767109403705 | 0.16938684504857 | -0.15598040432111 |
| H | -4.66942346648500 | -0.73430874049271 | -1.43325247105038 |
| H | -0.59078218334669 | -1.53625250519898 | 2.58136690787459 |
| H | 0.71458260606195 | -1.82372876180102 | 1.84274060835710 |
| H | 0.13672068595646 | 2.86996278936942 | -1.14243846145087 |
| H | 0.86001909893809 | 2.90419834975636 | 0.22790752382893 |
| H | -2.53086061400216 | 2.33530365694509 | -2.08461325608993 |
| H | -1.35343998038650 | 1.32592165958386 | -1.96830416370386 |
| H | -2.81584890519486 | -2.58341591599606 | 1.98859629309402 |
| H | -2.96689770606251 | -1.05633844429242 | 1.71758319955643 |
| H | -3.25188736490011 | 1.12923896487878 | 1.46462946290481 |
| H | -4.53781382937897 | 0.68420853395460 | 2.25947543498959 |
| H | -2.97864462794387 | -2.36391057177616 | -1.85097361932884 |
| H | -4.22057269092636 | -2.92368634214508 | -2.60935116664962 |
| H | -1.08716922738737 | -1.52675328408682 | -1.43245849996694 |
| H | -1.67269386369237 | -2.06675278410052 | -0.07360078564557 |
| H | -1.31426003396658 | 2.37360511735490 | 1.54431058761311 |
| H | -1.99054281317305 | 2.24851762105622 | 0.14302529428144 |
| H | 0.43894039020895 | -0.04955518663610 | -2.38263129126222 |
| H | -0.68318386782998 | -0.39375468547782 | -3.45284569711234 |
| H | 2.69637254691625 | 0.46598295891755 | 2.50425770025598 |
| H | 1.30876836349705 | -0.07562190926498 | 2.96467711057390 |
| H | 5.12104484399378 | 0.38829584540273 | -1.67561580511450 |
| H | 3.62232898074738 | 0.15282208316529 | -1.28177024083868 |
| H | 4.54916541573317 | 0.07166522866200 | 0.79833463455662 |
| H | 5.12788601711186 | 0.08698895555276 | 2.24090398629907 |
| H | 2.15010654778984 | -2.46275899468430 | 0.02117510064905 |
| H | 0.60652690062134 | -2.51468668368467 | -0.39034823995049 |
| H | 4.31823539740144 | -2.07471289104932 | -0.51647598801139 |
| H | 4.11705341684349 | -2.19308014114789 | 0.97891270843353 |
| H | 1.00214117282285 | 1.99845269262242 | 2.37191416217854 |
| H | 0.58435130244794 | 3.49004643476967 | 2.62752711428895 |
| H | 1.63171095036453 | -0.37165450201237 | -0.69616235588906 |
| N | -3.93809621532580 | 0.54246525275747 | -5.17024477562942 |
| N | -1.74947783369948 | -0.54927490450088 | -5.12704775878718 |
| C | -2.80429464838942 | -0.00777878715242 | -5.10491316456893 |
| H | -4.16844922663160 | 1.09314686917213 | -4.31997895258145 |

W18-b..HNCNH from W18-b..HNCN + H
| | | | |
|---|---|---|---|
| H | 1.60765961410515 | 1.16630920335910 | -1.02484537232435 |
| O | 1.05682845567785 | 2.75299520809151 | -0.68844817189798 |
| O | 0.40261274708751 | 2.71776888566570 | 2.05092577582593 |
| O | 1.88696027968807 | 0.73819548455164 | 2.94304926762886 |
| O | 0.40870772049663 | -1.57459622847874 | 2.77006938165950 |
| O | 4.53059287213663 | -0.05669786274852 | -1.09616935518121 |
| O | 4.37122612737369 | -0.11404320052302 | 1.67912441115415 |
| O | 3.97855347474700 | -2.63682725605788 | 0.16070709595418 |
| O | 1.31671967442365 | -2.02750622280764 | 0.14005437704602 |
| O | 1.88192362926430 | 0.27337743383755 | -1.35024661162282 |
| O | -0.65258765738136 | -0.18562396910788 | -2.34013718550589 |
| O | -1.36502640312760 | -2.40247402010402 | -0.84121197923196 |
| O | -3.58400237887409 | -2.58975961957897 | -2.42205061297752 |
| O | -1.62409099412843 | 2.33084698981684 | -1.62766553727005 |
| O | -2.21167711355567 | 2.24582953976475 | 1.07433868806131 |
| O | -3.89245645549185 | 0.27518863590374 | 1.25906476571877 |
| O | -2.32338086262035 | -1.87716562565979 | 1.77176922760530 |
| O | -4.17209018767535 | 2.28552487638542 | -2.88399032074230 |
| O | -4.60034230025523 | -0.01563769826729 | -1.53388518964755 |
| H | -4.88773411717495 | 2.89707827999106 | -2.70404545054070 |
| H | -4.38697380386386 | 1.44969484298663 | -2.40400140744561 |
| H | -4.36667483186741 | 0.01960957694689 | -0.59242373394763 |
| H | -4.33250392953750 | -0.88816869796926 | -1.86353871916534 |
| H | -0.54843811326432 | -1.54369227776444 | 2.63976823508631 |
| H | 0.75652267671657 | -1.83822247083723 | 1.90038187587818 |
| H | 0.20663594995881 | 2.89317804690299 | -1.13156545375692 |
| H | 0.88090406556144 | 2.87410128871462 | 0.26130594439182 |
| H | -2.43060089035692 | 2.54372147601490 | -2.12632632521777 |
| H | -1.37166656883725 | 1.42772777529236 | -1.90254087019812 |
| H | -2.79183663648828 | -2.59206138538802 | 2.20610390787935 |
| H | -2.95218030027789 | -1.12662011182940 | 1.68902570969948 |
| H | -3.24067115609496 | 1.03820791457127 | 1.24992140213185 |
| H | -4.63368324397961 | 0.55641167577461 | 1.79858297512820 |
| H | -2.79083439621647 | -2.69216196446028 | -1.84933157597311 |
| H | -4.07438192662869 | -3.41170156056001 | -2.37731529024043 |
| H | -1.12511738153099 | -1.55148643187552 | -1.26312001114954 |
| H | -1.68586082917737 | -2.19811042021851 | 0.05643339823609 |
| H | -1.36446429818983 | 2.33868087795746 | 1.54288877146161 |
| H | -2.00563406990109 | 2.33358871766499 | 0.12198973988743 |
| H | 0.31530812307372 | -0.06527065745519 | -2.31060428770815 |
| H | -0.92959991699042 | -0.44179262528705 | -3.23926629638832 |
| H | 2.71154565040330 | 0.49198806643277 | 2.49701135842995 |
| H | 1.35160637760873 | -0.08079491466030 | 3.00011868687885 |
| H | 5.07626387387458 | 0.40044043799366 | -1.73703967392594 |
| H | 3.58770645324610 | 0.13452576536171 | -1.32362297116504 |
| H | 4.53210173560845 | 0.11602923834762 | 0.74420426266374 |
| H | 5.14343204374003 | 0.16819198796115 | 2.17219150239963 |
| H | 2.18942782748687 | -2.46679274152695 | 0.06501096397316 |
| H | 0.63898553607602 | -2.54508754170256 | -0.31711784586893 |
| H | 4.33784670316624 | -2.05627940694936 | -0.52452330361997 |
| H | 4.16503113121931 | -2.15135769956014 | 0.97566452262507 |
| H | 0.97446309419943 | 1.97659321400277 | 2.39502767789132 |
| H | 0.52338026188103 | 3.45666497889646 | 2.64928357029588 |
| H | 1.62029568042065 | -0.39560418285096 | -0.68756433576283 |
| N | -3.53728670491105 | 0.88156858939636 | -5.41873021661945 |
| N | -2.23843339795558 | -0.98348987229489 | -4.55463868071501 |
| C | -2.93660482032959 | -0.05244392481835 | -4.94095890201429 |
| H | -3.84647230856153 | 1.60776471743518 | -4.77052125890535 |
| H | -2.73264773669608 | -1.77692882387957 | -4.14620027736173 |

W18-b..NH$_2$CN from W18-b..HNCN + H

| | | | |
|---|---|---|---|
| H | 1.62375005225496 | 1.28678784428303 | -1.01591401881981 |
| O | 0.99359436466915 | 2.83925478949201 | -0.68135812083303 |
| O | 0.41308702688769 | 2.72073555852388 | 2.06704990559650 |
| O | 1.80272989421502 | 0.63658974627735 | 2.86898513251822 |
| O | 0.28349811756251 | -1.62652361464165 | 2.52868985270295 |
| O | 4.56703496642091 | -0.02252502291963 | -1.03964433005804 |
| O | 4.30851819228886 | -0.27525566760435 | 1.70657353114026 |
| O | 3.82115529278017 | -2.65153436255119 | -0.01444593193776 |
| O | 1.17710872631691 | -1.99879603649537 | -0.11930425692453 |
| O | 1.94197490752631 | 0.40834243294910 | -1.34069192521611 |
| O | -0.42303711333639 | -0.03390844836058 | -2.67676318546229 |
| O | -1.57950027505077 | -2.11232408976163 | -1.23982838783190 |
| O | -4.08142012190564 | -2.12819882639293 | -2.30817836769351 |
| O | -1.67199135881212 | 2.24261619252678 | -1.64718756630394 |
| O | -2.19003369970168 | 2.25640899302468 | 1.07805992904354 |
| O | -3.98650688771802 | 0.45731946512634 | 1.57344345706147 |
| O | -2.41311037667289 | -1.75470997926970 | 1.47126388629460 |
| O | -4.30729229993373 | 2.13527878290180 | -2.72652513592304 |
| O | -5.11882965520039 | 0.22481319232927 | -1.02877808293168 |
| H | -4.93778486308208 | 2.85459642694372 | -2.78535408870581 |
| H | -4.65239731729457 | 1.49980653255882 | -2.05083956098677 |
| H | -4.77860027084099 | 0.27114641115160 | -0.11946815482933 |
| H | -4.86951528055549 | -0.64502377893874 | -1.37984527643132 |
| H | -0.67152384057814 | -1.56002551883301 | 2.39253771545505 |
| H | 0.62406453475666 | -1.86854024157587 | 1.65047569252224 |
| H | 0.13293900593851 | 2.94929482733335 | -1.11102479005941 |
| H | 0.83558988639628 | 2.95135281483092 | 0.27235291737701 |
| H | -2.51455779630864 | 2.40585996632648 | -2.10291883164847 |
| H | -1.31006636615925 | 1.42119509862709 | -2.03240335325691 |
| H | -2.86706983043279 | -2.55123773780108 | 1.75245159135542 |
| H | -3.03883793548749 | -1.00920951026661 | 1.59867982056244 |
| H | -3.28805687929284 | 1.16794620897839 | 1.45242453640273 |
| H | -4.55569247324700 | 0.74038527841395 | 2.29080088449211 |
| H | -3.15928326634690 | -2.22404451470646 | -1.96889095324919 |
| H | -4.44591181172756 | -3.01176746661581 | -2.37859260006879 |
| H | -1.17544135437455 | -1.33916252739991 | -1.68385048563711 |
| H | -1.78947161832782 | -1.85623433183170 | -0.32432385911488 |
| H | -1.33618886564764 | 2.36690830830059 | 1.53090895864929 |
| H | -1.99905667050132 | 2.25337054124993 | 0.11909601478308 |
| H | 0.52252506982702 | 0.10946366281290 | -2.48147584620073 |
| H | -0.53730870271815 | -0.14384580157111 | -3.63764471865949 |
| H | 2.64425948185615 | 0.37985526510669 | 2.46237071398527 |
| H | 1.23923156874457 | -0.16571805234422 | 2.85866873195646 |
| H | 5.14402707034319 | 0.44074955510036 | -1.64772151540918 |
| H | 3.63737444099281 | 0.22217004116655 | -1.26881575474584 |
| H | 4.53196989566980 | 0.01320940849140 | 0.80105863288552 |
| H | 5.06978224053833 | -0.08352535702291 | 2.25680235694620 |
| H | 2.05183563719804 | -2.43721104227368 | -0.18628873290493 |
| H | 0.52792020406624 | -2.47583885011566 | -0.65060459422518 |
| H | 4.24264058570888 | -2.04546973139527 | -0.63957835626069 |
| H | 3.99693324167277 | -2.23721330876857 | 0.84115546598251 |
| H | 0.95909831699128 | 1.94496705442513 | 2.37432944734375 |
| H | 0.54652019681126 | 3.42319433187290 | 2.70502463974097 |
| H | 1.60286030318458 | -0.28850498262225 | -0.74744518778921 |
| N | -3.81424388389826 | -0.14867656201291 | -4.51991339072417 |
| N | -1.47868035962208 | -0.27125082585472 | -5.35074183379156 |
| C | -2.56371614328457 | -0.20776991398403 | -4.95666350625113 |
| H | -4.08074461523622 | 0.75960727262950 | -4.13885469146092 |
| H | -4.06318157102275 | -0.92324067572300 | -3.90524929105085 |

W18-c..N$_2$

| | | | |
|---|---|---|---|
| H | 1.62261287857132 | 1.19692856849828 | -1.00456623391970 |
| O | 1.02504314740491 | 2.75957165123856 | -0.71924135159233 |
| O | 0.46150381429156 | 2.76061573582551 | 2.02573136789094 |
| O | 1.69331205175071 | 0.64531043734178 | 2.99676005101318 |
| O | 0.22886259692038 | -1.66428439657525 | 2.74143689226900 |
| O | 4.57980416931388 | -0.01986018198699 | -0.89380173719125 |
| O | 4.22937136347756 | -0.22185047032811 | 1.86604178506397 |
| O | 3.97082174976268 | -2.66775200538079 | 0.24663351795435 |
| O | 1.33034674882499 | -1.96310216757513 | 0.14016433367926 |
| O | 1.92736420233126 | 0.30620256931703 | -1.31438449954498 |
| O | -0.28466559561927 | -0.22043964818702 | -2.78634481349156 |
| O | -1.23545672928819 | -2.21639299737580 | -1.02941787177104 |
| O | -3.87992988285651 | -2.94738174486542 | -0.73255096243214 |
| O | -1.60191697386906 | 2.08799940237260 | -1.71746425773302 |
| O | -2.15526942514843 | 2.38041595745730 | 1.00358290161319 |
| O | -3.94194006411337 | 0.59492269087749 | 1.66403603743676 |
| O | -2.37781661231653 | -1.65813032923772 | 1.47662630888807 |
| O | -4.12159441189262 | 1.64330749338972 | -2.68412655699984 |
| O | -4.94847806312163 | -0.24121742454609 | -0.88316357989563 |
| H | -4.86193457071412 | 2.23521351449838 | -2.81906765040086 |
| H | -4.44192599240630 | 0.93346854399479 | -2.08485195037080 |
| H | -4.72075798812122 | 0.05954712025940 | 0.01118782842344 |
| H | -4.68331637841883 | -1.17245059649138 | -0.92452650468306 |
| H | -0.72000655626433 | -1.68509773654369 | 2.55143376259578 |
| H | 0.64428064161903 | -1.88535436204595 | 1.89138455020233 |
| H | 0.15812958901506 | 2.85025648198716 | -1.14225143198509 |
| H | 0.87446879800808 | 2.89806848983983 | 0.23319695533333 |
| H | -2.50076588977827 | 2.13564072908176 | -2.10345431884347 |
| H | -1.22390673434508 | 1.27493267145658 | -2.09008529308398 |
| H | -3.00762594063072 | -2.34779614142657 | 1.23540347719279 |
| H | -2.91947781152809 | -0.86013247640337 | 1.64120304643651 |
| H | -3.27786165463800 | 1.31669135467405 | 1.47750341954500 |
| H | -4.46700475573402 | 0.87318738650681 | 2.41536277084872 |
| H | -2.96455470450703 | -2.86827930187910 | -1.06465470367186 |
| H | -4.25059604590916 | -3.74973976053495 | -1.10181927624857 |
| H | -1.03239340277659 | -1.53730955819044 | -1.69657850149843 |
| H | -1.48424078059057 | -1.75208282228559 | -0.20812614729142 |
| H | -1.30749861448966 | 2.48518085111821 | 1.46673663851443 |
| H | -1.94106093229275 | 2.25427258536660 | 0.05912470999548 |
| H | 0.62268719204537 | -0.05152144582802 | -2.44448696167543 |
| H | -0.23753504890868 | -0.26363148516957 | -3.74265640872521 |
| H | 2.51906939937104 | 0.35773494938791 | 2.57577014670602 |
| H | 1.11612854689869 | -0.14774839348506 | 3.02599558002502 |
| H | 5.16750428699408 | 0.47096873572364 | -1.46928588202080 |
| H | 3.65754249901520 | 0.18099073567454 | -1.17762335350888 |
| H | 4.47036021113479 | 0.06832400663795 | 0.96729703384328 |
| H | 4.91115419299763 | 0.09097931307696 | 2.46422773770294 |
| H | 2.18539565563495 | -2.43788219773208 | 0.09104698393144 |
| H | 0.65025083548570 | -2.43094423549891 | -0.36791123417214 |
| H | 4.37199764442733 | -2.07096724171490 | -0.39929796335406 |
| H | 4.11944481045337 | -2.20979297566499 | 1.08544368548490 |
| H | 0.93235135100924 | 1.96507733114167 | 2.39934799248684 |
| H | 0.61695983994421 | 3.48492104316843 | 2.63357860566187 |
| H | 1.66152569786684 | -0.36863822484810 | -0.65432418997087 |
| N | 3.96243096963647 | 0.12616585899711 | 5.97577333353693 |
| N | 4.51619293846687 | 0.17202399387503 | 5.04075487514989 |

W18-c..C

| | | | |
|---|---|---|---|
| H | 1.50884260231360 | 1.06418547072546 | -1.29885675229087 |
| O | 1.17221173214922 | 2.65610384096788 | -0.80281775524762 |
| O | 1.16665522389195 | 2.91532744555688 | 1.95632632927410 |
| O | 2.53214675820697 | -0.23482493902401 | 1.55638568650562 |
| O | 1.21040135362603 | -2.34595788574745 | 1.96243621791408 |
| O | 4.41222955531717 | 1.01295049968967 | -1.72161400349220 |
| O | 4.78539000228099 | -0.76915678033269 | 0.21328002256631 |
| O | 4.11776838645799 | -3.00048355966375 | -1.47305800251478 |
| O | 1.55552910358485 | -2.35218569127393 | -0.78639208319599 |
| O | 1.75540489789438 | 0.26923262233857 | -1.83686197336444 |
| O | -0.69626098594381 | -0.00337914083864 | -2.92100257474539 |
| O | -1.23294797199936 | -2.17390098829607 | -1.23085964962266 |
| O | -3.63898559935648 | -2.98405789325003 | -0.13925438663216 |
| O | -1.65730338452533 | 2.16777612496355 | -1.28689164065305 |
| O | -1.51710437871678 | 2.22928149246022 | 1.52465263031464 |
| O | -2.87397701101226 | 0.25146826904430 | 2.52644709485846 |
| O | -1.52158677761340 | -1.91273084600050 | 1.57398035654182 |
| O | -4.36824617655001 | 1.78951661717458 | -1.42022047271075 |
| O | -4.65855144028534 | -0.28190090850486 | 0.35350164623098 |
| H | -5.08395924303740 | 2.39617221160026 | -1.22899358983456 |
| H | -4.49664670596378 | 1.01948577813759 | -0.82465300821964 |
| H | -4.16838368984978 | -0.08608939008271 | 1.16823660992601 |
| H | -4.44223313508583 | -1.19907424335479 | 0.13066144266016 |
| H | 0.24785634409805 | -2.22714341031965 | 2.06003349055170 |
| H | 1.32022693165931 | -2.62561559087756 | 1.03477861929919 |
| H | 0.25509494364848 | 2.84578641766554 | -1.04727772084271 |
| H | 1.23392622849792 | 2.83715079885483 | 0.15587684908602 |
| H | -2.63161106849771 | 2.22396704321373 | -1.36789966299465 |
| H | -1.41088180710759 | 1.41057041247895 | -1.84076362298926 |
| H | -2.22610885176693 | -2.55873986366860 | 1.43541708823150 |
| H | -1.95435931049237 | -1.14198616674039 | 2.00030216213693 |
| H | -2.34975716581733 | 1.04189605443165 | 2.20552679921795 |
| H | -3.10386845395527 | 0.41358041258080 | 3.44229516930339 |
| H | -2.90468192233862 | -2.86026319978617 | -0.77041777802453 |
| H | -4.13885364639852 | -3.74959611322724 | -0.42535148677320 |
| H | -1.20387113059588 | -1.43307517344543 | -1.86291966736664 |
| H | -1.20569912451216 | -1.78651319135379 | -0.33710105012409 |
| H | -0.60706789657346 | 2.45110955049717 | 1.79083495913746 |
| H | -1.52211941006148 | 2.17473100475684 | 0.55122205320706 |
| H | 0.26478706491872 | 0.10154904405765 | -2.73174894862039 |
| H | -0.82504078944580 | 0.08924274475253 | -3.86613140374771 |
| H | 3.36542984043502 | -0.50557491002416 | 1.08901675881281 |
| H | 1.98308052318852 | -1.05874707945920 | 1.84984979002245 |
| H | 4.35324543499500 | 1.94593332931138 | -1.50213653840262 |
| H | 3.49504794612307 | 0.74645781770612 | -1.94265986987067 |
| H | 4.77446752461248 | -0.03561701747478 | -0.45616280913186 |
| H | 5.54062318056756 | -0.62847837733303 | 0.78912180591617 |
| H | 2.32549798431654 | -2.77867211670676 | -1.20786573371836 |
| H | 0.72746648702262 | -2.62786376311956 | -1.20844920861585 |
| H | 4.55579631344882 | -2.87211935499333 | -2.31574226744921 |
| H | 4.49715099272592 | -2.34127705183970 | -0.86864909551121 |
| H | 1.78593038220097 | 2.17252769944353 | 2.27520765796554 |
| H | 1.42488046330009 | 3.70511784882523 | 2.43526505651680 |
| H | 1.71320293343428 | -0.52645435950895 | -1.27425299792649 |
| C | 2.80579194258614 | 0.79727345501285 | 2.61684345843642 |

W18-c..N$_2$ + C → W18-c..CNN

| | | | |
|---|---|---|---|
| H | 1.67677999141177 | 1.19455250870540 | -0.75686075029431 |
| O | 1.08702273290230 | 2.73273214375350 | -0.35750548636032 |
| O | 0.47238372639955 | 2.44517255758916 | 2.35478360411866 |
| O | 1.70074999438580 | 0.20001726624524 | 3.09404117669872 |
| O | 0.18205603227379 | -2.04810636487576 | 2.68698344923543 |
| O | 4.61628401398966 | -0.09288598411666 | -0.71676387133730 |
| O | 4.26503024583758 | -0.59004152668499 | 1.99573445018393 |
| O | 3.94909198581992 | -2.84343390573673 | 0.09274271128744 |
| O | 1.32027043284457 | -2.06838164935063 | 0.08187579623751 |
| O | 1.98053396908442 | 0.32714230217666 | -1.12915532249188 |
| O | -0.20435957202377 | -0.03947289535039 | -2.69469682203201 |
| O | -1.23020415297820 | -2.15487001922850 | -1.12841006378211 |
| O | -3.88210860242994 | -2.90461185533944 | -0.94777041391693 |
| O | -1.52726188414388 | 2.18472486563437 | -1.46260431531477 |
| O | -2.13107617341042 | 2.22977351406369 | 1.26704189704224 |
| O | -3.94240626237583 | 0.41643794409526 | 1.75398911994522 |
| O | -2.40840263630587 | -1.83316303294494 | 1.40169642558286 |
| O | -4.03884576027771 | 1.84128556456962 | -2.49416393847963 |
| O | -4.91325335422339 | -0.18087851767899 | -0.87247801382802 |
| H | -4.77055464835479 | 2.45172551148950 | -2.58694612912541 |
| H | -4.37461125167495 | 1.08688901037838 | -1.96182995479256 |
| H | -4.69889612519165 | 0.03853334775881 | 0.04854090973586 |
| H | -4.66462212041580 | -1.11041413272957 | -0.98770235756407 |
| H | -0.76444707563273 | -2.01759156945486 | 2.48541059346148 |
| H | 0.60042712632902 | -2.20076022942063 | 1.82349782882698 |
| H | 0.22825647944417 | 2.87488164523760 | -0.78276478024510 |
| H | 0.92432014100387 | 2.78950326352800 | 0.60160824320907 |
| H | -2.41912647762060 | 2.26883375874270 | -1.85818220323499 |
| H | -1.14572022042624 | 1.40227015332632 | -1.89221794493191 |
| H | -3.03679249205573 | -2.49531295713727 | 1.08989977099590 |
| H | -2.95041805479394 | -1.05079747323158 | 1.63024013960670 |
| H | -3.27251246624254 | 1.14711552518593 | 1.63081623548690 |
| H | -4.48009220787875 | 0.64435065451851 | 2.51340068249450 |
| H | -2.96229005440118 | -2.80016680281177 | -1.25870277750342 |
| H | -4.25398566940039 | -3.66277840670312 | -1.39993279740523 |
| H | -1.01341029795917 | -1.41604566522887 | -1.72373790953572 |
| H | -1.49738279742141 | -1.77123865108960 | -0.27177568426813 |
| H | -1.28955541278288 | 2.27261571906025 | 1.75145684268159 |
| H | -1.90268225076939 | 2.19387976655093 | 0.31844236783515 |
| H | 0.69715851752804 | 0.08446302763418 | -2.32084688347990 |
| H | -0.13686673718352 | -0.00426738697905 | -3.65006983400341 |
| H | 2.52382014016141 | -0.08303413044070 | 2.66636471244407 |
| H | 1.10359826306456 | -0.57895493495735 | 3.07701817753738 |
| H | 5.21805672335392 | 0.43290883015579 | -1.24496063948560 |
| H | 3.69952584079658 | 0.15618554584660 | -0.98216138664945 |
| H | 4.52033670270985 | -0.22117496057514 | 1.12957744014006 |
| H | 4.96550551128177 | -0.37669459657225 | 2.61538408347276 |
| H | 2.16272776387521 | -2.55363150596898 | -0.03083785427872 |
| H | 0.62883248796494 | -2.45251079691724 | -0.47947921595084 |
| H | 4.36155668102218 | -2.17199179838891 | -0.46740783618956 |
| H | 4.10219829956364 | -2.50621267649660 | 0.98548117301799 |
| H | 0.91920876503928 | 1.59755657600277 | 2.61225813905225 |
| H | 0.65508846026292 | 3.06207570228981 | 3.06903784888286 |
| H | 1.68483071954070 | -0.39662859948143 | -0.53754320877472 |
| N | 3.41864397450951 | 2.12128256082281 | 4.84672354454524 |
| N | 4.33206660381967 | 1.35659709799610 | 4.87843173254053 |
| C | 2.47449594645406 | 2.90062301023404 | 4.79329416215663 |

W18-c..c-CNN

| | | | |
|---|---|---|---|
| H | 1.64029869536989 | 1.23201137682080 | -0.89172725909092 |
| O | 1.01163446252501 | 2.78468482225158 | -0.57894064267049 |
| O | 0.41932537774698 | 2.69656661182859 | 2.17851860280605 |
| O | 1.68056321148878 | 0.51444202935349 | 3.04692292738522 |
| O | 0.20516486225018 | -1.73895009890780 | 2.77420542611892 |
| O | 4.62226032620505 | 0.04321071545061 | -0.78075361876499 |
| O | 4.18995103670644 | -0.25970943642987 | 1.95156023038113 |
| O | 4.00192515634038 | -2.65229810869010 | 0.24929756881188 |
| O | 1.36023992048823 | -1.95251428016553 | 0.19666707172465 |
| O | 1.96058605665420 | 0.35166894416715 | -1.21286654708367 |
| O | -0.23203444362676 | -0.15731899823962 | -2.72378725839047 |
| O | -1.19510667655933 | -2.17457877921527 | -1.00266335731895 |
| O | -3.82321655476688 | -2.97245582675396 | -0.73906071170079 |
| O | -1.58967053829222 | 2.10593694592853 | -1.61622157724775 |
| O | -2.19001098690783 | 2.33713923728171 | 1.09878804395228 |
| O | -3.98660513673498 | 0.53626757608491 | 1.70128384025735 |
| O | -2.38379676747268 | -1.68814769566460 | 1.50510727787485 |
| O | -4.08611750511667 | 1.65593217840514 | -2.63282601321368 |
| O | -4.92109921883018 | -0.27467502948063 | -0.88382413682399 |
| H | -4.83297510749353 | 2.23793351004548 | -2.77475671828556 |
| H | -4.40798045374591 | 0.92906448499644 | -2.05560206627658 |
| H | -4.72046029444388 | 0.01476856226966 | 0.02052015230301 |
| H | -4.64979380354926 | -1.20370066544208 | -0.93109587486248 |
| H | -0.74283665814147 | -1.73995716300940 | 2.57520123329297 |
| H | 0.62553276288347 | -1.96554592429333 | 1.92737023595501 |
| H | 0.14547194753015 | 2.87391813615686 | -1.00443858097151 |
| H | 0.85669547423966 | 2.90787119262890 | 0.37338637418099 |
| H | -2.48142415650908 | 2.15374513815761 | -2.01874601612242 |
| H | -1.19584893753825 | 1.30636528403820 | -2.00174673046383 |
| H | -2.99191569368424 | -2.38611519497632 | 1.23278113271638 |
| H | -2.94778231237402 | -0.90772403280258 | 1.67818544615328 |
| H | -3.32569869636080 | 1.26513769769886 | 1.53694257896985 |
| H | -4.53502294878704 | 0.80378131403099 | 2.43976106509455 |
| H | -2.90955656731263 | -2.87336816147170 | -1.06967558161212 |
| H | -4.17790453508194 | -3.78031114702449 | -1.11222879665652 |
| H | -0.99485351191413 | -1.48019512945326 | -1.65476456863069 |
| H | -1.46475168261945 | -1.72908892506611 | -0.17795994900769 |
| H | -1.35054182848016 | 2.43070871665791 | 1.57723844242431 |
| H | -1.96135821859647 | 2.23344864575863 | 0.15477489008243 |
| H | 0.67043503679516 | 0.00873207771491 | -2.36854117535519 |
| H | -0.17446180359118 | -0.18488137383133 | -3.68009986999087 |
| H | 2.48348830844502 | 0.24701184855075 | 2.56451834533833 |
| H | 1.10054715868972 | -0.28354427874954 | 3.07186812107978 |
| H | 5.21410276476818 | 0.55271577173257 | -1.33544990521272 |
| H | 3.70295182928966 | 0.24533556777735 | -1.06911916987961 |
| H | 4.48369009253803 | 0.07060738651543 | 1.08322175004442 |
| H | 4.80153439028389 | 0.07106508975494 | 2.61611521146088 |
| H | 2.22089975654667 | -2.41460740074327 | 0.11907896717875 |
| H | 0.68604055533336 | -2.39763530154843 | -0.33939417105937 |
| H | 4.40983811629854 | -2.03110467994482 | -0.36877809357365 |
| H | 4.15023866370964 | -2.23243234411373 | 1.10772582999114 |
| H | 0.88620790123723 | 1.88982315243337 | 2.51627167055542 |
| H | 0.57006413841532 | 3.39412069320034 | 2.81818497581411 |
| H | 1.69220686401466 | -0.33640780025255 | -0.56796401473915 |
| N | 3.61844642182140 | -0.10480695800741 | 5.77203389538424 |
| N | 4.32818097533893 | 0.94603053980202 | 4.81908829180446 |
| C | 3.09327628887707 | 0.91010583448390 | 5.18467766906859 |

W18-c..NCN

| | | | |
|---|---|---|---|
| H | 1.66881223910920 | 1.22110712813571 | -0.69713831438451 |
| O | 1.06146264365416 | 2.75644052827330 | -0.27759607975167 |
| O | 0.43381705272548 | 2.43660116903957 | 2.43675126718199 |
| O | 1.67608073354357 | 0.18917532271212 | 3.08677268944364 |
| O | 0.17669478011183 | -2.06449512633143 | 2.67388909381340 |
| O | 4.61716315669000 | -0.06976751451447 | -0.72074536980010 |
| O | 4.23083393324708 | -0.59791292366044 | 1.98608499506699 |
| O | 3.96385223865891 | -2.83029481345332 | 0.07296550567836 |
| O | 1.33336325943308 | -2.06298710978695 | 0.07996820211559 |
| O | 1.97288927166596 | 0.36180250453552 | -1.08513345812999 |
| O | -0.19420709220500 | 0.01031453290109 | -2.67531931153757 |
| O | -1.20903681691376 | -2.14087000513946 | -1.15450547817028 |
| O | -3.85035324787929 | -2.92057452393022 | -0.99892744010231 |
| O | -1.53832867333081 | 2.19599806568683 | -1.40925144880427 |
| O | -2.16037363007385 | 2.18976431545223 | 1.31297778528029 |
| O | -3.96766303050533 | 0.35609058612265 | 1.75519495946059 |
| O | -2.40583727699686 | -1.87243405687035 | 1.37361129473181 |
| O | -4.03538553923070 | 1.84845496872472 | -2.46952600897474 |
| O | -4.90987933394258 | -0.20826287222161 | -0.89139812461783 |
| H | -4.77243354156576 | 2.45234407184671 | -2.56325917269288 |
| H | -4.37062546548379 | 1.08232019390785 | -1.95376989360744 |
| H | -4.70713670469022 | -0.00108680413456 | 0.03493109195406 |
| H | -4.64893874494553 | -1.13268746218207 | -1.01988648289108 |
| H | -0.76804247994074 | -2.04413323323437 | 2.46313457700925 |
| H | 0.60607918953077 | -2.21788112470214 | 1.81585187192826 |
| H | 0.20163646853044 | 2.88963068283953 | -0.70396692601504 |
| H | 0.89550717190422 | 2.80861276117138 | 0.68036577308421 |
| H | -2.42782384398816 | 2.28235633566007 | -1.80992248477548 |
| H | -1.14957558804124 | 1.42334379013516 | -1.85038829207877 |
| H | -3.02328481364240 | -2.53719170304712 | 1.04584890015544 |
| H | -2.95833513162794 | -1.10020816802496 | 1.61001048036675 |
| H | -3.30090958243568 | 1.09158591359787 | 1.65114125753831 |
| H | -4.51398566957325 | 0.56703224668043 | 2.51328520485674 |
| H | -2.93002142757714 | -2.80357337879202 | -1.30439584224708 |
| H | -4.21464629075450 | -3.67477651908479 | -1.46369980599638 |
| H | -0.99461065929966 | -1.39169598432947 | -1.73764904323044 |
| H | -1.48365264264458 | -1.77074138825292 | -0.29438387000384 |
| H | -1.32102788474562 | 2.22928347753548 | 1.80064016033164 |
| H | -1.92819955798733 | 2.16971358188481 | 0.36450863284992 |
| H | 0.70109892290729 | 0.13574274412341 | -2.28676323248903 |
| H | -0.11450400215797 | 0.06327282901421 | -3.62890798833369 |
| H | 2.49180275070168 | -0.07907105840301 | 2.63419670861457 |
| H | 1.08151604087674 | -0.59216327597048 | 3.06073418545105 |
| H | 5.21722930682493 | 0.46567007878253 | -1.24098985341069 |
| H | 3.69962832092634 | 0.18696933111437 | -0.97279041142743 |
| H | 4.51428828476139 | -0.22108189336289 | 1.13313424889604 |
| H | 4.86060576963615 | -0.30859584592502 | 2.65107135998514 |
| H | 2.18055425489344 | -2.53928688172708 | -0.03896732415963 |
| H | 0.64947125695112 | -2.44044728487111 | -0.49479797080149 |
| H | 4.37085351846525 | -2.14946532537957 | -0.47984073712926 |
| H | 4.11938703880876 | -2.50551161363521 | 0.97001118492550 |
| H | 0.89601467226167 | 1.59118947138098 | 2.68096251534456 |
| H | 0.56464571881016 | 3.04099016686396 | 3.17035698993403 |
| H | 1.68891248183633 | -0.37399091705063 | -0.50279844513839 |
| N | 2.58208475383690 | 2.84384578902052 | 4.89273186096357 |
| N | 4.48809722185435 | 1.38189818243104 | 4.45901775408044 |
| C | 3.52892517072699 | 2.10067095398775 | 4.67979853488856 |

W18-c..HNNC from W18-c..CNN+H

| | | | |
|---|---|---|---|
| H | 1.75362 | 1.14070 | -0.96466 |
| O | 1.22360 | 2.68386 | -0.54718 |
| O | 0.81074 | 2.50092 | 2.22506 |
| O | 2.02093 | 0.17328 | 3.06477 |
| O | 0.43725 | -1.95350 | 2.60582 |
| O | 4.65782 | -0.17106 | -0.98892 |
| O | 4.45602 | -0.56503 | 1.75196 |
| O | 4.03281 | -2.90316 | -0.11716 |
| O | 1.40981 | -2.10467 | -0.03442 |
| O | 2.02188 | 0.26090 | -1.33803 |
| O | -0.26633 | -0.09751 | -2.75841 |
| O | -1.20866 | -2.15746 | -1.07487 |
| O | -3.84601 | -2.88002 | -0.66597 |
| O | -1.46107 | 2.15881 | -1.44968 |
| O | -1.86975 | 2.25193 | 1.31197 |
| O | -3.69801 | 0.50467 | 1.95437 |
| O | -2.21557 | -1.76246 | 1.52499 |
| O | -4.03629 | 1.84777 | -2.31145 |
| O | -4.83170 | -0.13285 | -0.59863 |
| H | -4.76496 | 2.46594 | -2.37132 |
| H | -4.34757 | 1.10744 | -1.74592 |
| H | -4.56451 | 0.10373 | 0.30389 |
| H | -4.60479 | -1.06812 | -0.70873 |
| H | -0.52214 | -1.92796 | 2.47018 |
| H | 0.79612 | -2.15394 | 1.72426 |
| H | 0.33906 | 2.83405 | -0.91336 |
| H | 1.13775 | 2.77457 | 0.41824 |
| H | -2.37604 | 2.26028 | -1.78421 |
| H | -1.12822 | 1.36577 | -1.89971 |
| H | -2.87052 | -2.41997 | 1.26054 |
| H | -2.73082 | -0.97084 | 1.78216 |
| H | -3.01741 | 1.21285 | 1.77664 |
| H | -4.18105 | 0.75986 | 2.74135 |
| H | -2.95672 | -2.79984 | -1.06046 |
| H | -4.26389 | -3.65002 | -1.05359 |
| H | -1.02833 | -1.43947 | -1.70693 |
| H | -1.41549 | -1.74362 | -0.21605 |
| H | -1.00088 | 2.32844 | 1.73868 |
| H | -1.70541 | 2.19480 | 0.35133 |
| H | 0.66037 | 0.02044 | -2.45216 |
| H | -0.26841 | -0.07221 | -3.71652 |
| H | 2.82106 | -0.08439 | 2.57346 |
| H | 1.38210 | -0.57394 | 2.96013 |
| H | 5.25482 | 0.34671 | -1.53053 |
| H | 3.73757 | 0.08137 | -1.24048 |
| H | 4.63948 | -0.26116 | 0.84197 |
| H | 5.22365 | -0.33655 | 2.27924 |
| H | 2.23870 | -2.60442 | -0.17725 |
| H | 0.68436 | -2.48468 | -0.55452 |
| H | 4.41877 | -2.22825 | -0.69248 |
| H | 4.23934 | -2.58292 | 0.76980 |
| H | 1.24834 | 1.67244 | 2.52281 |
| H | 1.00074 | 3.16259 | 2.89862 |
| H | 1.75269 | -0.44586 | -0.71535 |
| N | 2.20172 | 2.28542 | 5.41292 |
| N | 2.58386 | 1.11220 | 5.76064 |
| C | 1.86790 | 3.38710 | 5.15057 |
| H | 2.42734 | 0.52047 | 4.92139 |

W18-c..HCNN from W18-c..CNN+H

| | | | |
|---|---|---|---|
| H | 1.68025006855680 | 1.18751598147843 | -0.81513791678496 |
| O | 1.11334292143279 | 2.73712607651203 | -0.45300396341573 |
| O | 0.52009519562148 | 2.51410545580332 | 2.26787885866711 |
| O | 1.66883640503111 | 0.24228763928925 | 3.11021665861988 |
| O | 0.16516500896623 | -2.01732218744247 | 2.73833213013254 |
| O | 4.60563974450403 | -0.08908528466774 | -0.69515801817019 |
| O | 4.26504511161879 | -0.44500185736140 | 2.05028395873928 |
| O | 3.97596237641868 | -2.79460131088559 | 0.22794287622433 |
| O | 1.33978688668813 | -2.04529334408589 | 0.14850671559612 |
| O | 1.97608793725056 | 0.30644288509326 | -1.16256466255889 |
| O | -0.22707690315868 | -0.09415037641879 | -2.68715793784489 |
| O | -1.22401973754518 | -2.17194301885156 | -1.05981821412614 |
| O | -3.88210466215143 | -2.88100469526745 | -0.84916182736437 |
| O | -1.51782520391639 | 2.17836288531118 | -1.50754731175978 |
| O | -2.10474086040065 | 2.32624573060287 | 1.22396190756732 |
| O | -3.90430862962121 | 0.51302265149449 | 1.75926408365602 |
| O | -2.40960503617619 | -1.77213143223032 | 1.46029606600909 |
| O | -4.03378606268814 | 1.82759638484825 | -2.52619491119103 |
| O | -4.90235265378123 | -0.14936271129699 | -0.84503095829071 |
| H | -4.76577293594861 | 2.43584599240395 | -2.63067227924276 |
| H | -4.36716197418209 | 1.08806713659191 | -1.97200728896608 |
| H | -4.67536156304530 | 0.09390151429125 | 0.06695572292547 |
| H | -4.65477637435414 | -1.08147076756966 | -0.93924415737687 |
| H | -0.77893763431436 | -1.97343613557193 | 2.52661714260012 |
| H | 0.59286757456870 | -2.17259832833348 | 1.87983873977562 |
| H | 0.24907125457644 | 2.87571590543859 | -0.86829856820231 |
| H | 0.96101712726668 | 2.81302948189707 | 0.50692821454092 |
| H | -2.41009993220756 | 2.26068769388301 | -1.90243711588151 |
| H | -1.14766179011222 | 1.37990684189529 | -1.91730969899754 |
| H | -3.04965087752511 | -2.42293792971654 | 1.14724463826678 |
| H | -2.93386013253582 | -0.97313799293451 | 1.67097336226562 |
| H | -3.23531983702685 | 1.24056951361517 | 1.61457002381551 |
| H | -4.44432912808874 | 0.76499433143655 | 2.50934203402367 |
| H | -2.96189944494139 | -2.79862756660355 | -1.16695063258299 |
| H | -4.26331885698711 | -3.65336736137445 | -1.26837242370190 |
| H | -1.01097424625706 | -1.44885164833441 | -1.67573795432634 |
| H | -1.48177141198769 | -1.76490331016087 | -0.21150167606274 |
| H | -1.26156957167527 | 2.39398040909259 | 1.70285358222270 |
| H | -1.88051795809650 | 2.26019889957184 | 0.27605919409332 |
| H | 0.67985544955981 | 0.03722132386541 | -2.32884312538596 |
| H | -0.17206958445354 | -0.08453732271676 | -3.64398007717483 |
| H | 2.50391854239146 | -0.01984343420840 | 2.69014532644636 |
| H | 1.09132332496131 | -0.55166329661136 | 3.09825552169251 |
| H | 5.21657156743508 | 0.42456474825299 | -1.22480603708145 |
| H | 3.69360735278286 | 0.14478211649609 | -0.98843395673596 |
| H | 4.48915117662284 | -0.12558624462660 | 1.15646797950304 |
| H | 4.91181082113381 | -0.06771948664991 | 2.65067185911226 |
| H | 2.19354434074973 | -2.51894991714489 | 0.07402543288055 |
| H | 0.66848807011011 | -2.46123635384134 | -0.41336961700475 |
| H | 4.38935859003034 | -2.12485631201251 | -0.33459092083579 |
| H | 4.14783231489757 | -2.46854464250498 | 1.12054087809604 |
| H | 0.91255718305562 | 1.64236876410902 | 2.52510210944356 |
| H | 0.79804966050897 | 3.11632259832359 | 2.96768207338851 |
| H | 1.68736447440996 | -0.39468336601031 | -0.54142497355302 |
| N | 3.32758201513521 | 2.28915229463278 | 4.67802670044174 |
| N | 4.39307341785771 | 2.07307371294722 | 4.33946308262815 |
| C | 2.13505533405324 | 2.65990554307275 | 4.92257939941332 |
| H | 1.62151379398235 | 2.02816918568449 | 5.64596064093194 |

W18-c..c-HNNC from W18-c..HNNC
| | | | |
|---|---|---|---|
| H | 1.73963819057643 | 1.17093888651993 | -0.91338693014286 |
| O | 1.18596102887260 | 2.71225882833424 | -0.49969561852381 |
| O | 0.75055084101083 | 2.44086185254953 | 2.26532891781900 |
| O | 1.95786034022637 | 0.06575799946974 | 3.07263975337127 |
| O | 0.41324491488748 | -2.05854313541532 | 2.60904990938197 |
| O | 4.65569707047860 | -0.13372523919271 | -0.86491486661516 |
| O | 4.41809592498772 | -0.62208447220588 | 1.86091928168987 |
| O | 4.02363248840019 | -2.88721348951043 | -0.11388267488139 |
| O | 1.40453104541161 | -2.09256539100638 | -0.02402665892537 |
| O | 2.02709905897324 | 0.29807172108186 | -1.28711731498249 |
| O | -0.23500634526714 | -0.05438742532795 | -2.75736363309359 |
| O | -1.19522027336729 | -2.13080634120966 | -1.10338169593605 |
| O | -3.81946791967352 | -2.90251463352022 | -0.70619157164229 |
| O | -1.48038703111641 | 2.16328686670111 | -1.42893648280296 |
| O | -1.92554096470146 | 2.24490894810428 | 1.32740959354306 |
| O | -3.72564734570019 | 0.45845310718039 | 1.94536034994480 |
| O | -2.21634227722334 | -1.78730917252012 | 1.50242704636672 |
| O | -4.04108237985503 | 1.83242739366686 | -2.31889624111819 |
| O | -4.83508057348664 | -0.16365889695949 | -0.62293274759350 |
| H | -4.77388229839909 | 2.44504828015756 | -2.38547272493460 |
| H | -4.35341385223787 | 1.08726646984419 | -1.76046376729764 |
| H | -4.57541482368390 | 0.06598325483309 | 0.28367130675106 |
| H | -4.60294926687542 | -1.09711747073980 | -0.73717197813203 |
| H | -0.54420143346286 | -2.00309259683151 | 2.46446043917312 |
| H | 0.77626803767712 | -2.22622240975076 | 1.72183977950551 |
| H | 0.30187540994736 | 2.85828383084983 | -0.86881056078820 |
| H | 1.09487210853108 | 2.80455752355242 | 0.46444967481397 |
| H | -2.39292156265616 | 2.25435605790396 | -1.77365946390322 |
| H | -1.12736729690591 | 1.38372266737137 | -1.88685716535603 |
| H | -2.85958100133962 | -2.44904668151784 | 1.21946131174011 |
| H | -2.74597486272811 | -1.00729623119191 | 1.76638337972363 |
| H | -3.05861734966866 | 1.18155223366905 | 1.77625826305631 |
| H | -4.22268328013168 | 0.70080771667197 | 2.72768819462786 |
| H | -2.93293925465349 | -2.80548915333402 | -1.10268465841051 |
| H | -4.22882489810613 | -3.67332197350944 | -1.10127802214827 |
| H | -1.00952315344995 | -1.40845210321049 | -1.72866758166265 |
| H | -1.41342014177013 | -1.72386253709255 | -0.24433738658813 |
| H | -1.06133468284746 | 2.30400158472390 | 1.76615149874133 |
| H | -1.74896861612310 | 2.19452756349499 | 0.36851847006796 |
| H | 0.68650451829620 | 0.06404154674624 | -2.43744865145226 |
| H | -0.22506726649406 | -0.01214995966492 | -3.71485743487285 |
| H | 2.78688603717777 | -0.19478244990030 | 2.62947629976925 |
| H | 1.33382974922113 | -0.70120192791659 | 2.99176425908017 |
| H | 5.26855236269951 | 0.39466502930992 | -1.37797234961080 |
| H | 3.74371111276500 | 0.12290491611586 | -1.14015497094576 |
| H | 4.59218036980973 | -0.26843810390648 | 0.96720267673816 |
| H | 5.17477799510521 | -0.39137968653005 | 2.40332508397512 |
| H | 2.23825684397318 | -2.58200472887653 | -0.17779001442247 |
| H | 0.68429116328401 | -2.46362938615265 | -0.55803330542959 |
| H | 4.41884016248335 | -2.17944151415474 | -0.64207311246974 |
| H | 4.22639264212928 | -2.62716073573658 | 0.79299140742450 |
| H | 1.16339199579263 | 1.57635799941993 | 2.46727299781407 |
| H | 0.99711991180192 | 3.01544738241023 | 3.00113107644209 |
| H | 1.74389470358628 | -0.41884072799620 | -0.68283012492950 |
| N | 2.09488992394778 | 2.98295159349079 | 4.92220412524675 |
| N | 2.30819815414515 | 1.65759674036737 | 5.25349124095462 |
| C | 3.23946687369366 | 2.65393691318050 | 5.52822878121461 |
| H | 2.20004849433232 | 0.87417189026037 | 4.57103242313461 |

W18-c..HNCN from W18-c...c-HNNC

| | | | |
|---|---|---|---|
| H | 1.76413012694520 | 1.14346774939536 | -0.96940964667720 |
| O | 1.23083257934858 | 2.68018543865633 | -0.53137722007764 |
| O | 0.82662467502880 | 2.44422268286397 | 2.23883113542376 |
| O | 2.05859199577293 | 0.11972115637463 | 3.11518863246960 |
| O | 0.46365070943439 | -1.97643712785859 | 2.58895988548579 |
| O | 4.66866222383993 | -0.17644768766428 | -0.95797613380344 |
| O | 4.46999969341471 | -0.63538882168661 | 1.77808384446117 |
| O | 4.01426721944260 | -2.91475822219554 | -0.13846408559118 |
| O | 1.40366574756383 | -2.09413179457644 | -0.06119763796925 |
| O | 2.03962853998711 | 0.26903075518221 | -1.34967151831913 |
| O | -0.24499716848882 | -0.08181813793622 | -2.78219924416155 |
| O | -1.21276493433698 | -2.13336325724720 | -1.09701030345468 |
| O | -3.83931543461506 | -2.88162792658139 | -0.66548702877291 |
| O | -1.44885599467050 | 2.15276671159919 | -1.44605808792372 |
| O | -1.85248719428120 | 2.22034440989961 | 1.31568696814802 |
| O | -3.69155723246951 | 0.48990120690094 | 1.97002736305309 |
| O | -2.19194003604777 | -1.76145540142353 | 1.51739964935513 |
| O | -4.02952545162996 | 1.84328771164672 | -2.28879507428221 |
| O | -4.83452359606355 | -0.13688713775623 | -0.57979698148716 |
| H | -4.75361526271542 | 2.46694740926183 | -2.34793670441987 |
| H | -4.34657563959822 | 1.10488091196434 | -1.72391353149351 |
| H | -4.56563254363151 | 0.09586370693273 | 0.32332901316146 |
| H | -4.60816950966302 | -1.07217681856324 | -0.69143283481160 |
| H | -0.49661391865087 | -1.93035517362066 | 2.46579600595752 |
| H | 0.80842635758814 | -2.14870760396060 | 1.69555970524400 |
| H | 0.34392989347890 | 2.83488020071676 | -0.88963697514089 |
| H | 1.15131469286661 | 2.76095074374140 | 0.43534196236906 |
| H | -2.36714542033675 | 2.24922489674790 | -1.77345413402584 |
| H | -1.11004658270217 | 1.36870887813557 | -1.90706454678947 |
| H | -2.83978031146902 | -2.42878389662529 | 1.26017523208039 |
| H | -2.71718631666304 | -0.97904962856871 | 1.78255206622348 |
| H | -3.00941004624311 | 1.19560611245419 | 1.78851138622625 |
| H | -4.16672210500720 | 0.74336654020377 | 2.76231202888281 |
| H | -2.95541575776125 | -2.79250407497880 | -1.06967062972997 |
| H | -4.25944515169569 | -3.64720580137591 | -1.05940307711849 |
| H | -1.02560068427688 | -1.41792422750784 | -1.72963628205781 |
| H | -1.41791099370208 | -1.71836173004337 | -0.23838658097359 |
| H | -0.98153132075265 | 2.27626182397715 | 1.74157330171562 |
| H | -1.69168020691150 | 2.17047304531357 | 0.35399827276307 |
| H | 0.68164794837258 | 0.03508642217072 | -2.47615908711603 |
| H | -0.24881518536200 | -0.04616467115366 | -3.73994819724310 |
| H | 2.87100709926036 | -0.12909614845871 | 2.63963613641481 |
| H | 1.43323602273366 | -0.63957245870558 | 3.00699432876766 |
| H | 5.27394919603620 | 0.34399477575851 | -1.48777568968254 |
| H | 3.75285691350662 | 0.08155505556223 | -1.22048895667799 |
| H | 4.63225243161584 | -0.29223763598746 | 0.87805714419048 |
| H | 5.26577699324668 | -0.46691327834326 | 2.28596263929245 |
| H | 2.22792501961250 | -2.60116135499446 | -0.20751567398140 |
| H | 0.67012540204648 | -2.47197911098747 | -0.57187681401153 |
| H | 4.41201421948399 | -2.23555909959633 | -0.70048968400877 |
| H | 4.21033731435002 | -2.60490169725685 | 0.75477700288100 |
| H | 1.26490961460138 | 1.61117768439354 | 2.51699692303170 |
| H | 1.00952569659566 | 3.08602833712399 | 2.93526333581179 |
| H | 1.75771727732814 | -0.44538939221946 | -0.74101855442921 |
| N | 1.70502047456988 | 3.57155821376998 | 5.06306361645290 |
| N | 2.41567184703163 | 1.31268174400814 | 5.69767116568899 |
| C | 2.04308982496515 | 2.47149803315439 | 5.34715598120395 |
| H | 2.33617557197728 | 0.66009518306397 | 4.89319402197497 |

W18-c..HNCNH from W18-c..HNCN + H

| | | | |
|---|---|---|---|
| H | 1.77240329026294 | 1.12678776384821 | -0.87090340152508 |
| O | 1.25439896102910 | 2.66428739560276 | -0.41213804274681 |
| O | 0.79942636972248 | 2.37909139993812 | 2.33998500838423 |
| O | 1.97860400285481 | 0.01683551505115 | 3.11942477149133 |
| O | 0.38983996716937 | -2.12305336575220 | 2.62687222472256 |
| O | 4.67502505109978 | -0.19794681493839 | -0.91983931557681 |
| O | 4.42470093361503 | -0.69113093992014 | 1.80691731951462 |
| O | 4.01307395877509 | -2.93879176358641 | -0.10552804598911 |
| O | 1.39843254129589 | -2.12969862199067 | -0.01727544115530 |
| O | 2.03708354664202 | 0.25622341970973 | -1.26691413855281 |
| O | -0.23194584071165 | -0.03654334746550 | -2.72206510561846 |
| O | -1.20659615579823 | -2.13714994353256 | -1.10838846507836 |
| O | -3.85142941475670 | -2.84920599503613 | -0.76705918104562 |
| O | -1.42777485761913 | 2.19249098112284 | -1.37082659669018 |
| O | -1.86222098426597 | 2.21490387102721 | 1.38710238307429 |
| O | -3.70217288254619 | 0.45652391704488 | 1.96004960475290 |
| O | -2.23968201959293 | -1.81908608828788 | 1.49299383117340 |
| O | -3.99777589611435 | 1.91337051554512 | -2.26333887041960 |
| O | -4.82770581962351 | -0.10623238466634 | -0.61436546032476 |
| H | -4.72195748886998 | 2.53691759727831 | -2.32125768175732 |
| H | -4.32342595667092 | 1.16097905179196 | -1.72225801766570 |
| H | -4.55955016986731 | 0.10419662042119 | 0.29440127889642 |
| H | -4.60235496862012 | -1.03913212136546 | -0.74796371685519 |
| H | -0.56474667596706 | -2.05652668215075 | 2.47534676779179 |
| H | 0.75947953974213 | -2.26766155227839 | 1.73932998659541 |
| H | 0.37469186963550 | 2.84061226011946 | -0.77734683323456 |
| H | 1.16186322167689 | 2.71957438895545 | 0.55594782300381 |
| H | -2.33926925980038 | 2.30122097703381 | -1.71211298375748 |
| H | -1.08863018089359 | 1.41236655051185 | -1.83859487169499 |
| H | -2.89375247997216 | -2.46596977137218 | 1.20196845227562 |
| H | -2.75524261544966 | -1.03175460537046 | 1.76148828312303 |
| H | -3.01888215549937 | 1.16823379746878 | 1.80554574019841 |
| H | -4.19524786447580 | 0.69484004705202 | 2.74593644352461 |
| H | -2.95261780152375 | -2.75985259967155 | -1.13831221159805 |
| H | -4.26587006976259 | -3.59939557925196 | -1.19491363412046 |
| H | -1.00695591022574 | -1.40311123055489 | -1.71577643963902 |
| H | -1.42014556203827 | -1.74706414652905 | -0.24024387511033 |
| H | -0.99310254424780 | 2.24526255050982 | 1.82112368041361 |
| H | -1.69250363445728 | 2.18740114422449 | 0.42586877751198 |
| H | 0.69146195522075 | 0.06311154833543 | -2.39805914185772 |
| H | -0.21797429569330 | 0.01066955829359 | -3.67919154303674 |
| H | 2.76922189403030 | -0.23427604499379 | 2.61117811362240 |
| H | 1.35595205950036 | -0.74447474711159 | 3.05272222453581 |
| H | 5.27717380072037 | 0.33719058365262 | -1.43827147512295 |
| H | 3.75821289713509 | 0.06552609888138 | -1.16994891682899 |
| H | 4.60872767528903 | -0.31998564240363 | 0.92313000228918 |
| H | 5.16104323263752 | -0.44720224477523 | 2.37098949187035 |
| H | 2.22986483116993 | -2.62449091572334 | -0.16699089159774 |
| H | 0.67937404575771 | -2.49841901361835 | -0.55354790456458 |
| H | 4.41369009943203 | -2.26308899359735 | -0.66952045461009 |
| H | 4.21500170530661 | -2.62673044885800 | 0.78623463422737 |
| H | 1.21318156845410 | 1.52528578973120 | 2.59991176274270 |
| H | 1.01792383012688 | 3.01278944641088 | 3.03555070655466 |
| H | 1.75660819290167 | -0.46518959407948 | -0.66585876065730 |
| N | 1.70638875786087 | 3.64737509109275 | 5.06537594308931 |
| N | 3.00732084976004 | 1.60784435556690 | 5.30382995924870 |
| C | 2.30585607070898 | 2.58807320929573 | 5.22388728873943 |
| H | 2.70030511304108 | 0.78177918376533 | 4.77989102414690 |
| H | 0.98976776038940 | 3.88448778139870 | 5.73729005051724 |

W18-c..NH₂CN from W18-c..HNCN + H
| | | | |
|---|---|---|---|
| H | 1.76044626356578 | 1.14825808667545 | -0.90504410119281 |
| O | 1.23387364635795 | 2.67216318876685 | -0.44929633388098 |
| O | 0.79831538876075 | 2.39067172616493 | 2.29779519551907 |
| O | 2.01981569095675 | 0.02828359866682 | 3.14469318375744 |
| O | 0.42469688014517 | -2.07965134191253 | 2.60151465597774 |
| O | 4.66710213929535 | -0.15605487093095 | -0.91284564786198 |
| O | 4.46168972670236 | -0.66279832987579 | 1.81346206021226 |
| O | 4.02996164150961 | -2.90599930360314 | -0.12597164127776 |
| O | 1.41031784452359 | -2.11356485985905 | -0.03576427544067 |
| O | 2.03535843588627 | 0.27679264233220 | -1.29553204633833 |
| O | -0.23100905962250 | -0.04774615331927 | -2.75315884207152 |
| O | -1.19859691023159 | -2.12391003002191 | -1.10546450124798 |
| O | -3.82889692787131 | -2.87033673747213 | -0.71489686866591 |
| O | -1.44762730391889 | 2.17053811823448 | -1.39635278204957 |
| O | -1.86433444503764 | 2.20814098252220 | 1.36693757429627 |
| O | -3.69890182114142 | 0.46464787454320 | 1.97138636662131 |
| O | -2.21303099015676 | -1.79035202372002 | 1.50576284433362 |
| O | -4.02476480448491 | 1.87469617906270 | -2.26557428745143 |
| O | -4.83772679760810 | -0.13011427204054 | -0.58686996407419 |
| H | -4.74467132552167 | 2.50438696901410 | -2.31001030497939 |
| H | -4.34491379534905 | 1.12831016916615 | -1.71327550648198 |
| H | -4.56415456138400 | 0.09167235689818 | 0.31779559722839 |
| H | -4.60538947463294 | -1.06208522241435 | -0.71263416483267 |
| H | -0.53117156717273 | -2.00813585972109 | 2.45629993421246 |
| H | 0.78657197662343 | -2.23482841119218 | 1.71181714430803 |
| H | 0.35239527876645 | 2.84215436598918 | -0.81300369217441 |
| H | 1.14425797309132 | 2.73266879333060 | 0.51936492781566 |
| H | -2.36303058049028 | 2.26930921825987 | -1.73026591393637 |
| H | -1.10332304405762 | 1.39360421939621 | -1.86502086877784 |
| H | -2.86127416452508 | -2.44567276207291 | 1.22000737409067 |
| H | -2.73667769067948 | -1.00866558327831 | 1.77677771142355 |
| H | -3.01537827474910 | 1.17485088896613 | 1.80431226837283 |
| H | -4.18623728723240 | 0.71233165464801 | 2.75799055777198 |
| H | -2.93937228900457 | -2.77340600941428 | -1.10547125291698 |
| H | -4.24242172738614 | -3.62991072282238 | -1.12683893317422 |
| H | -1.00641933982080 | -1.39910871103919 | -1.72613576432657 |
| H | -1.40707658316038 | -1.72071958550627 | -0.24234800915207 |
| H | -0.99512395370120 | 2.25058183403071 | 1.80051365420824 |
| H | -1.69653201299396 | 2.17562229500835 | 0.40597385256389 |
| H | 0.69194159409932 | 0.06359937443320 | -2.43283771217376 |
| H | -0.22259448417399 | 0.00448866159037 | -3.71005612178176 |
| H | 2.83334342461115 | -0.20247521717650 | 2.66308796118865 |
| H | 1.39966985030663 | -0.73049091709451 | 3.02084385116847 |
| H | 5.26933450295727 | 0.38532999578228 | -1.42474348967737 |
| H | 3.74980602609955 | 0.10033737453502 | -1.17139845403067 |
| H | 4.61748459088328 | -0.29903825625384 | 0.92039366941806 |
| H | 5.22982961410948 | -0.43678891402626 | 2.34116545526857 |
| H | 2.24256860519822 | -2.60538087649976 | -0.18943367658462 |
| H | 0.68900278975888 | -2.48265514893179 | -0.56898194245545 |
| H | 4.42353855538867 | -2.21999933008698 | -0.68265679193412 |
| H | 4.23318848107406 | -2.60566014462586 | 0.76916779387803 |
| H | 1.22585161258746 | 1.54898848360670 | 2.56170215666416 |
| H | 1.01598660283511 | 3.02035097185284 | 2.99881677973307 |
| H | 1.75646237068904 | -0.44317264552877 | -0.69295970067262 |
| N | 1.86356915307712 | 3.50983097389655 | 4.94078533780617 |
| N | 2.33355502001857 | 1.20019124961945 | 5.72529362286300 |
| C | 2.05743739170296 | 2.44143513945814 | 5.33248221462958 |
| H | 2.28495114983835 | 0.53267775979817 | 4.94843830850315 |
| H | 1.88612106958854 | 0.90848188559157 | 6.57927680418155 |